\documentclass[pre,twocolumn,showpacs]{revtex4-1}

\usepackage{amsmath}
\usepackage{amssymb}
\usepackage{graphicx}
\usepackage{color}

\graphicspath{{./}{./figures/}}
\definecolor{labelkey}{cmyk}{.4,.2,0,0}
\definecolor{MidnightBlue}{cmyk}{0.98,0.13,0,0.43}
\definecolor{DarkGreen}{rgb}{0,0.7,0.1}

\newcommand{\fig}[2]{\includegraphics[width=#1]{./figures/#2}}

\newcommand{\Fig}[1]{\includegraphics[width=8.7cm]{./figures/#1}}
\newcommand{\rme}{\mathrm{e}}
\newcommand{\rmd}{\mathrm{d}}

\newcommand{\nn}{\nonumber}

\renewcommand{\log}{\ln}
\newcommand{\remarka}[1] {}

\def\be{\begin{equation}}
\def\ee{\end{equation}}

\def\bea{\begin{eqnarray}}
\def\eea{\end{eqnarray}}

\newcommand{\dif}{{\mathrm d}}

\def\XXint#1#2#3{{\setbox0=\hbox{$#1{#2#3}{\int}$}
\vcenter{\hbox{$#2#3$}}\kern-.5\wd0}}

\date{\today}

\begin{document}

\title{Distribution of velocities and acceleration for a particle in  Brownian correlated disorder: inertial case}

\author{Pierre Le Doussal, Aleksandra Petkovi\'{c}, and Kay J\"org Wiese}

\affiliation{Laboratoire de Physique Th\'{e}orique-CNRS, Ecole Normale Sup\'{e}rieure, 24 rue Lhomond, 75005 Paris, France}

\begin{abstract}
We study the motion of an elastic object driven in a disordered environment in presence of both dissipation and inertia. We consider random forces with the statistics of random walks and reduce the problem to  a single degree of freedom. It is the extension of the mean field ABBM model in presence of an inertial mass $m$. While the ABBM model can be solved exactly, its extension to inertia exhibits complicated history dependence due to oscillations and backward motion. The characteristic scales for avalanche motion are studied from numerics and qualitative arguments.
To make analytical progress we consider two variants which coincide with the original model whenever the particle moves only forward. Using a combination of analytical and numerical methods together with simulations, we characterize the distributions of instantaneous acceleration and velocity, and compare them in these three models. We show that for large driving velocity, all three models share the same large-deviation function for positive velocities, which is obtained analytically for small and large $m$, as well as for $m=6/25$. The effect of small additional thermal and quantum fluctuations can be treated within an approximate method.
\end{abstract}

\maketitle

\section{Introduction}

The dynamics of a large class of classical and quantum systems can be modeled within the description of an elastic manifold driven by an applied external force through a disordered medium \cite{Fisher,brazovski,LeDoussalGiamarchi1997}. Some examples are domain walls in magnetic systems in the presence of time-dependent magnetic fields \cite{review}, flux-line lattices in type-II superconductors driven by an applied transport current \cite{fluxlines}, charge-density waves in solids in an
electric field \cite{Gruner1994,cdw}, pinned or driven Wigner crystals
\cite{Perruchot2000,Reichhardt2001,ChitraGiamarchi2004,CugliandoloGiamarchiLeDoussal2006},
dislocations in metals \cite{dislocations}, interface between two fluids in a porous medium \cite{fluids}, earthquakes \cite{Fisher}, and crack fronts in brittle materials \cite{cracks}. In all these systems, the competition between elastic forces, quenched disorder and external driving shapes the dynamics. As a result, the response is usually complicated.

If the driving force is sufficiently small, the system is trapped due to disorder in a metastable state. When increasing the external driving, some weakly pinned parts will start moving. They will be stopped by elastic forces that describe interactions between weakly and strongly pinned regions of the manifold. Further increase of the driving usually results in jumps of a segment of the system, and avalanche motion.

One example occurs in soft magnets. When smoothly increasing  the magnetic field ($H$), the magnetization ($M$) changes in an  irregular way. This process can be explained by considering the motion of domain walls separating regions of opposite magnetization. The derivative of the magnetization with respect to the magnetic field, $\partial M/\partial H$, is known as Barkhausen noise and can be related to avalanche motion \cite{UrbachMadisonMarkert1995,KimChoeShin2003}. An important step
towards modeling the dynamics in these systems was made by Alessandro, Beatrice, Bertotti and Montorsi (ABBM).
On a phenomenological basis they introduced \cite{ABBM1,ABBM2} a Langevin equation for the velocity of a single
degree of freedom, i.e.\ a particle, which represents the center of mass of the domain wall. It is simple enough to allow
for an exact solution. Their approach is known as ABBM model. It was successful in explaining the distribution of sizes and the duration of pulses in the Barkhausen signal, both for an extremely small and for a finite increase-rate of the external field \cite{CizeauZapperiDurinStanley1997,ZapperiCizeauDurinStanley1998,DurinZapperi2000,review}.
In the ABBM model, the probability of the instantaneous  domain-wall velocity is found to be $P(\dot{u})\sim \dot{u}^{-\alpha} \exp(-\dot{u}/\dot{u_0})$, where $\alpha=1-v$. Here $v$ is proportional to the rate of increase of the field and $\dot{u}_0$ is some characteristic cutoff. The avalanche sizes $S$ are distributed according to $P(S)\sim S^{-\tau}$ up to some large-scale cutoff, with $\tau=(3-v)/2$. For vanishing rate $v\to 0^+$, different samples of different materials were found to be characterized by universal exponents $\alpha$ and $\tau$, regardless of the specific microscopic details about the sample structure.

In their phenomenological theory, ABBM assumed that the random-force landscape seen by the particle has the long-range correlations of a Brownian motion, while the original disorder seen by the domain wall is of short-ranged nature. This approximation, made in order to explain experiments, turns out to be justified in some cases. First, in the limit of an interface with infinite-ranged interactions (i.e.\ a fully connected lattice model) it was shown that the ABBM model becomes exact, i.e. it describes exactly the center-of-mass motion \cite{ZapperiCizeauDurinStanley1998}. It is  believed to provide a mean-field model, which should be valid in particular to describe domain walls in situations where the long-ranged dipolar forces generate long-ranged elasticity and puts the system at its upper critical dimension \cite{cizeau1}.

Recently, two of us have developed a field theoretic approach to describe and compute avalanche-size distributions and velocity distributions for elastic interfaces of internal dimension $d$ in short-ranged disorder \cite{LeDoussalWiese2008c,LeDoussalWiese2011a,LeDoussalWieseinprep2012}. It was shown that in the quasi-static limit $v\to 0^+$ the velocity of the center of mass is indeed described by the ABBM model at and above the critical dimension $d_c$, with corrections subdominant in the spring constant (parameter $\mu$ below). Deviations become important for $d<d_c$, and were computed in a $d=d_c - \epsilon$ expansion, where $d_c=4$ for short-range elasticity, and $d_c=2$ in case of dipolar forces. The theory also allows to predict the spatial dependence of avalanches \cite{LeDoussalWiese2011a,LeDoussalWieseinprep2012} which cannot be obtained from the ABBM model. It also provides an independent exact solution of the ABBM model at any driving velocity \cite{DobrinevskiLeDoussalWiese2011} based on the Martin-Siggia-Rose (MSR) formalism via the solution to a non-linear saddle point equation, called the instanton equation.

The ABBM model, and the subsequent field theoretical approach, provides a good descriptions of avalanche motion in
classical systems evolving with the simplest over-damped dynamics. One would like to extend these theories to describe elastic
systems with a more general dynamics, including inertial and retardation effects, and to describe avalanche dynamics in quantum systems. Studies of classical models with stress overshoots \cite{SchwarzFisher2001,SchwarzFisher2003} has shown that although the depinning transition may be not too much affected in the thermodynamic limit, the avalanche-size distributions can be quite different.

Retardation effects are important for instance in magnets. Apart from universal power laws discussed above (characterized by the exponents $\alpha$ and $\tau$), pulses of different durations in Barkhausen noise are expected to collapse on the same curve after proper rescaling \cite{ref1}. However, in some experiments on ferromagnetic alloys the pulse shape is found to be asymmetric \cite{spasojevic,durin,ref1}. This asymmetry was explained to be a transient effect of eddy currents \cite{durin2,naturelett}. Namely, the domain wall motion generates eddy currents. The response is not immediate, but instead finite-time delays exist after the corresponding wall displacement is made. These  effects of retardation can be taken into account by introduction of a negative mass of the domain wall \cite{durin2,naturelett}.

Although the above-mentioned effects are important in some samples, inertial effects are always present in the domain wall dynamics. A domain wall is characterized by the so-called D\"{o}ring mass, which is due to gyromagnetic effects \cite{doering}. However,  inertial effects are often neglected with respect to a larger damping present in the system, and simplified models excluding the mass are studied. In many other systems the dynamics is only weakly dissipative and inertial effects can be important. Some examples are geological faults, motion of contact lines of a droplet on a dirty rough surface and crack fronts in brittle materials.
Domain walls with an internal degree of freedom also exhibit a non-trivial dynamics reminiscent of inertial effects \cite{Lecomte2009}.

The description of avalanches in quantum systems with quenched disorder is also a challenge. There is great current interest in non-equilibrium quantum systems and their full counting statistics \cite{BernardDoyon2012,Levitov2002,Beenakker2003,RocheDerridaDoucot2005,Klich2002}. Higher moments of the noise have been measured in avalanche processes and exhibit some resemblance to their classical counterparts \cite{GabelliReulet2009}. Out of equilibrium elastic quantum systems in presence of disorder and a bath have been studied in the thermal,  and quantum-creep regimes \cite{GorokhovFisherBlatter2002,NattermannGiamarchiLeDoussal2003} where the driving force is small and the dynamics is slow and governed by the time scales set by by thermal or quantum tunneling over barriers. For a fixed driving force above the depinning threshold however, there are no barriers. To study avalanches, it is convenient to drive the system with an external spring at fixed but small velocity $v$ \cite{LeDoussalWiese2011a}. Then the effective driving force changes in time, and the spring provides a restoring force that keeps the system near the depinning transition. In the stationary state the system is temporarily pinned, then unpins and jumps to the next metastable state, and so on. In that situation, while thermal or quantum fluctuations may help trigger an avalanche (see e.g.\ p.\ 319 in Refs.\ \cite{review} or \cite{RepainBauerJametFerreMouginChappertBernas2004}) they should be less important during the avalanche process itself, which usually involves much faster time scales than that of barrier crossing. During the avalanche the system is rolling down the potential hill, with possible overshoots and oscillations due to inertia. Thus a semi-classical equation of motion keeping only inertia and damping into account should be a reasonable starting point.

Given these motivations, in this paper we study the ABBM model in presence of inertia. We consider the motion of a particle representing the center-of-mass position of an interface, that is driven by a spring  at velocity $v>0$ in a Brownian-correlated random-force landscape. The feature which makes the ABBM model  solvable is that the motion is always forward. In presence of inertia this property is lost. The disorder thus generates non-trivial correlations in time when the particle visits the same positions several times. To make progress we thus consider two  variants of the model.

One variant is a model ``on a tree", i.e. such that when the particle changes the direction of motion, it experiences a different Brownian disorder. Although it may seem artificial, it could in fact be of  relevance for interfaces since different parts of an interface are exposed to different disorder potentials, and in presence of inertia the backward motion of the center-of-mass does not have to involve the same segments of the system as the forward one. The advantage of this model is that it can again be studied using a Fokker-Planck equation, which however does not appear to be exactly solvable. We determine the joint distribution of velocity $\dot{u}$ and acceleration $a$: (i) in perturbation theory at small and large mass, (ii) for large driving velocity $v, $ and (iii) numerically. We then compare with a numerical solution of the original model, i.e.~the ABBM model with inertia.

The second variant  we call the $\sqrt{\dot u}$ model. It is the model for which the method developed in \cite{LeDoussalWiese2011a} naturally extends. The  non-linear instanton equation is now a differential equation of second order in the time variable. It is the saddle point equation of the MSR action for the ABBM model with inertia under the assumption that the particle moves in the direction of the drive only. We are unable to solve it exactly for generic values of the mass. We solve it: (i) in perturbation at small and large mass, and (ii) for a magic value of the mass where exact solutions exist, related to the Abel equation. It is also easy to solve numerically and from it we obtain, for that model, the Fourier-Laplace transform of the velocity distribution. Also, we calculate exactly the moments characterizing the distribution function ($\overline{a^k\dot{u}^n}$) for arbitrary $v$ and mass. The only unpleasant feature of this model is that due to backward motion, complex velocities appear. As long as they have a small probability, e.g. for large $v$ or small mass, it gives the correct physics. In fact, by comparing with numerics, we find that
this model provides quite interesting approximations to the ABBM model even for not so small values of mass and velocities.

Although the three models, namely the original one, the tree model and the $\sqrt{\dot u}$ model, do correspond to different ways to treat the negative velocities, it appears that they share the same {\it large-deviation function} at positive instantaneous velocity.
The latter describes the large $v$ limit (i.e. driving velocity) of the probability distribution of the (instantaneous) velocity and acceleration. Hence we conjecture that we have obtained in this paper the exact large-deviation function for the ABBM model with inertia in the positive-velocity domain.
This conjecture is explained and argued for in details, and supported by numerics. We find that the large-deviation function is determined by the nonlinear instanton  equation and we obtain its analytical form: (i) in perturbation for small $m$ and for large $m$; (ii) for the magic value of the mass. In addition we discuss for all three models the probability that the particle, starting with  given acceleration and velocity, reaches zero velocity before or at time  $t$ . We refer to the latter as the exit probability.

Although our calculation is performed at zero temperature and $\hbar$, the distribution of velocities and accelerations in an avalanche is expected to be robust, and should survive at low temperatures, as well as in the presence of quantum fluctuations.  More precisely, from the above discussion, it should be valid as as long avalanche durations remain small compared to barrier-crossing time-scales, which is the regime studied in this paper.
A more complete theory however, yet to be worked out, would need to  incorporate several additional effects: (i) the renormalization of disorder by fluctuations; (ii) in the under-damped limit the total avalanche duration may be notably increased as the system oscillates before settling into the next metastable state; (iii) the scale dependence of these effects. Although the present approach is only a first step, it is expected
to capture some of the effects of inertia in classical and quantum avalanches. In particular at the end of the paper we show how to
incorporate some of the thermal and quantum effects in the moving system by studying the $\sqrt{\dot u}$ model in presence of
an additional thermal or quantum noise.

The paper is organized as follows.
First, in Sec.~\ref{sec:ABBM}, we rederive the distribution of velocities for the ABBM model and make the connection with the MSR formalism employed in Ref.~\cite{LeDoussalWiese2011a}. Then we start analyzing inertial effects. In Sec.~\ref{sec:model} we introduce the ABBM model with inertia and  analyze the results of a numerical simulation. Then, in Sec.~\ref{sec:tree} we consider a particle on ``the tree". The model is introduced in Sec.~\ref{sec:deftree}. In Sec.~\ref{sec:perturbation} we solve the corresponding Fokker-Planck equation perturbatively in the inertia, and determine the corresponding probability distribution. The large-$v$ limit is discussed in Sec.~\ref{sec:largev}. In Sec.~\ref{sec:largemass} we solve the Fokker-Planck equation perturbatively in $1/m$ and give a solution  for the $\sqrt{\dot{u}}$-model in the same limit. In Sec.~\ref{sec:numerics} we solve the Fokker-Planck equation numerically, and finally  compare analytical and numerical results in Sec.~\ref{sec:comparison}. Then, in Sec.~\ref{sec:complex} we consider the $\sqrt{\dot{u}}$-model.
We start with the definition and its basic properties in Sec.~\ref{sec:defcomplex}. Then we  connect the instanton  and Fokker-Planck approaches in Sec.~{\ref{sec:instanton}}. In Sec.~\ref{sec:mainmoments},
we calculate exactly the moments characterizing the distribution function.
We solve the instanton equation perturbatively in the mass and from that we find a perturbative expansion of the distribution function in Sec.~\ref{sec:perturbation2}, while  in Sec.~\ref{sec:exactm625} we solve it exactly for the ``magic''  value of the mass. In Sec.~\ref{sec:mcritical} we analyze in more detail one fixed value of the mass.  In Sec.~\ref{sec:largedeviation} we introduce and discuss the large-deviation function. Its perturbative expansion in small and large $m$ is given in Sec.~\ref{sec:smallm} and in Sec.~\ref{sec:largem}, respectively, while in Sec.~\ref{sec:exact} we give its exact result for the magic value of the mass. Supplementary material is  relegated to appendices \ref{app:perturbationtheory} to \ref{s:convergenceF}.

\section{\label{sec:ABBM}ABBM model}

Before we start considering inertial and dissipative effects together, we first review the ABBM model \cite{ABBM1,ABBM2} that neglects inertia, rederive its velocity distribution, and recall the connection to the saddle point (instanton equation) approach of Ref.~\cite{LeDoussalWiese2011a,LeDoussalWieseinprep2012,DobrinevskiLeDoussalWiese2011}.

We study an elastic interface at zero temperature, whose center-of-mass position is given by the equation of motion
\begin{align}\label{eq:ofmotion}
\eta  \dot{u}(t) = F(u(t))+\mu^2 \left[vt-u(t)\right],
\end{align}
where $\dot{u}=\dif u/\dif t$ and $F$ is the disorder force. It is Gaussian distributed with correlations
\begin{align}\label{eq:force}
\overline{[F(u)-F(u')]^2}=2\sigma|u-u'|.
\end{align}
In Eq.~(\ref{eq:ofmotion}), $\eta$ measures dissipation and $v$ is the driving velocity. For the specific realization  in magnetic samples, the interface describes a domain wall and the term $\sim\mu^2 u$ models the demagnetizing field generated by free magnetic charges on the boundary of the sample \cite{review}. In general, this term is  a restoring force and $\mu^2$  the spring constant  by which the particle (representing the center-of-mass) is driven.

For $v>0$, the Middleton theorem \cite{middleton} states that the particle always moves forward in the steady state (and for all $t>0$ if its initial velocity at $t=0$ is positive). The above equations can be solved via the Fokker-Planck equation\begin{align}\label{eq:Fokkerm=0}
\frac{\partial P(\dot{u},t)}{\partial t}=&
-\frac{\partial}{\partial \dot{u}}j(\dot{u},t).
\end{align}
Here $j(\dot{u},t)$ is the probability current
\begin{align}\label{eq:current}
j(\dot{u},t)=& \left( \frac{\mu^2 v}{\eta}-\frac{\mu^2 \dot{u}}{\eta} \right)P(\dot{u},t)-\frac{\partial}{\partial \dot{u}}\left( \frac{\sigma \dot{u}}{\eta^2}P(\dot{u},t)\right).
\end{align}
The first term is the contribution from the drift  and the second one from the diffusion.

Before proceeding further let us recall the main scales for the ABBM model, and introduce the
appropriate dimensionless units to be used in this paper. Times will be measured in units of
the relaxation time of the quadratic well,
\bea\label{eq:time}
\tau_\mu = \frac{\eta}{\mu^2}\ .
\eea
Displacements (i.e. the $u$ direction) will be measured in units of:
\bea \label{space}
S_\mu=\sigma/\mu^4
\eea
where $S_\mu$ gives an estimate of the large-size cutoff for the distribution of avalanche sizes, as
defined in \cite{LeDoussalWiese2008c}. Velocities will be thus measured in terms of a velocity scale
set by the disorder,\bea\label{eq:velocity}
v_\mu = \frac{\sigma}{\eta \mu^2} = \frac{S_\mu}{\tau_\mu}
\ ,\eea
With these units of time and space
the ABBM model contains only one  parameter, the driving velocity $v : = v/v_\mu$
in dimensionless units. Below, we will mostly use these units, keeping the freedom to
restore dimensionfull units when needed.

Let us now discuss the steady state of the ABBM model. In that case $\partial_t P=0$ and $j(\dot{u})=const$. However, from the condition that the particle always moves forward follows that $j(\dot{u})=0$. Solving  Eq.~(\ref{eq:current}) with this constraint, one obtains
(in dimensionless units):
\begin{align}\label{eq:distribution}
P(\dot{u})=\frac{ e^{-\dot{u}}
    \dot{u}^{v-1}}{\Gamma (v)}\theta(\dot{u}).
\end{align}
It is important to note the dramatically different behavior of $\lim_{\dot{u}\to 0}P(\dot{u})$ for $v<1$ and $v>1$. In the former it is divergent while in the latter it tends to zero. The value $v=1$ separates the regime of intermittent motion, where the particle is most of the time at rest, from the regime where it moves continuously.

Now we briefly discuss an alternative way of solving the Fokker-Planck equation and make connection with the approach introduced in Ref. \cite{LeDoussalWiese2011a} based on the instanton equation, and further studied in \cite{DobrinevskiLeDoussalWiese2011}. After performing the Laplace transform we find
\begin{align}
-\frac{\partial \hat{P}}{\partial t}+\lambda \frac{\mu^2 v}{\eta}\hat{P}(\lambda)-\frac{\mu^2}{\eta}\lambda\partial_{\lambda}\hat{P}+
\frac{\sigma}{\eta^2}\lambda^2\partial_{\lambda}\hat{P}+\mbox{BT}=0.
\end{align}
Here $\hat{P}(\lambda)=\int_{0}^{\infty}P(\dot{u})e^{\lambda \dot{u}}\mathrm{d}\dot{u}$ and the boundary terms (BT) are
\begin{align}
\mbox{BT}=& - \left[ j(\dot{u},t)e^{\lambda \dot{u}}\right]\Big|_{0}^{\infty} -\frac{\sigma \lambda}{\eta^2} \left[\dot{u} P(\dot{u})e^{\lambda \dot{u}}\right]\Big|_{0}^{\infty}.
\end{align}
We ignore for the moment the boundary terms and look for a solution of the form
\bea
\hat{P}(\lambda,t) =e^{v Z(\lambda,t)}
\eea
with $Z(0,t)=0$ since $\hat{P}(\lambda=0,t) =1$. After introducing dimensionless quantities $\lambda'={\lambda \sigma}/{\eta \mu^2}$, and $Z'={Z\sigma}/{\eta\mu^2}$ and omitting primes, we find \cite{LeDoussalWiese2011a}
\begin{align} \label{eqZ}
\frac{\partial Z(\lambda,t)}{\partial t}+\frac{\partial Z (\lambda,t)}{\partial \lambda}(\lambda-\lambda^2)=\lambda.
\end{align}
This equation admits a time-independent solution $Z (\lambda,t) = Z(\lambda)$ with $Z(0)=0$:
\bea \label{resold}
Z(\lambda)=-\log(1-\lambda)
\eea
Hence we recover the result of \cite{LeDoussalWiese2011a} for the steady state.
Doing the inverse Laplace transform of $\hat{P}(\lambda)=(1-\lambda)^{-v}$ we obtain Eq.~(\ref{eq:distribution}).
Now, we can check that the boundary terms indeed vanish for $\lambda<1$, i.e.\ in the domain
in which \(Z(\lambda)\) is defined.

In addition one can make the connection to the instanton equation as follows.
The equation (\ref{eqZ}) can be solved by the method of characteristics: Define a
function $\lambda(t)$ which obeys the  differential equation
\begin{align}\label{eq:instanton}
\frac{\dif \lambda(t)}{\dif t}&=\lambda(t) -\lambda^2(t)\ .
\end{align}
Further define $Z(t) := Z(\lambda(t),t)$. Then the total derivative, using (\ref{eqZ}) is
\begin{align}\label{eq:instanton2}
\frac{\dif Z(t)}{\dif t}&=\lambda(t).
\end{align}
The equation (\ref{eq:instanton}) is nothing but the instanton equation
of Ref.\
\cite{LeDoussalWiese2011a}, with here \(\lambda(t)=\tilde u(t)\) there.
It admits the solution
\bea \label{soluabbm}
\lambda(t) =\frac{\lambda}{\lambda+(1-\lambda)e^{-t}}
\eea
with  boundary condition $\lambda(-\infty)=0$. In addition,
\begin{align}\label{eq:Z}
Z(t)=\int_{-\infty}^t \lambda(t')\,  \dif t',
\end{align}
where we have defined $\lambda(0)=\lambda$. Hence
if we express $Z(\lambda) := Z(t=0)$ as a function of $\lambda=\lambda(0)$
we obtain precisely (\ref{resold}). In Ref.\ \cite{LeDoussalWiese2011a,LeDoussalWieseinprep2012} Eqs.~(\ref{eq:instanton}) and (\ref{eq:Z}) were obtained not as
the Laplace transform of the Fokker-Planck equation, but by a completely different route using
the Martin-Siggia-Rose (MSR) dynamical action.
This will be explained in more details below in the   case
where inertial terms are allowed.

Eq. (\ref{eqZ}) is  solved for any initial condition $Z(\lambda,t=0)=Z_0(\lambda)$ as
\cite{LeDoussalWieseinprep2012}\
\be
Z(\lambda,t) = - \ln(1-\lambda+\lambda e^{-t}) + Z_0\left(\frac{\lambda}{\lambda + (1-\lambda) e^t} \right)
\ .\ee
Hence,
\be \label{decay}
\hat P(\lambda,t) = (1-\lambda+\lambda e^{-t})^{-v} \hat P_0\left(\frac{\lambda}{\lambda + (1-\lambda) e^t} \right)
\ .\ee
Using $\overline{\dot{u}^n(t)}=\partial^n_{\lambda}\hat{P}(\lambda,t)|_{\lambda=0}$, we obtain the decay to the steady state after a change in the driving velocity, $v=v_0+\theta(t)(v-v_0)$:
\bea \label{eq:decay}
&& \overline{\dot u(t)} = v (1-e^{-t}) + e^{-t} \overline{\dot u(0)} \\
&& \overline{\dot u(t)^2}^c = v (1-e^{-t})^2 + 2 \overline{\dot u(0)} e^{-t} (1-e^{-t})
+ e^{-2 t} \overline{\dot u(0)^2}^c \nn,
\eea
where the symbol $c$ denotes connected correlation functions. This is in agreement with the results of Ref.~\cite{DobrinevskiLeDoussalWiese2011}, where it was obtained within the MSR approach. Note that for any $t>0$ Eq.~(\ref{decay}) behaves as $\hat P(\lambda,t) \sim A(t) (-\lambda)^v$ with
$A(t) = (1-e^{-t})^{-v} \hat P_0(- \frac{1}{e^t-1})$, hence $P(\dot u,t) \sim A(t) \dot u^{v-1}/\Gamma(v)$
and the current at the origin vanishes, which justifies ignoring the boundary terms above
\footnote{The next order at small $u$ should be examined with the same conclusion.}.

\section{ABBM model with inertia\label{sec:model}}
\subsection{Definition of the model}
In this section, we consider the generalization of the ABBM model to include the effect of inertia.
The equations of motion in the laboratory frame are\begin{align}\label{12}
\frac{\mathrm{d}\dot u(t)}{\mathrm{d} t}&=a(t),\\\label{13}
m\frac{\mathrm{d}a(t)}{\mathrm{d} t}&=\mu^2[v-\dot{u}(t)]-{\eta}a(t)+
\partial_t F\big(u(t)\big),
\end{align}
where $F(u)$  is a function  of $u$, and as for the ABBM model (\ref{eq:force})\begin{equation}
\label{14}
\overline{[F(u)-F(u')]^2}=2\sigma|u-u'|.
\end{equation} In the limit $m\to 0$ the model simplifies to the ABBM model considered
in the previous section.

For later use we note that {\em if} $u(t)$ is monotonously increasing, then the correlator
of the random force \(\partial_t F\big(u(t)\big)\) in eq.\ (\ref{13}) is
\begin{eqnarray}\label{3}
\overline{ \partial_t F\big(u(t)\big)  \partial_{t'} F\big(u(t')\big)} &=&
2 \sigma \dot u(t) \delta(t-t') \nn\\
&=& 2 \sigma |\dot u(t)| \delta(t-t')
\end{eqnarray}
Note that {\em if} $\dot u(t)\ge 0$, both versions with and without the absolute
value are equivalent.
Under this assumption an alternative way to model (\ref{14}) is to replace
\begin{eqnarray} \label{4}
\partial_t F\big(u(t)\big)&\to &  \sqrt{\dot u(t)} \, \xi(t)\nn\\
&=&  \sqrt{ |\dot u(t)|}\, \xi(t)
\end{eqnarray}
with \begin{equation}\label{5}
\left< \xi(t) \xi(t')\right> = 2 \sigma \delta(t-t').
\end{equation}
We will get back to these formulations shortly. First, let us discuss
 the units. Keeping the same units for time and velocity (and space)
as given by (\ref{eq:time}) and (\ref{eq:velocity}), {\it the inertial model depends on two dimensionless parameters}, the driving
velocity $v:=v/v_\mu$ as before, and the dimensionless mass
\bea
m : = \frac{m \mu^2}{\eta^2} = \frac{\tau_m}{\tau_\mu} = \frac{\tau_0^2}{\tau_\mu^2}
\eea
Indeed two new time scales can be defined:
\bea \label{newtime}
\tau_m=m/\eta \quad , \quad \tau_0 = \sqrt{\tau_m \tau_\mu}= \sqrt{m}/\mu
\eea
where $\tau_m$ is the damping time beyond which damping overcomes inertia
and $\tau_0$ is the characteristic oscillation time of the harmonic oscillator
in the absence of damping. These time scales are not independent,
so there is really only one new time scale. The units of acceleration are $S_\mu/\tau_{\mu}^2 = \sigma/\eta^2$.

Below we study the model as a function of these two parameters $m$ and $v$.
Note that there are various limits of interest.
We will in particular study the limit of small and large $m$, as well as the limit of large $v$. The large-$m$ limit can be rephrased as
the limit where disorder $\sigma$ and damping $\eta$ are both small. Note that the weak-disorder
limit is more general since it is such that only $\sigma$  is small.

Note a remarkable property of the ABBM model without inertia: while the space unit (\ref{space}) depends on disorder
the characteristic relaxation time (\ref{eq:time}) remains {\it independent of the disorder}. Although at small $v$
there is a broad distribution of time scales, the characteristic time remains $\tau_\mu$. In the language of RG it
means that the friction $\eta$ is not corrected by disorder \cite{LeDoussalWiese2011a,DobrinevskiLeDoussalWiese2011}: it is a consequence
of the Brownian force landscape which can be proved using the Middleton theorem. In presence of inertia
the oscillation and damping time scales $\tau_0$ and $\tau_m$ given in (\ref{newtime}) are
the bare ones (i.e. in the absence of disorder) and it remains to be understood whether
characteristic oscillation and damping times are affected by disorder.

\subsection{Phenomenology of the inertia model}

The model defined by Eqs.~(\ref{12})-(\ref{14}) is difficult to analyze analytically, since the particle may change its direction of  motion.
Let us therefore start with a numerical simulation
and some qualitative considerations. We first describe how the typical trajectories change as
$m$ is increased, and then we show some numerical results for the distribution of instantaneous velocities.

\subsubsection{Qualitative features of trajectories}

\begin{figure*}
\includegraphics[width=\linewidth]{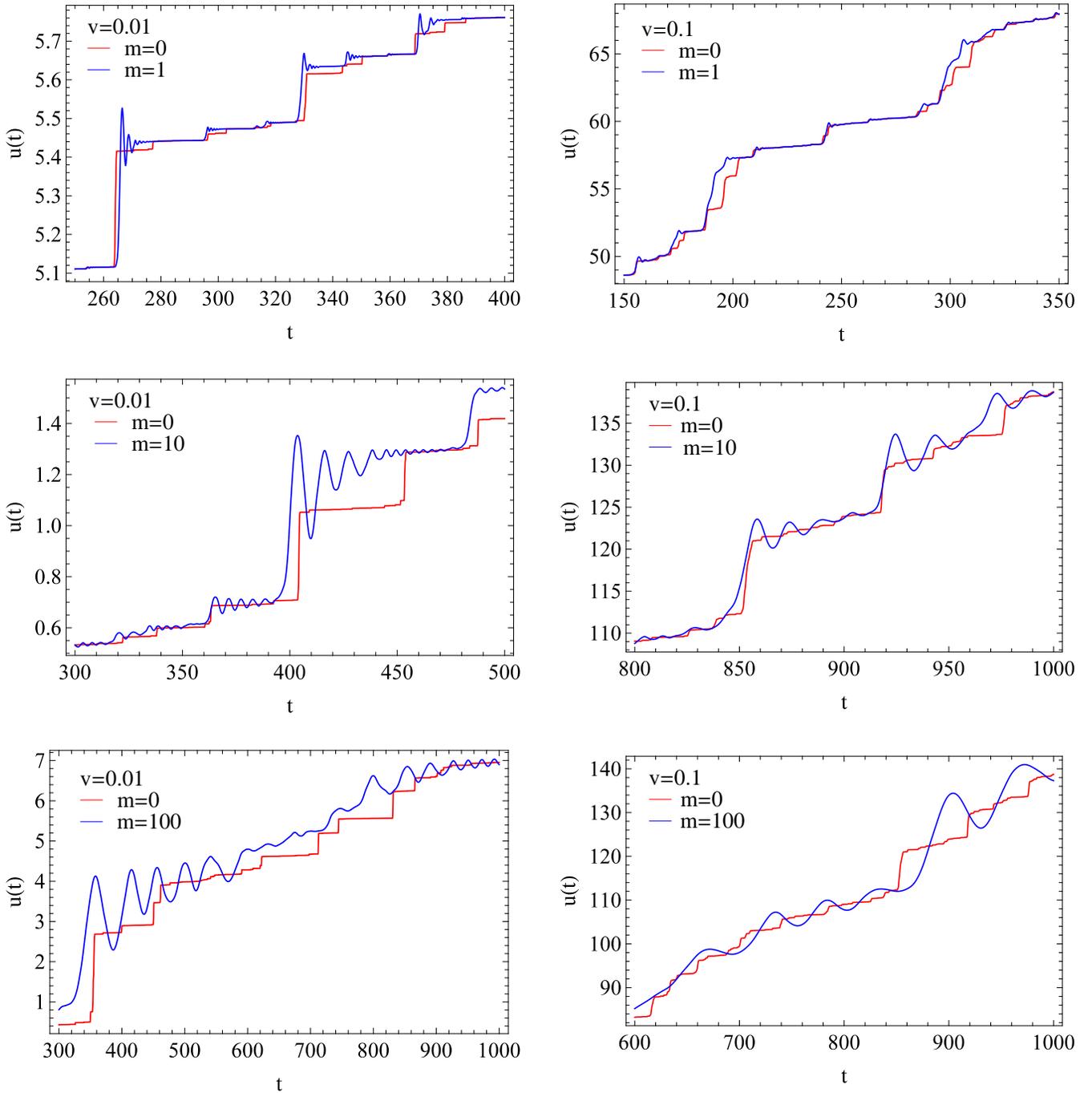}\\
\caption{Each figure shows two trajectories in the same disorder realization and for the same driving velocity, but different values of the mass, as denoted.}
\label{fig:all}
\end{figure*}

Let us start with a numerical simulation of Eqs.~(\ref{12})-(\ref{14}). We show in Fig.~\ref{fig:all} some examples of typical trajectories $u(t)$, for different values of the mass and the driving velocity in the same realization of disorder, such that we can see how the trajectories are correlated with the disorder. In the first set we choose a small driving velocity  $v=0.01$ and in the second
a larger one $v=0.1,$ so we can see the evolution from the avalanche regime (at small
$v$) to the faster driven regime (at larger $v$).

For small $v$ we see that upon increasing the mass (up to moderate values): (i) time windows where the particle is pinned in a metastable state at position $u_i$ (at zero or almost zero velocity) still exist but are shorter since the particle oscillates before coming to rest
\footnote{in the limit $v=0^+$ metastable states can still be defined}. (ii) Hence avalanches (from one metastable
state to another one) can still be defined. (iii) Due to inertia, the avalanche starts more slowly,
however the particle overshoots and may not settle in the next metastable state (as for $m=0$) but in
one farther away. These metastable states $u_i$ are a subset of the metastable states for $m=0$. As the mass increases, more and more of the metastable states at $m=0$ get eliminated. Thus as the mass increases the smaller $m=0$ avalanches
(i.e. with smaller barriers to the next metastable state) are ``eaten up" or merge. The larger $m=0$ avalanches,
with larger barriers, remain, although the dynamics is quite different.
One notes that the frequency of oscillation increases as the particle settles to the new metastable state
under the action of damping.

For larger $v$ as $m$ increases the avalanche structure disappears and one enters into a regime
better described by oscillations in the co-moving frame, see Fig.~\ref{fig:all}. However there remains some correlation
with the $m=0$ avalanche structure, the larger $m=0$ avalanches seem to induce the largest oscillations (Fig.~\ref{fig:all}).

While for small mass the motion remains under-damped and the time scale remains $O(\tau_{\mu})$, in the larger-mass
regime $\tau_m \gg \tau_{\mu}$ ($m \gg 1$ in dimensionless units) the motion occurs on larger time scales. Let us describe qualitatively the
avalanches in that regime. For that it is
useful to rewrite the equation of motion as:
\bea
&& \frac{\rmd}{\rmd t} E = \mu^2 v t \dot u - \eta \dot u^2 \\
&& E = \frac{1}{2} m \dot u^2 + V(u) + \frac{\mu^2}{2} u^2\ ,
\eea
where $F(u)=-V'(u)$ and the disorder-potential fluctuations
typically grow as $V(u)-V(0) \approx \sqrt{\sigma} u^{3/2}$.
Balancing disorder with the quadratic well gives the avalanche cutoff size $u \sim S_\mu$.
Hence for $u \ll S_\mu$ the disorder term dominates.
Consider an avalanche starting at $t=0$. For small $v$ and $\eta$, as the previous metastable
state $u(t=0)=u_i$ becomes unstable, the particle will first oscillate with amplitude $\Delta u$
between $u_i$ and the smallest root of $V(u)-V(u_i) + \frac{\mu^2}{2} (u^2-u_i^2)$,
and the total energy $E$ can be considered  constant during a period
\footnote{In the limit $\eta, v=0^+$ one can define the quasi-static motion of the edges of the oscillation
interval $[u_{\min}(t),u_{\max}(t)]$. It should also proceed by jumps.}.
There is clearly a distribution of amplitude $\Delta u$ from the disorder, but we
can estimate the typical oscillation time $\tau_{\rm osc}$ as a function of the amplitude $\Delta u$
as
\be \label{tauosc}
\tau_{\rm osc} \sim m^{1/2} (\Delta u)^{1/4}/\sigma^{1/4}\ ,
\ee
by balancing the kinetic energy
with the disorder. The quadratic well controls
the scale of the largest amplitudes $\Delta u \sim S_\mu$, which correspond to
a time scale $\tau_0$. The total energy $E$ will decay on much larger time scale
$\tau_m \gg \tau_0,$ and the particle will settle in one of the available metastable states within the
range of the first oscillation. From the above estimate (\ref{tauosc}) one sees
that the frequency of oscillation will indeed increase as $\Delta u$ decreases to zero.
This picture requires that the quadratic well has moved by less than $S_\mu$
during the avalanche time hence $v < v_m := S_\mu/\tau_m$ (in dimensionless units this is $v < 1/m$).
If $v>v_m$ the particle has no time to converge to a metastable state and the definition
of an avalanche becomes less clear. In the regime $v_m < v < v_0=S_\mu/\tau_0$
the motion still remains quite correlated to the disorder and plateaus would still
be visible by averaging over oscillations. Finally for $v>v_0$ (in dimensionless units this is $v > 1/\sqrt{m}$)
multiple oscillations are not visible and the trajectory becomes smoother.

Note that $V(u)$ is a random acceleration process, since $V''(u)$ is a white noise.
Hence in the large inertial mass limit, $\Delta u$ can be seen as the first return to the
origin of a random acceleration process. This leads to $P(\Delta u) \sim \Delta u^{-5/4}$
for small $\Delta u \ll S_\mu$, the distribution being cutoff around $S_\mu$.
Similar arguments, although for a slightly different model, were made in
\cite{SchwarzMaimon2001,SchwarzFisher2003} using earlier results \cite{Sinai,Burkhardt}.

It is also interesting to note that in any time window $[t_i,t_f]$ where $\dot u(t)>0$ one can parametrize trajectories as
function of the position $u(t)=\int_{t_i}^t \rmd t' \dot u(t')$ and rewrite either model as a stochastic equation for $\dot u(u)$ as
\be
 \frac{\rmd^2}{\rmd u^2} \left(\frac{1}{2} m \dot u^2\right) + \frac{d}{du} (\eta \dot u) = \mu^2 \left(\frac{v}{\dot u} -1 \right) + F'(u)\ .
\ee
In the limit of $m \to 0$ one recovers the standard ABBM stochastic equation \cite{ABBM1,ABBM2} and $\dot u(u)$ can be mapped
to the radial coordinate of a Brownian motion in dimension $d=1+v/v_\mu$ (see e.g. Section VI B in \cite{LeDoussalWiese2008a})
leading to the distribution of avalanches sizes $P(S) \sim S^{-\tau}$ with $\tau=\frac{3}{2} - \frac{v}{2 v_\mu}$ for $v<v_\mu$,
for avalanches much smaller than the cutoff, $S \ll S_\mu$, for which the quadratic well does not play a role.
In presence of inertia the above equation can be used between two zeroes of the velocity and
for small $\eta,\mu$ leads again to an exit time for the random acceleration problem.
Some considerations about exit times are given in App.~\ref{sec:exit}.

\subsubsection{Velocity distributions}

\begin{figure}[t]
\includegraphics[width=\linewidth]{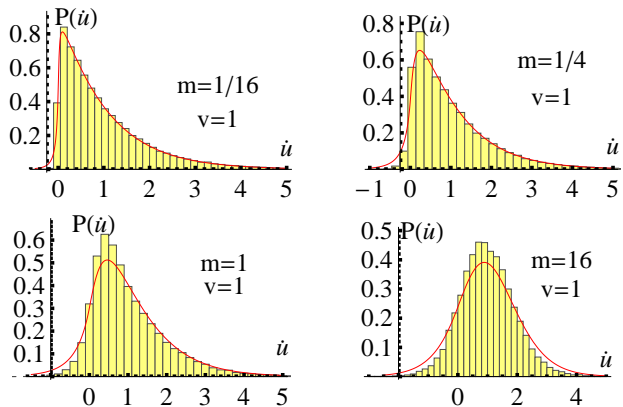}\\
\caption{Probability density distribution of the instantaneous velocity $\dot
u$ for the ABBM model with inertia in the form of a histogram for
$v=1$ and different masses.
The red full line denotes the probability density obtained in  Sec.~\ref{sec:numerics}
for the tree model. Note that the axis is vertical axis is shifted from $\dot{u}=0$
to the left.}\label{fig:numerics2}
\end{figure}

\begin{figure}[t]
\includegraphics[width=\linewidth]{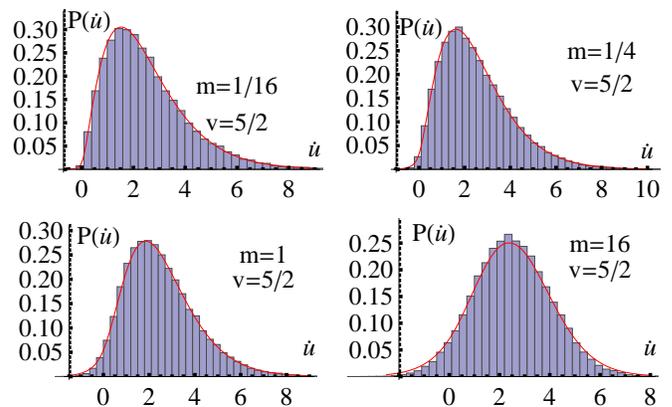}\\
\caption{Histogram of the probability density distribution for the ABBM model
with inertia for $v=5/2$ and different
masses.
in units
The red
full line denotes the probability density obtained in  Sec.~\ref{sec:numerics}
for the tree model. Note that the axis is vertical axis is shifted from $\dot{u}=0$
to the left.}\label{fig:numerics2v}
\end{figure}

Consider on figure \ref{fig:numerics2} the histograms   of the distribution of the instantaneous velocity $\dot u$.
We see that for driving velocity $v=1$, by increasing the mass $m$ from \(1/16\) to \(16\), the distribution becomes more and more symmetric, and peaked around $v=1$. The same happens for larger driving velocities, $v=5/2$, see figure \ref{fig:numerics2v}. Note that all histograms have a non-vanishing tail for negative \(\dot u\), but that this tail gets smaller when decreasing the mass.
(Attention: the axis on the figures is shifted from $\dot{u}=0$ to the left).
By comparing  Figs.~\ref{fig:numerics2} and \ref{fig:numerics2v}, we see that when increasing the driving velocity $v$ for a fixed mass, the probability for negative velocities decreases. This is clearly seen on the plots of Fig.\ \ref{f:asdfasdf}, page \pageref{f:asdfasdf}, for mass $m=1/4$, and $v=1/10$, $v=1/2$, and $v=5$. The red lines (solid on Figs.\  \ref{fig:numerics2}, \ref{fig:numerics2v},
and dashed on Fig.~\ref{f:asdfasdf}) represent different approximations to be discussed later.

This negative tail
renders the analytical analysis difficult, not per se due to the negative velocities, but since the particle moves backwards through the same disorder, thus the disorder it sees becomes {\em correlated in time}, and the system has  {\em memory}. Since we possess currently no powerful tool to tackle this situation, we will treat two different {\em local, memory-free} variants as explained in the introduction: (i) the particle on the tree model that can be formulated by using for the random-force correlator  the second line of Eqs.~(\ref{3}) or (\ref 4); and (ii) the $\sqrt{\dot{u}}$ model given by the first line of Eqs.~(\ref{3})
or (\ref 4).

There is a strong motivation to consider these two variants. Let us look at Figs.~\ref{fig:numerics2} and \ref{fig:numerics2v}. The red full lines represent the numerical solution for a particle on the tree. We see that
when increasing $v$ or decreasing $m$, the ``particle on the tree'' becomes a good {\em approximation} of the ABBM model with inertia. The same holds (not shown here) for the \(\sqrt{\dot u}\) model. The physical reason is simple: At higher and higher driving velocities or smaller and smaller mass, the particle will less and less often move backward, thus these events will lose their importance for $P(\dot u)$.

We will indeed prove a much stronger statement.
Consider the large-deviation function, defined (supposing the limit exists) by \begin{equation}
F(x) :=- \lim_{v\to \infty} \frac{\log[ P(x v)]}{v}\ .
\end{equation}
We will show that $F(x)$ indeed exists for all three models, and that for \(x>0\) all three large-deviation functions coincide. We expect, but cannot prove, that for \(x<0\) these functions will differ.

In the next section \ref{sec:tree}, we start with the particle on the tree, which may be expected to be the most physical one. We then continue with the \(\sqrt{\dot u}\) model in section \ref{sec:complex}, for which we have the most analytical results.

\section{Tree model}
\label{sec:tree}

\subsection{Definition of the model\label{sec:deftree}}

In this section we examine the motion of a particle, with finite mass $m$,  in a Brownian-correlated disorder force. Since the Middleton theorem does not hold, we assume that when the particle changes  direction, it  experiences a different disorder potential,  uncorrelated with that experienced previously. The model describes the motion on a tree with constraint that the particle always chooses a different branch when changing its direction of  motion. On each branch the disorder satisfies Eq.~(\ref{eq:force}). As mentioned in the introduction, the model may be relevant to describe
systems which do not visit the same microscopic configuration twice, while the center of mass is oscillating back and forth, i.e. systems
with large deviations from the Middleton theorem. Note that  the tree is not defined from the start, but is generated dynamically, thus there is one tree associated to each trajectory or history. This may however be captured by the limit of a fixed tree with high branching rate.

The equation of motion in the laboratory frame is:
\begin{align}\label{eq:motiona}
\frac{\mathrm{d}\dot{u}}{\mathrm{d} t}&=a,\\
\label{eq:motionb}
m\frac{\mathrm{d}a}{\mathrm{d} t}&={\mu^2}\left[v-\dot{u}(t)\right]-{\eta}a+
{\partial_tF_t[u]},
\end{align}
where $F$  is a functional of $u(t)$. Then, the effective disorder correlator becomes
\begin{align}\label{eq:force1}
&& \partial_t \partial_{t'} \overline{(F_{t}[u] - F_{t'}[u])^2}=- 4  \sigma|\dot{u}(t)|\delta(t-t'). \\
&& \overline{\partial_t F_{t}[u] \partial_{t'} F_{t'}[u]}=2 \sigma|\dot{u}(t)|\delta(t-t').
\end{align}
In the limit of $m\to 0$, the model simplifies to the ABBM model considered in the previous section. Note that this tree model has the pecularity, that the system
reaches  a stationary state without being pinned, even in the  {\em absence} of driving. The reason is that  every time it changes the direction the
disorder is renewed. Only if the system starts at $a=\dot u=0$ or arrives there ``by accident'', it remains there for ever.

Now the probability distribution depends on two variables, velocity $\dot{u}$ and acceleration $a$. The corresponding Fokker-Planck equation is a parabolic differential equation and has the form
\begin{align}\label{eq:Fokker}
\frac{\partial P(\dot{u},a,t)}{\partial t}=&-a\frac{\partial P}{\partial \dot{u}}+\frac{\partial^2}{\partial a^2}\left( \frac{\sigma |\dot{u}|P(\dot{u},a,t)}{m^2}\right)
\notag\\&
-\frac{\partial}{\partial a}\left\{\left( -\frac{\mu^2}{m}\dot{u}-\frac{\eta}{m}a+\frac{\mu^2}{m}v\right)P(\dot{u},a,t)\right\}
\end{align}
In the following two sections we analyze this equation both analytically and numerically.
There are several limits which can be studied analytically, namely small  or large $m$ at fixed $v$, and large $v$
at fixed $m$.

\subsection{\label{sec:perturbation}Perturbation expansion in small \(m\)}

Equation (\ref{eq:Fokker}) is  complicated and an exact analytic solution is not known. There are many systems \cite{review} where the mass term is small and could be treated as perturbation with respect to the other terms. Therefore, we start with perturbation theory in $m$,
i.e. $m\mu^2/\eta^2\ll 1$ in dimensionfull units.

In dimensionless units the Fokker-Planck equation simplifies, and effectively depends only on the two parameters, $m$ and $v$:
\begin{align}\label{eq:Fokkerdimnesionless}
\frac{\partial P(\dot{u},a,t)}{\partial t}=&-a\frac{\partial P}{\partial \dot{u}}+\frac{\partial^2}{\partial a^2}\left( \frac{ |\dot{u}|P(\dot{u},a,t)}{m^2}\right)
\notag\\&
-\frac{\partial}{\partial a}\left\{\left( -\frac{\dot{u}(t)}{m}-\frac{a}{m}+\frac{v}{m}\right)P(\dot{u},a,t)\right\}.
\end{align}
We are interested in the stationary situation \(\partial_t P(\dot u,a,t)=0\).

In the limit $m\to 0^+$ the acceleration is divergent since in the ABBM model the disorder generates a white noise with no small-scale cutoff
(see Section \ref{sec:largev} below).
Analyzing the structure of the Fourier-transformed Fokker-Planck equation and the moments that follow from it (similarly to Sec.~\ref{sec:mainmoments}), we conclude that one has to introduce a reduced acceleration $\tilde{a}=\sqrt{m} a$ in order to be able to organize the perturbation theory in $m$. Then
in the region $|\tilde{a}|\gtrsim \sqrt{m}$ and $\dot{u}\gtrsim m>0$ (we will refer to it as the region 1) we find:
\begin{align}\label{eq:distributionperturbative}
P^{(1)}(\dot{u},a)=\sqrt{m}e^{-\frac{\tilde{a}^2}{2\dot{u}}-\dot{u}}\sum_{n=0}^{\infty}
\frac{F_n(\tilde{a},\dot{u})}{\Gamma(v)\sqrt{2\pi}}\dot{u}^{v-\frac{3+4n}{2}}m^{n/2}.
\end{align}
The index $(1)$ denotes that the expression is valid in  region $1$. Here $F_n$ satisfies the recursion for $n \geq 0$ (and $F_{-1}=0$):
\begin{align}\label{eq:recursion}
&\tilde{a} \partial_{\tilde{a}} F_n - \dot{u} \partial_{\tilde{a}}^2 F_n + \dot{u}^2 (v-\dot{u}) \partial_{\tilde{a}} F_{n-1} + \tilde{a} \dot{u}^2 \partial_{\dot{u}} F_{n-1} \notag\\&+ \frac{1}{2} (\tilde{a}^3 - (4 n-1) \tilde{a} \dot{u}) F_{n-1} = 0.
\end{align}
We start solving Eq.~(\ref{eq:recursion}) from the smallest $n=0$. Then, using the solution for $F_0$ we solve the next equation for $n=1$ and find $F_1$. The procedure develops further in the same way. However, if we want to determine $F_i$ we have to solve all the differential equations (\ref{eq:recursion}) with $n\leq i+2$. The reason is that we are interested in a distribution that has all finite moments $\overline{a^k\dot{u}^j}$, and therefore  decays faster than algebraically for large $\dot{u}$ and $a$. This condition has to be satisfied for any mass, hence for any order
in the expansion. Taking this into account when analyzing the solution in $(i+2)$nd order allows to discard some solutions to the equation appearing in the $i$th order, i.e.~in $F_i$.  The details of the calculation and some intermediate results are given in  App.~{\ref{app:perturbationtheory}}.

For brevity we state her only the first three terms; the further terms are lengthy, and given in  App.~\ref{app:perturbationtheory}. \begin{align}\label{eq:F0text}
F_0&=1,\\\label{eq:F1text}
F_1&=\frac{1}{2} \left(\tilde{a}
   \dot{u}-\frac{\tilde{a}^3}{3}\right)+c_3 \dot{u}^2,\\\label{eq:F2text}
F_2&=\frac{\tilde{a}^6}{72}-\frac{5 \tilde{a}^4
   \dot{u}}{48}-\frac{1}{6} \tilde{a}^3 c_3
   \dot{u}^2-\frac{1}{4} \tilde{a}^2 \dot{u}^2
   (\dot{u}-v)+\frac{1}{2} \tilde{a} c_3 \dot{u}^3\notag \\ &+c_5
   \dot{u}^4-\frac{1}{48} \dot{u}^3 \left[24
   \dot{u}^2-48 \dot{u} v \log (\dot{u})-24 v^2+36
   v-5\right].
\end{align}
There still remain undetermined constants $c_3$ and $c_5$. They have to be fixed such that $\int\dif a\,\dif\dot{u\,}P(\dot u,a)=1$ for {\em all} $m$. This task can not be done now, since region 2 may also contribute. It is further complicated by the fact that also negative velocities may contribute, and our expansion does not give a result for those. We discuss this issue in the next section, see  Eqs.~(\ref{eq:constants1}) and (\ref{eq:constants2}), when comparing the analytical results with the numerical solution of Eq.~(\ref{eq:Fokker}).

Integrating out velocities from $P^{(1)}(\dot{u},a)$ one obtains the conditional probability distribution of acceleration when $\dot{u}>m$. Leading two contributions in the region 1 are
\begin{align}\label{eq:distributiona}
P^{(1)}(a)\rmd a =& \rmd \tilde a \bigg[ \frac{2^{\frac{3}{4}-\frac{v}{2}}
      }{\sqrt{\pi } \Gamma(v)}\left|\tilde{a}\right|^{v-\frac{1}{2}}K_{\frac{1}{2}-v}\left(\sqrt{2}
   \left|\tilde{a}\right|\right)\notag\\
   &+ \sqrt{m} \frac{
   2^{-\frac{1}{4}-\frac{v}{2}}
      }{3 \sqrt{\pi } \Gamma
   (v)}\left|\tilde{a}\right|^{v-\frac{3}
   {2}}\notag\\ &\times \Big(3 \sqrt{2}
   \left(\tilde{a}+c_3 (2 v-3)\right)
   K_{\frac{3}{2}-v}\left(\sqrt{2}
   \left|\tilde{a}\right|\right)\notag\\&-2
   \left|\tilde{a}\right|
   \left(\tilde{a}-3 c_3\right)
   K_{\frac{5}{2}-v}\left(\sqrt{2}
   \left|\tilde{a}\right|\right)\Big)\bigg]
\end{align}
Here $K_n(z)$ is the modified Bessel function of the second kind. Similarly, by integrating out $a$ from $P^{(1)}(\dot{u},a)$ we obtain the conditional probability distribution of velocities  in  region 1, for $|\tilde{a}|>\sqrt{m}$, of which we state the first   terms. Higher order terms are discussed in App.~\ref{app:perturbationtheory}.
\begin{align}\label{eq:distributionm}
P^{(1)}(\dot{u})=&\frac{e^{-\dot{u}} \dot{u}^{v-1}}{\Gamma
   (v)}
   +\frac{c_3 \sqrt{m} e^{-\dot{u}}
   \dot{u}^{v-1}}{\Gamma (v)}\notag
   +\frac{m
   e^{-\dot{u}} \dot{u}^{v-2} }{4 \Gamma
   (v)} \\& \times \left[\left(4
   c_5-1\right) \dot{u}-2 \dot{u}^2+4 \dot{u} v \log
   (\dot{u})+2 (v-1) v\right].
\end{align}
In the limit of $m\to 0$ only the first term survives and we recover the result given by Eq.~({\ref{eq:distribution}}).
A note of caution concerning the distributions (\ref{eq:distributiona}) and (\ref{eq:distributionm}) is in order:
They have been obtained by integrating the joint distribution (\ref{eq:distributionperturbative})  over
all values of $a$, and over positive $\dot u$, while (\ref{eq:distributionperturbative}) is valid in a more
restricted range. These distributions may acquire some correction from  other regions, which  should be
small (and maybe even subdominant to the correction given above but we cannot prove it).
In any case, it cannot affect the leading-order result, i.e.\ the first line in (\ref{eq:distributiona}), which is the (normalized) distribution
of (reduced) acceleration $\tilde a$ in the presence of a small inertia, and which is a novel exact result.

Next we discuss  region 2, defined as $|\tilde{a}|\lesssim \sqrt{m}$ and $0<\dot{u}\lesssim m$. If one naively assumes that the results obtained in  region 1 are valid in  region 2, one finds  non-integrable divergences in $P$ for small velocities $\dot{u}$. In order to organize the perturbation theory in  region 2, we have to introduce $\tilde{\dot{u}}=\dot{u}/m$ and use $a=\tilde{a}/\sqrt{m}$.
Then,
\begin{align}\label{eq:expansionboundary}
P^{(2)}(\dot{u},a)=f(m,v)\sum_{n=0}^{\infty}
\tilde{\tilde{P}}_n^{(2)}(a,\tilde{\dot{u}})m^n,
\end{align}
where $f$ is an undetermined function. Plugging this equation into the Fokker-Planck Eq.~(\ref{eq:Fokkerdimnesionless}), we obtain the equations that determine $\tilde{\tilde{P}}_n^{(2)}$\footnote{The function $f$ does not enter them, since the Fokker-Planck equation is linear equation in $P$.}.
Unfortunately, these equations are as difficult to solve as the full Fokker-Planck equation, and we did not succeeded in solving them analytically. However, in App.\ref{app:matching} we demonstrate that Eq.~(\ref{eq:expansionboundary}) has the correct form and that the perturbation theory is properly organized. We show that $P^{(2)}(\dot{u},a)$ matches at the boundary between regions 1 and 2 the distribution function found in  region 1.

\subsection{Large driving velocity $v \gg v_\mu$\label{sec:largev}}

A naive argument is that in that limit the noise $\sqrt{|\dot u|} \xi(t)$ can be replaced by $\sqrt{v} \xi(t)$, hence becomes Gaussian. One can then
 directly solve the Langevin equation (in dimensionless units) in frequency space,
\bea
\dot u_\omega = \frac{\sqrt{v} \xi_\omega +2\pi v\delta(\omega)}{- m \omega^2 + 1 + i \omega}\ .
\eea
This leads to a Gaussian distribution for $\dot u$ and $\tilde a$ with equal-time correlations in the steady state:
\bea
&& \overline{(\dot u(t)-v)^2} = \int_\omega \frac{2 v}{(1-m \omega^2)^2 + \omega^2} = v \\
&& \overline{\tilde a(t)^2} =  - \int_\omega \frac{2 v m \omega^2}{(1-m \omega^2)^2 + \omega^2} = v\ ,
\eea
with $\overline{\dot u(t) \tilde a(t)}=0$ and $\int_\omega=\int \frac{d\omega}{2 \pi}$. We used $\overline{\xi_{\omega}}=0$. Thus we find
\begin{align}\label{eq:GaussianV}
P_{v \gg 1}(\tilde{a},\dot{u}) \approx \frac{1}{2\pi v} \exp\left(-\frac{1}{2 v} \left[\tilde{a}^2 + (\dot{u} - v)^2\right]\right).
\end{align}
Note that since the dynamics being the same for positive $\dot u$ this large $v$ result
should hold for all three models.

Integrating over the acceleration, this can be identified with a Boltzmann distribution
\bea
 P_{v\gg 1}(\dot{u}) \sim \sqrt{\frac{m}{2 \pi T_{\rm eff}}} e^{- \frac{ m (\dot u-v)^2}{2 T_{\rm eff}} }
\eea
in the moving frame for the Hamiltonian of a free particle of mass $m$. The effective temperature (in
dimensionfull units) is
\bea
T_{\rm eff} = m v_\mu v  = m v \frac{\sigma}{\eta \mu^2}\ .
\eea
Hence at large driving velocity the disorder, the quadratic well, and the damping act together in the moving frame as
a thermal noise \footnote{although this could be interpreted by saying that the effect disorder in the moving frame
is the same as an equilibrium white noise of variance $2 \eta T_{eff}$ for a particle in a quadratic well with damping $\eta$
(i.e. satisfying usual fluctuation-dissipation relations), this would lead to an incorrect value for the noise hence it is
is not a valid interpretation. There is an underlying Hamiltonian system but it is non-standard and involves the acceleration, see
Sec.~\ref{sec:largem}. This is because the starting equation of motion can only be written in a Langevin form in terms of velocity.}.

From the above result we could guess that the probability of a negative velocity decays as
$e^{-v/(2 v_\mu)}$ at large $v$. We will see below that this is not quite accurate. Indeed,
we will go beyond the above argument and show that there is a large-deviation function
which describes the deviations from the Gaussian at large driving velocity. These deviations  appear in the far tails
at $|\dot{u}-v| \sim v$, see Sec.~\ref{sec:largem}.

\subsection{Large mass $m \gg 1$\label{sec:largemass}}

At large mass the oscillation time $\tau_0=\sqrt{m}/\mu$ increases but becomes much smaller
than the damping time $\tau_m = m/\eta$ (necessary for damping to
overcome inertia). Hence the system is in the underdamped limit and the particle
oscillates many times before it comes to rest. It can be seen by rewriting the
Langevin equation in dimensionless units:
\bea
&& \frac{\rmd \dot u}{\rmd t'} = \tilde a \\
&& \frac{\rmd \tilde a}{\rmd t'} = - \dot u + v + \frac{\sqrt{|\dot u|}}{m^{1/4}} \xi(t') - \frac{\tilde a}{\sqrt{m}}\ ,
\eea
where we have used the reduced acceleration and
defined the reduced time as $t=\sqrt{m} t'$ in units of the oscillation time. At large $m$
we see that to leading order we have a Hamiltonian system with $p=\tilde a$ and $q=\dot u-v$ and $H(p,q)=\frac{1}{2} (p^2 + q^2)$,
i.e.\ a harmonic oscillator, weakly perturbed by (i) a noise, and (ii) the damping. Although
these terms are small they will select the steady state as we show now.

For large $m$ the Fokker-Planck equation has  a well-defined limit if one scales the probability as
\bea \label{eq:tildeP}
P(\dot u,a,t) = \sqrt{m} \tilde P(\dot u,\tilde a = a \sqrt{m}, t'=t/\sqrt{m})\ .
\eea
The scaling function $\tilde P$ satisfies
\be
 \partial_{t'} \tilde P = [ (\dot u-v) \partial_{\tilde a}  - a \partial_{\dot u} ] \tilde P + \frac{1}{m^{1/2}} [1 + \tilde a \partial_{\tilde a} + D(\dot u)  \partial_{\tilde a}^2 ] \tilde P,
\ee
where we have allowed for a general noise function $D(\dot u).$ It is equal
to $D(\dot u)=|\dot u|$ for the tree model that we study now, and $D(\dot u)=\dot u$
for the $\sqrt{\dot u}$ model studied below.

This equation is well suited to analyze the
time regime $t \sim \tau_0$, i.e. $t' =O(1),$ when the system oscillates. However at
even larger times it will be damped and will reach a steady state. For the latter we
are interested in the limit
\bea
\lim_{m \to \infty} \lim_{t' \to \infty} \tilde P(\dot u, \tilde a , t')\ .
\eea
We start by searching for the steady state as a time-independent solution in the form
\bea
\tilde P(r,\theta) = \sum_{n=0}^\infty m^{-n/2} P_n(r,\theta)\ .
\eea
It is convenient to use the action-angle variables of the harmonic
oscillator,
\bea
&& \dot u = v + r \sin \theta \\
&& \tilde a = r \cos \theta\ .
\eea
In these variables the Fokker-Planck equation becomes
\bea
 \partial_\theta \tilde P &=& \frac{1}{\sqrt{m}} [ 1 +  r \cos \theta O +  D(v+r \sin \theta) O^2 ] \tilde P\qquad  \\
 O &:=& \cos \theta \partial_r - \frac{\sin \theta}{r} \partial_\theta\ .
\eea
This yields the recursion
\bea
&&  \partial_\theta P_0 = 0 \\
&& \partial_\theta P_{n+1} =  [ 1 +  r \cos (\theta)\, O +  D(v+r \sin \theta) O^2 ] P_{n}  , \  n \geq 0 \nn
\eea
Note that the operators $O$ and $D(v+r \sin \theta)$ commute. To leading order the general solution is
\bea \label{eq:P0}
P_0(r,\theta) = P_0(r)\ ,
\eea
where $P_0$ it to be normalized as\bea \label{normP0}
\int_0^\infty 2 \pi r  P_0(r)\,\rmd r = 1\ .
\eea
The higher-order terms satisfy
\bea
\int_0^\infty r \rmd r \int_0^{2 \pi} \rmd\theta\, P_n(r,\theta)=0, \quad n \geq 1\ .
\eea
The function $P_0$ is selected by the next order equation, as we now discuss:
\bea \label{next}
 \partial_{\theta}  P_1(r,\theta,t) = \phi(r,\theta)
\eea
with \bea
 \phi(r,\theta) &=& P_0(r) + a(r,\theta) P_0'(r) + b(r,\theta) P_0''(r)\qquad \\
 a(r,\theta) &=& r \cos^2(\theta) + \frac{1}{r} D(v+ r \sin \theta) \sin^2(\theta) \nn \\
 b(r,\theta) &=& \cos^2(\theta)  D(v+ r \sin \theta)\ .  \nn
\eea
The general solution of (\ref{next}) is
\bea
P_1(r,\theta)  = \int_0^\theta \rmd\theta' \phi(r,\theta') + P_1(r,0)
\eea
Now, we observe that one can integrate (\ref{next}), $\int_0^{2 \pi} \rmd \theta$ and
obtain a condition which must be satisfied, in order for $P_1$ to be meaningful,
i.e.\ a single-valued function in $e^{i \theta}$,
\bea
P_1(r,2 \pi) - P_1(r,0) = 0 = \int_0^{2 \pi} \rmd\theta \phi(r,\theta)\ .
\eea
This leads to a condition which determines the steady state
$P_0(r)$ to leading order as
\bea \label{eqdiff0}
P_0(r) + a(r) P_0'(r) + b(r) P_0''(r) = 0\ .
\eea
We have defined
\bea
&& a(r) = \frac{r}{2} + \frac{1}{r} \int_0^{2 \pi} \frac{\rmd\theta}{2 \pi} D(v + r \sin \theta) \sin^2(\theta)  \\
&& b(r) = \int_0^{2 \pi} \frac{d\theta}{2 \pi} D(v + r \sin \theta) \cos^2(\theta)\ .
\eea
The first-order correction reads
\bea
 P_1(r,\theta)  &=& \int_0^\theta \rmd \theta' \,[a(r,\theta') - a(r)] P_0'(r) \\
&& + \int_0^\theta \rmd \theta' \,[b(r,\theta')-b(r)] P_0''(r) + P_1(r,0)\ , \nn
\eea
where $P_1(r,0)$ is determined from the next-order equation using again
that $P_2$ should be single valued and the normalization condition.
Let us now analyze this equation for the various models.

\subsubsection{$D(\dot u) = \dot u$ model}

Consider now the simpler choice $D(\dot u)=\dot u$, which, anticipating a bit, is the
$\sqrt{\dot u}$ model defined below. In that case
\bea
&& a(r) = \frac{r}{2} + \frac{v}{2 r} \\
&& b(r) = \frac{v}{2}\ .
\eea
Solving the differential equation (\ref{eqdiff0}) we find two solutions. One is the Gaussian
\bea \label{eq:GauusianM}
P_0(r) = \frac{1}{2 \pi v} e^{- r^2/(2 v)}\ .
\eea
The other one decays as $1/r^2$ at large $r$ and thus cannot satisfy the normalization condition (\ref{normP0}).
Going back to variables  $\dot u$,$\tilde a$, this identifies with the Gaussian distribution (\ref{eq:GaussianV})
also found in the large-velocity limit.

The analysis can be continued to higher orders. Writing\bea
P_n = \frac{e^{-r^2/(2 v)}}{2 \pi v} Q_n(r,\theta)\ ,
\eea
where the $Q_n$ are polynomials in $r$, $\cos \theta$, and $\sin \theta$, which  satisfy the recursion\bea
&& Q_0 = 1 \\
&& \partial_\theta Q_{n+1} =  [ 1 +  r \cos \theta \hat O +  (v+r \sin \theta) \hat O^2 ] Q_{n}  \ , \quad n \geq 0. \nn
\eea
We have defined
\bea
\hat O= e^{r^2/(2 v)} O e^{-r^2/(2 v)} = \cos \theta\left ( -\frac{r}{v} + \partial_r \right) - \frac{\sin \theta}{r} \partial_\theta\ . \nn
\eea
One finds
\bea
&& Q_1 = -\frac{r \cos (\theta ) \left(r^2 \cos (2
   \theta )+r^2-6 v\right)}{6 v^2} = -\frac{a^3-3 a v}{3 v^2} \nn \\
   &&
   \eea
The second order $Q_2$ is given in Appendix \ref{app:Q}. The moments $\overline{\dot u^n a^m}$ of
the $\sqrt{\dot u}$ model are computed below in Sec.~\ref{sec:mainmoments} by another method and we have checked
that they agree with the ones obtained here from $P$ to the considered order ($1/m$).

\subsubsection{Tree model $D(\dot u)=|\dot u|$}

For the tree model one finds
\bea
&& a(r) = \frac{r}{2} + \int_0^{2 \pi} \frac{d\theta}{2 \pi} \Big|\frac{v}{r} + \sin \theta \Big| \sin^2(\theta)  \\
&& b(r) = r \int_0^{2 \pi} \frac{d\theta}{2 \pi} \Big|\frac{v}{r} + \sin \theta \Big| \cos^2(\theta)\ .
\eea
Explicit calculations give\bea
&& a(r) = \frac{r}{2} {+} \frac{1}{\pi} \frac{v}{r} {\rm arcsin}\left(\frac{v}{r}\right) - \frac{1}{3 \pi} \Big(\frac{v^2}{r^2}{-}4\Big) \sqrt{1{-} \frac{v^2}{r^2}} ,\  r > v \nn \\
&& a(r) = \frac{r}{2} +  \frac{v}{2 r}  , \quad r < v \label{ab} \\
&& b(r) =  \frac{v}{\pi} {\rm arcsin}\left(\frac{v}{r}\right) + \frac{r}{3 \pi} \left(2+\frac{v^2}{r^2}\right) \sqrt{1- \frac{v^2}{r^2}}   , \quad r > v \nn \\
&& b(r) = \frac{v}{2}  , \quad r < v
\eea
We see that for large $v$ one recovers the result of the $\sqrt{\dot u}$ model, namely the Gaussian. For
$v$ of order one, the solution $P_0$ of (\ref{eqdiff0})  can be computed numerically (see Fig.~\ref{fig:largem}).
For $v=0$ one can obtain an analytical expression using
$a(r) = {r}/{2} + {4}/{(3 \pi)}$ and $b(r) = 2 r/(3 \pi)$:
\bea \label{eq:P0m}
P_0(r) = \frac{9 \pi}{32} e^{-\frac{3 \pi}{4} r} \ .
\eea
The other solution is excluded, since it decays as $1/r^2$ at large $r$. For the leading correction in $m$
at $v=0$  we find
\bea
&& P_1(r,\theta) = P_0(r) Q^+_1(r,|\theta|) {\rm sgn}(\theta) \quad , \quad -\pi <\theta < \pi \\
&& Q^+_1(r,\theta) = \frac{1}{64} \Big[4 (\pi -2 \theta ) (3 \pi
   r-8) \\
   && -2 \pi  \cos (\theta )\Big(12 r \sin
   (\theta )+(3 \pi  r+4) \cos (2 \theta )+3
   \pi  r-20\Big)\Big] \nn
\eea
We can check that it has zero angular integral and vanishes at all $\theta=n \pi/2$. The moments are
\bea
 \overline{\tilde a^2} = \overline{\dot u^2} = \frac{16}{3 \pi^2} + O(1/m)\ ,
\eea
whereas they would vanish in the limit of $v=0$ for the $\sqrt{\dot{u}}$ model, where the Gaussian (\ref{eq:GauusianM}) is valid at zeroth order in $1/m$.

The Laplace-transform of Eq.~(\ref{eq:P0m}) gives, to leading order in $1/m$ \begin{eqnarray}\label{85}
\hat{P}(\lambda) &=&\overline {\rme^{{\lambda \dot u}}} \approx \int_{0}^{2\pi}\rmd \theta \int_{0}^{\infty}\rmd r\,  r P_{0}(r) \,\rme^{\lambda r \sin \theta} \nn\\&=& \frac{27 \pi ^3}{\left(9 \pi ^2-16
   \lambda ^2\right)^{3/2}}
\end{eqnarray} This function has a branch-cut singularity starting at $\lambda_c^\pm=\pm \frac{3\pi}{4}=\pm 2.35619$. It is interesting
to note that from (\ref{eqdiff0}) and (\ref{ab}) one can conclude directly that for any $v$, $P_0(r) \sim e^{- \frac{3\pi}{4} r}$ at large $r \gg v$. Hence
the tree model has $\lambda_c^\pm=\pm \frac{3\pi}{4}$ at $m=\infty$ independent of $v$. Similar branch cut singularities for the $\sqrt{\dot u}$
model will be discussed in Sec.~\ref{sec:defcomplex}.

\begin{figure}
\Fig{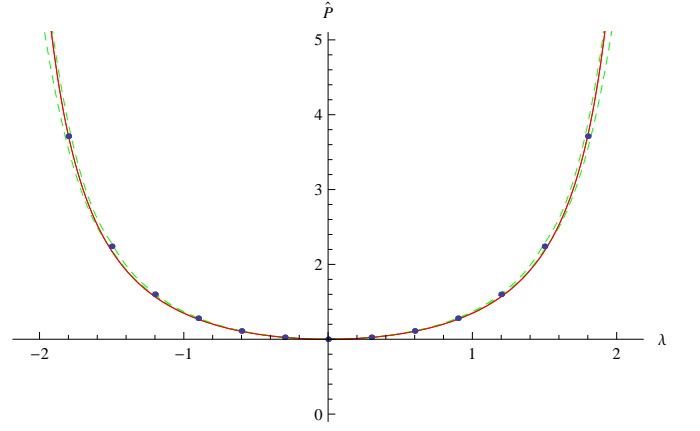}
\caption{Numerical simulation of  the tree model for $v=0.1$, $m=100$, $\mu=\eta=1$, and $5 \times 10^{7}$ data points. In order to eliminate the small asymmetry coming from nonzero velocity, we show symmetrized $\hat{P}$ in $\lambda$, i.e. $[\hat{P}(\lambda)+\hat{P}(-\lambda)]/2$. These are the data shown (blue dots), together with a $1\sigma$ estimate of their statistical errors (green, dashed) and the analytical curve (\ref{85}) (red/solid). Note that there is no adjustable parameter. For larger values of $|\lambda|$ (not shown) the statistical errors grow and one observe small deviations whose origin could be corrections from $1/m$ or small but finite $v$.
}\label{f:Zmtoinfinity}
\end{figure}

In figure \ref{f:Zmtoinfinity}, we show a numerical simulation of the equation of motion given by Eqs.~(\ref{eq:motiona}), (\ref{eq:motionb}), and (\ref{eq:force1}) for $v=0.1$, $m=100$, $\mu=\eta=1$, and $5 \times 10^{7}$ data points. The agreement of the simulations with the analytical result (\ref{85}) is excellent.

\subsection{Numerical solution of the Fokker-Planck equation}
\label{sec:numerics}

In this section we solve Eq.~(\ref{eq:Fokkerdimnesionless}) numerically using the discretization scheme proposed by Scharfetter and Gummel \cite{discretization} and analyze the probability distribution  for different values of the driving velocity $v$ and mass $m$.

\begin{figure}[bt]
\includegraphics[width=\columnwidth]{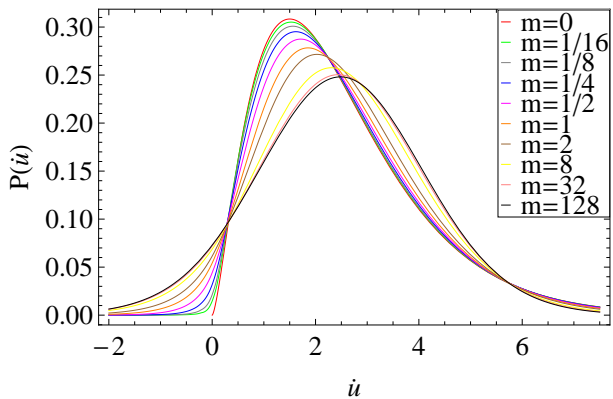}\\
\caption{(color online) Velocity probability distribution for a fixed driving velocity $v=5/2$.
Different curves correspond to different masses, see the explanation in the main text.}\label{fig:pfor2t5}
\end{figure}

The probability distribution of velocities for different masses and fixed driving velocity $v=5/2$ is shown in Fig.~\ref{fig:pfor2t5}. For $m=0$ there are no negative velocities $\dot{u}$ and the probability reaches its maximum at a  higher value than all  other curves shown in Fig.~\ref{fig:pfor2t5}. The general tendency is an increase of the probability for negative $\dot{u}$ when increasing the mass as well as a decrease of the maximum of the  probability. At the same time the maximum gets shifted towards higher velocities. We see that for  $m=32$ and $m=128$ there is only a small difference in the probability
distribution, and for large  mass the probability converges to a master curve at $m=\infty$, whose behaviour we discuss in more detail in the next section. Also note that there are two remarkable points where all  curves  intersect, a feature which remains to be understood.

\begin{figure}[hbt]
\includegraphics[width=\columnwidth]{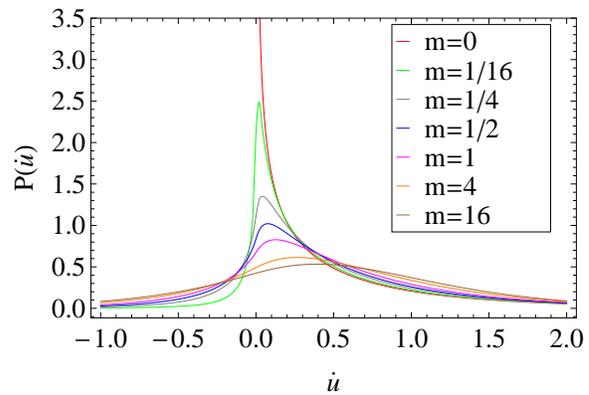}\\
\caption{(color online) Velocity probability distribution for a fixed driving velocity $v=1/2$.
Different curves correspond to different masses, see the explanation in the main text.}\label{fig:v0t5}
\end{figure}

Fig.~\ref{fig:v0t5} shows $P(\dot{u})$ for $v=1/2$ and different masses. We see that the divergence  for $m=0$ seems to disappear in the presence of a small mass, and that at the same time the probability for negative $\dot{u}<0$ becomes finite. A  small mass  changes the positive tail of the distribution only slightly. Inertia has the tendency to decrease the maximum of $P(\dot{u})$ as was noticed in Fig:~\ref{fig:pfor2t5} where $v>1$. This is also manifest for $v<1$.

\begin{figure}[b]
\includegraphics[width=\columnwidth]{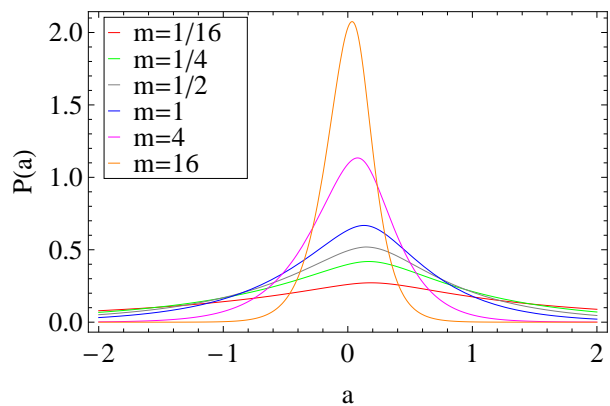}\\
\caption{(color online) Probability distribution of accelerations for a fixed driving velocity $v=1/2$.
Different curves correspond to different masses, see the explanation in the main text.}\label{fig:av0t5}
\end{figure}

Next we analyze the probability distribution of accelerations.  Fig.~\ref{fig:av0t5}  shows $P(a)$ for $v=1/2$ and different masses. The larger the mass is, the more symmetric the distribution becomes, in agreement with results from Sec.~\ref{sec:largemass}. Decreasing the mass the maximum of $P(a)$ gets lowered, while the distribution  broadens and the average of $a^2$ increases. A similar behavior is observed for $v>1$.

Apart from similarities between the ABBM model and the tree model (discussed in Sec.~\ref{sec:model}) there are also differences. By looking at Figs.~\ref{fig:numerics2} and \ref{fig:numerics2v}, we see that when the probability for negative velocities becomes considerable, a difference between the tree model and the ABBM model with inertia becomes visible. The probability distribution for the tree model is characterized by a smaller peak and larger tails than the ABBM model with inertia.

\subsection{\label{sec:comparison}Comparison of the numerical solution of the Fokker-Planck equation with perturbation expansion in \(m\) and \(1/m\)}

Here we compare numerical and analytical results from the  two previous sections focusing on the behaviour at small and large $m$. We start with small $m$.
We now have to determine the currently undetermined constants entering the distribution function, Eqs.~(\ref{eq:F0text})--(\ref{eq:F2text}). For sufficiently small $m$ and sufficiently large $v$, negative velocities appear with   small probability. Then, we can neglect their contribution to $\int\dif a\,\dif\dot{u\,}P=1$ as well as the contribution from the narrow region 2, since the main contribution comes form  region 1.  We find
\begin{align}\label{eq:constants1}
c_3=&0,\\\label{eq:constants2}
c_5=&\frac{1}{4}- v \psi(v),
\end{align}
where $\psi(x)$ is the digamma function.
In this region of parameters, in  App.~\ref{app:perturbationtheory} we state higher-order terms in $P(\dot{u})$ and additionally we calculate exactly moments characterizing the distribution function and some other correlations in Secs.~\ref{sec:mainmoments}, \ref{sec:exactm625} and App.~\ref{app:moments}.

\begin{figure}[hbt]
\includegraphics[width=\columnwidth]{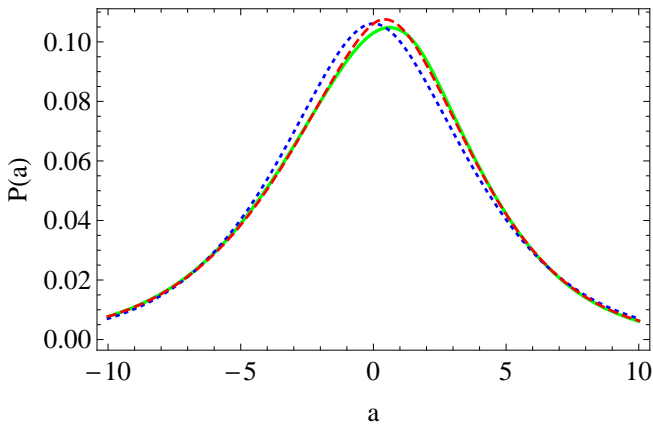}\\
\caption{(color online) The full green line represents the numerical solution for the joint probability distribution of acceleration and
velocity, integrated over positive velocities for $m=1/8$ and $v=5/2$. The dashed red line is the result of the second order perturbation theory (\ref{eq:distributiona}), while the dotted blue line is the first order of perturbation theory. }\label{fig:perturbationa}
\end{figure}

The numerical solution for the joint probability distribution of acceleration and
velocity, integrated over positive velocities for $m=1/8$ and $v=5/2$ is shown in Fig.~\ref{fig:perturbationa} by the full line. The  dotted line is the first-order, and the dashed line  the second-order perturbation theory result of Eq.~(\ref{eq:distributiona}).
We see  good agreement between analytical and numerical solutions. Also, the agreement increases with increasing  order of  perturbation theory. The small remaining difference may come from the approximation made when fixing $c_3$ and $c_5$, i.e.~due to neglecting the contributions from negative velocities and from the region $2$.

\begin{figure}[tb]
\includegraphics[width=\columnwidth]{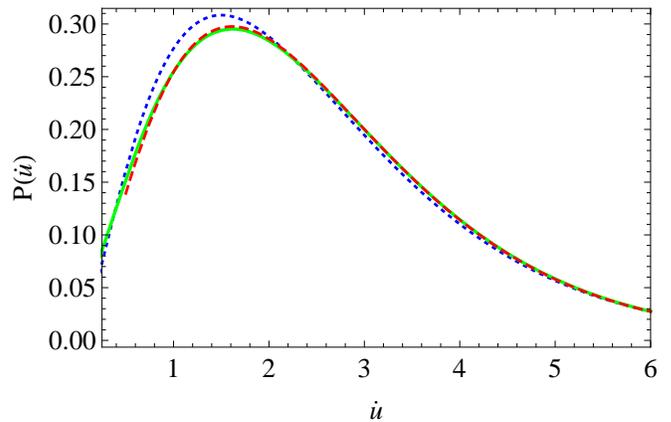}\\
\caption{(color online) The full green line represents numerical solution for velocity distribution for $m=1/4$ and $v=5/2$.  Dashed red line is the result of perturbation theory, Eq.~(\ref{eq:distributionmm}). Dotted blue line is $m=0$ curve.}\label{fig:perturbation}
\end{figure}

Let us now compare the distribution of velocities.  Fig.~\ref{fig:perturbation} presents $P(\dot{u})$ for $m=1/4$ and $v=5/2$. The full line is the numerical solution, the dotted line is $m=0,$ and the dashed line is the perturbation theory result of Eq.~(\ref{eq:distributionmm}). For $P(\dot{u})$ in the region $1$ there is very good agreement between  perturbation theory and numerical solution.

\begin{figure}[hbt]
\includegraphics[width=\columnwidth]{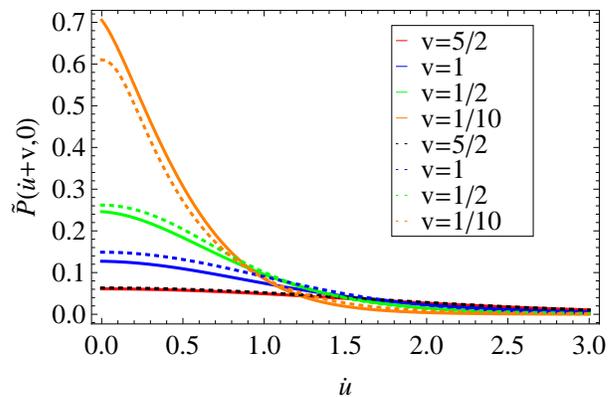}\\
\caption{(color online) The full lines represent the numerical solution for $\tilde{P}(\dot{u}+v,\tilde{a}=0)$  for $m=100$ and different values of $v$.  The dotted lines are the zeroth order of perturbation theory, Eq.~(\ref{eqdiff0}).}\label{fig:largem}
\end{figure}

Next we compare the perturbation expansion in $1/m$ studied in Sec.~\ref{sec:largemass} with the numerical solution. The distribution function becomes rotationally invariant in the $\dot{u}-\tilde{a}$ plane around  the point $(v,0)$  in the limit $m\to \infty$, see Eq.~(\ref{eq:P0}). Therefore, in Fig.~\ref{fig:largem} is shown $\tilde{P}(\dot{u}+v,\tilde{a}=0)$ for fixed mass $m=100$ and different driving velocities. The definition of $\tilde{P}$ is given by Eq.~(\ref{eq:tildeP}). Numerical solutions are shown by full lines, while dotted lines denote the zeroth order in the expansion given by solutions of the differential Eq.~(\ref{eqdiff0}) for the tree model. We see good agreement, especially for larger $v$.

\section{\(\sqrt{\dot u}\) model}
\subsection{Strategy}
In this section, we define a model which
\begin{itemize}
\item[(i)] is equivalent to the ABBM model for all trajectories such that the particle only moves forward;
\item[(ii)] is an exact saddle point of the MSR dynamical action hence allows for some analytical results.
In particular the generating function of the one-time velocity distribution is
given by
\begin{equation}\label{x1}
\hat P(\lambda) := \overline {\rme^{\lambda \dot u(t_{0})} } = \rme^{v
Z (\lambda)}\ ,
\end{equation}
where $Z (\lambda)$ does not depend on $v$.
\item[(iii)]   on the down-side $\dot{u}$ may become complex.
\end{itemize}
Although the MSR saddle point method for the ABBM model is exact only for $m=0$, its extension to
$m>0$ provides a natural approximation of the ABBM model in presence of inertia. For instance
let us define $P_{\mathrm{approx}}(\dot u)$ as the inverse Laplace transform of $\hat
P(\lambda)$, on the {\em real} $\dot{u}$ axis as:
\begin{equation} \label{approxP}
\hat P(\lambda) = \int_{-\infty}^{\infty} \rmd \dot u\, P_{\mathrm{approx}}(\dot u) \,\rme^{\lambda \dot u}\ .
\end{equation}
since $Z (\lambda)^{*} = Z(\lambda^{*})$, the function $P^{v}_{\mathrm{approx}}(\dot u)$ is real
and since $Z(0)=0$ (see below) it integrates to unity.
As velocities can become complex, using this function as an approximation of a probability
distribution (for the ABBM model) makes senses if, and  only if
\begin{itemize}
\item[(iv)] the  function $P_{\mathrm{approx}}(\dot u)$ is positive.
\end{itemize}
Although not an obvious fact, we have strong numerical evidence that this is indeed the case. Finally,
we will show in Section \ref{sec:largedeviation} that the approximation becomes exact for large $v$.

\label{sec:complex}
\begin{figure}\Fig{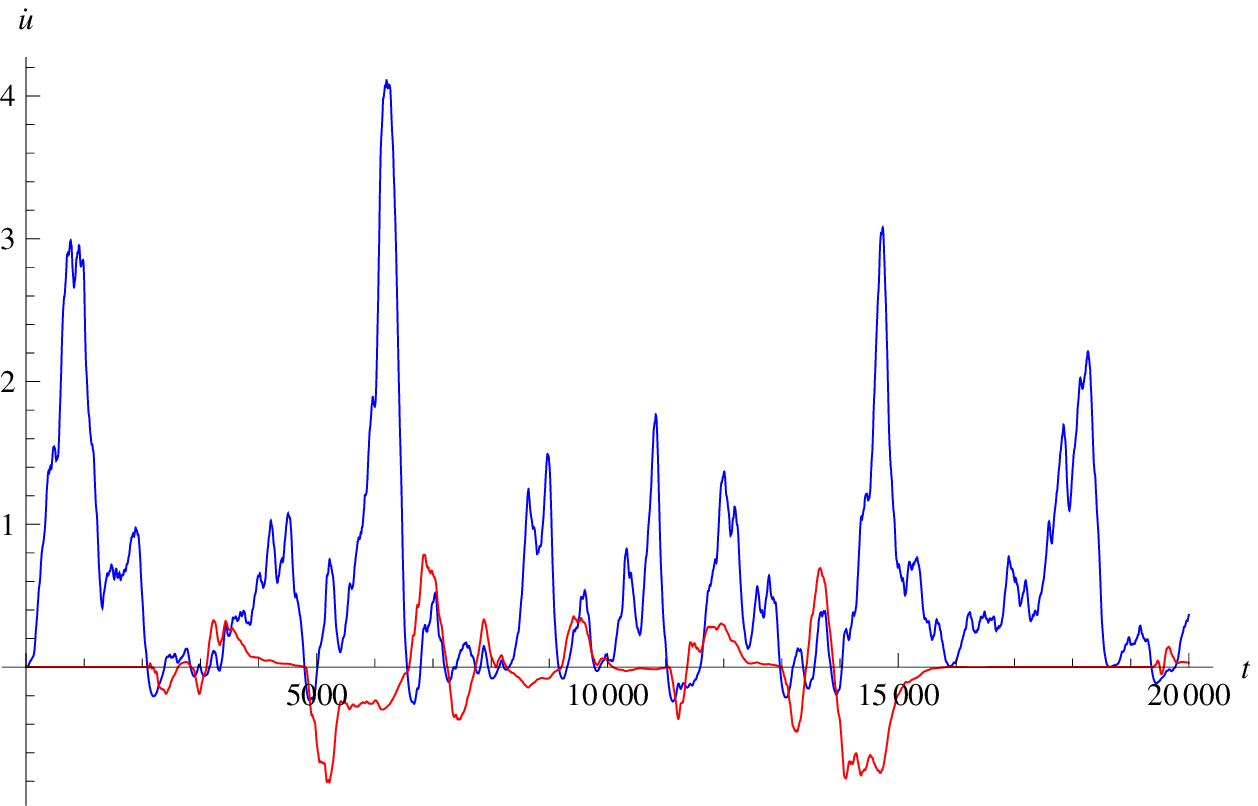}\\
\Fig{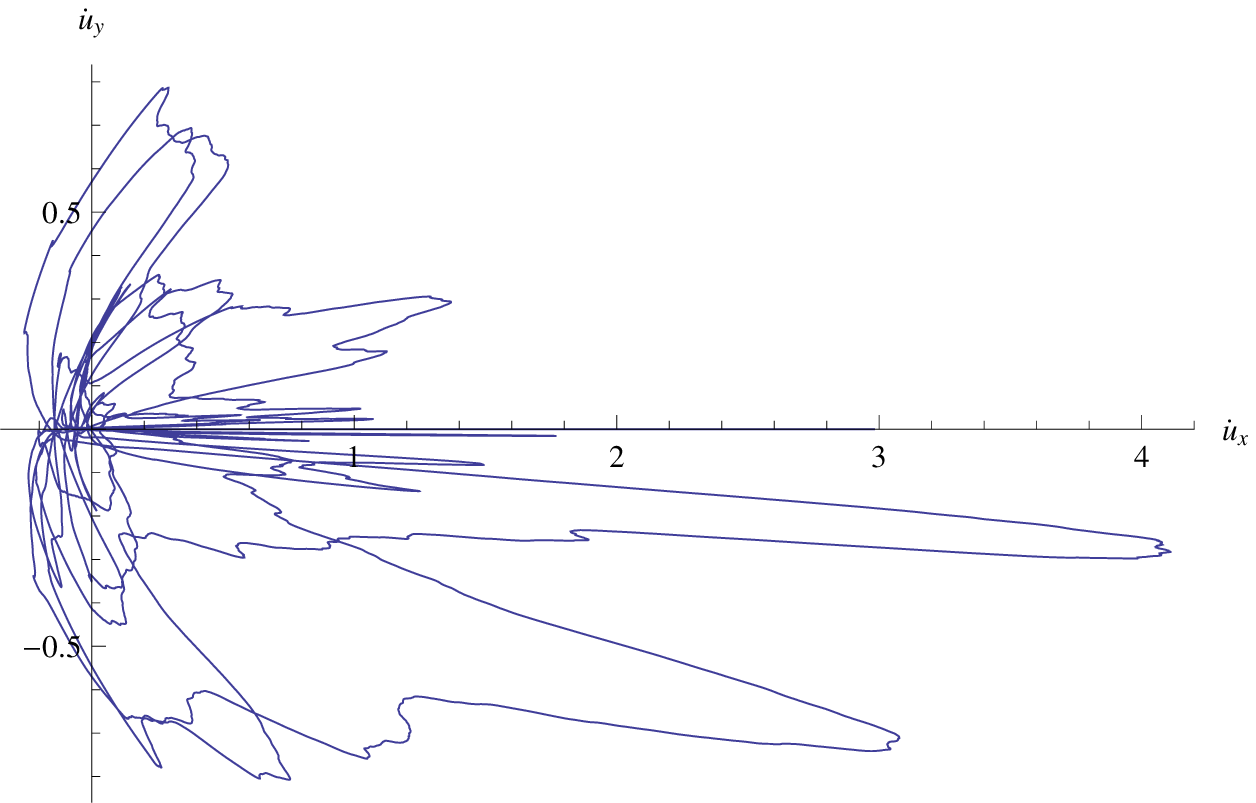}
\caption{Top: Real (blue) and imaginary (red) part of the velocity for one realization of the disorder, and \(v=0.5\). Bottom: phase portrait, i.e.\ trajectories in the complex plane \(\dot u = \dot u_x + i \dot u_y\). Note the different scale in real and imaginary directions.}\label{f:8}
\end{figure}

\subsection{Definition and basic
properties\label{sec:defcomplex}}
We define the model in the laboratory
frame,
\begin{align}\label{35}
\frac{\mathrm{d}\dot u(t)}{\mathrm{d} t}&=a(t),\\\label{36}
m\frac{\mathrm{d}a(t)}{\mathrm{d} t}&=\mu^2[v-\dot{u}(t)]-{\eta}a(t) + \sqrt{\dot u(t)} \, \xi(t)\\
\left<\xi(t) \xi(t')\right> &= 2 \sigma\, \delta(t-t') \label{37}
\end{align}
As long as the velocity is positive, the noise \( \sqrt{\dot
u(t)} \, \xi(t)\) is real, and the velocity remains real. However when
$\dot u$ becomes negative, complex velocities arise. This can be seen
on figure \ref{f:8}, for one given disorder realization, and driving
velocity \(v=0.5\) and $m=1/4$. Increasing the driving velocity, the
trajectories at the bottom of figure \ref{f:8} get closer to the real
axis, and less negative events arise. In addition,
one sees from (\ref{35}) that near the positive real velocity axis a
small imaginary part $i m \dot u = \epsilon$ experiences a linear force $- (\mu^2  - \frac{\xi(t)}{2 \sqrt{\dot u}}) \epsilon$ which on average brings it back towards the real axis except for small velocities.

Therefore, this model is expected to be a reasonable approximation to
the ABBM model with inertia for large $v$ and small $m$, see section \ref{sec:largedeviation}. Moreover, in
the same limit of parameters this model  works also well  for the
particle on the tree studied in the previous section. It will be discussed
below in section \ref{s:sudvABBM}  how good an approximation it can provide beyond this range of parameters.

While the velocities in this model can become complex, as we will show now, the generating function for the velocities is given by a single, \(v\)-{\em independent} function $Z(\lambda)$
\begin{equation}
\overline {e^{\lambda \dot u(t_0)}} = e^{v Z(\lambda)}.\label{38}
\end{equation}
The proof is based on the MSR formalism. Replacing $a(t)$ in the second line of the above Langevin equation by $\ddot u(t)$, the expectation (\ref{38}) can be written as
\cite{LeDoussalWiese2011a,LeDoussalWieseinprep2012,DobrinevskiLeDoussalWiese2011}
\begin{equation}\label{39}
\overline {e^{\lambda \dot u(t_0)}} =   \int {\cal D}[\dot u] {\cal D}[\tilde u]e^{\lambda \dot u(t_0) - {\cal S}[\dot u,\tilde u]}
\end{equation}
with the dynamical action
\begin{align}
\label{eq:MSRActionGen2}
&{S[\dot u,\tilde u] =} & &\nn \\
&  \int_t  \tilde u(t)\!\left[ m \partial_t^2 \dot u(t)+\eta\partial_t \dot u(t)+  \mu^2(\dot{u}(t)-v)  \right] - \sigma \dot u(t) \tilde u(t)^2 .\nn \\
\end{align}
In the square-brackets, we recognize the equation of motion without disorder, enforced by the response field $\tilde u(t)$. The last term  results
from the average over the disorder, using (\ref{37}).

Since the action is {\em linear} in \(\dot u\), the saddle-point w.r.t.\ this variable is {\em exact}, leading to the instanton equation
\begin{equation}\label{41}
 m \partial^2_t {\tilde u}(t)- \eta   \partial_t \tilde u(t)+\mu^2\tilde u(t)-\sigma \tilde u(t)^2 =\lambda \delta(t-t_0)
\end{equation}
This has to be supplemented with the
boundary conditions $\tilde{u}(\pm\infty)=0$. For the case $m>0$
studied here it turns out to be equivalent to requesting,
\begin{eqnarray} \label{prop1}
\tilde u(t)&=&0 \quad \forall \ t \geq t_0 \ ,
\end{eqnarray}
which implies $\partial_t \tilde u(t) =0$ for all $t>t_0$.
\footnote{note however that for $m<0$ there are solutions which vanish at infinity
and do not satisfy (\ref{prop1})}.
Supposing that (\ref{41}) holds, the only term which survives in (\ref{39}) is  the term of order $v$ in (\ref{eq:MSRActionGen2}), which yields (\ref{38}) with
\begin{equation}\label{42}
Z(\lambda) = \mu^2 \int_t \tilde u(t).
\end{equation}

\begin{figure}[t]
\Fig{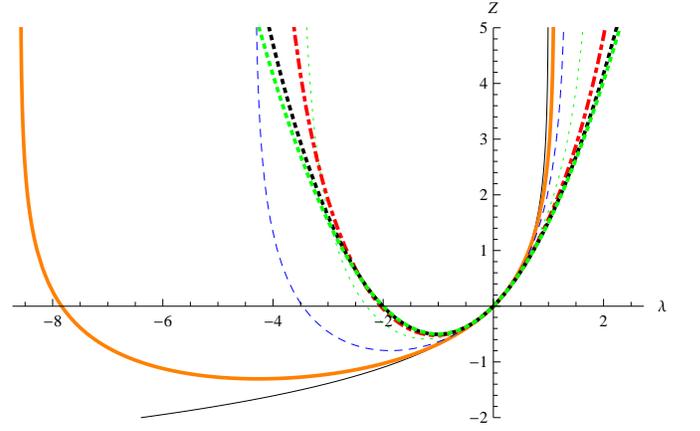}
\caption{The generating function \(Z(\lambda)\) for masses \(m=0\) (thin, solid, black), \(m=1/4\) (thick orange), \(m=1\) (blue, dashed), \(m=4\) (green, dotted), \(m=16\) (red, dot-dashed, thick), \(m=64\) (black, dotted, thick), \(m=256\) (green, dotted, thick). All functions diverge at \(\lambda=\lambda_c^+\) and  \(\lambda=\lambda_c^-\).}
\label{f:Z-diff-mass}
\end{figure}

\begin{figure}[b]
\Fig{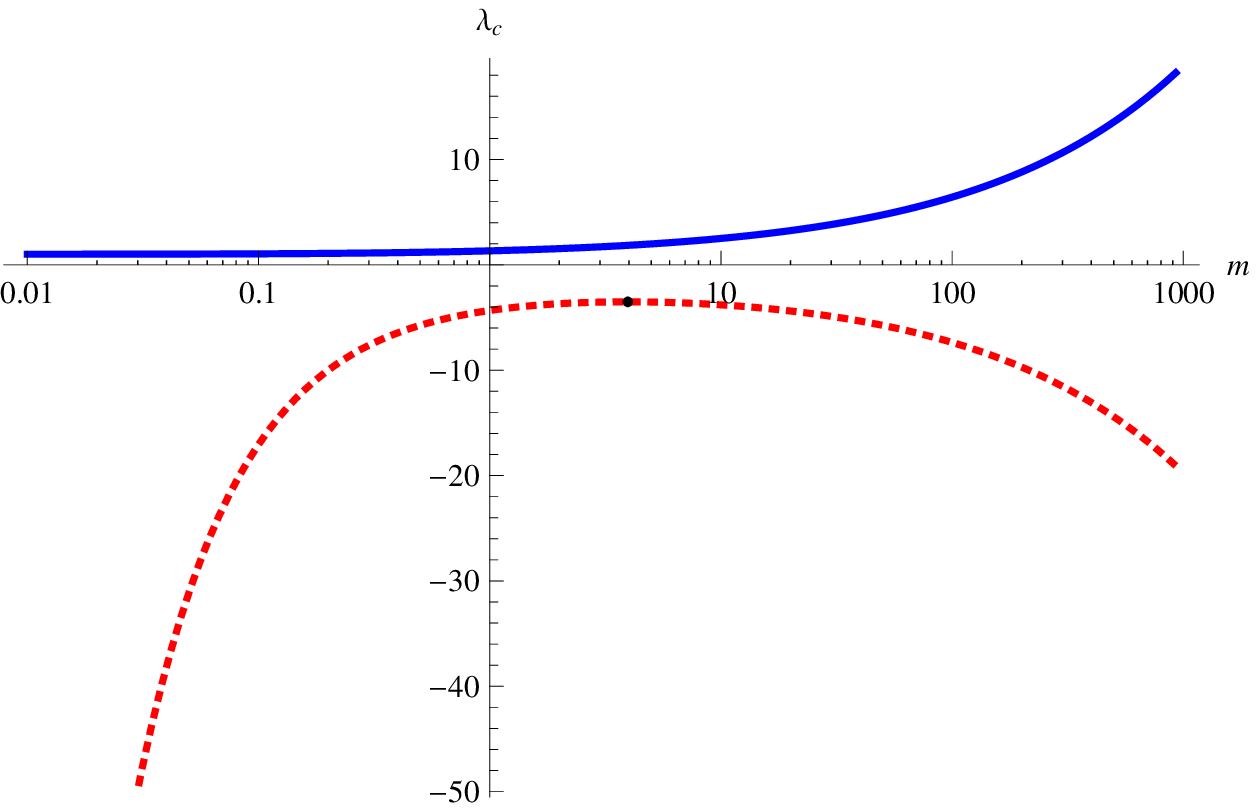}\\
\Fig{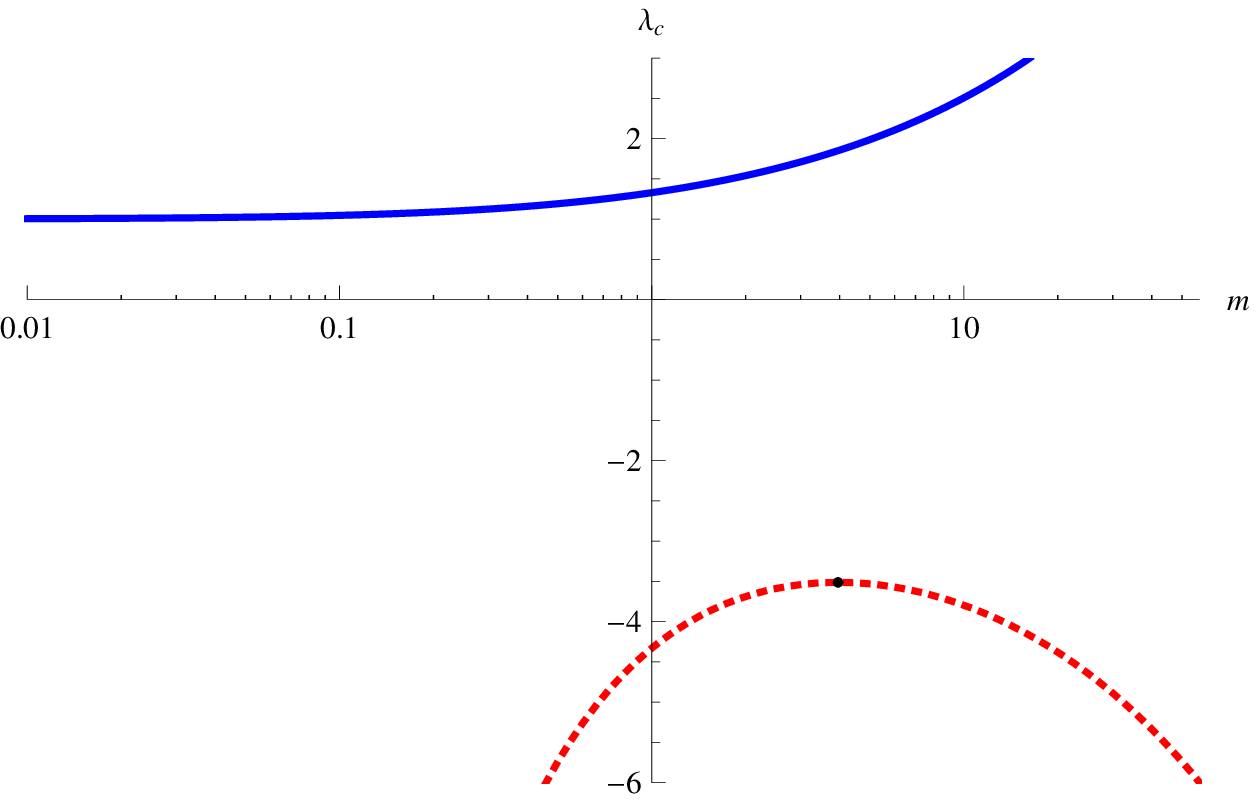}
\caption{Top: Location of the branch-cut singularity in \(Z(\lambda)\), as function of \(m\). Lower curve (red, dotted) is \(\lambda_c^-\), upper curve (blue) is \(\lambda_c^+\). The maximum of $\lambda_c^-$ is at $m=3.9540$, $\lambda_c^-=-3.5124$, marked by a dot. Bottom: Blow-up of the same curve.}
\label{f:lambdacofm}
\end{figure}

While analytical results for equations (\ref{41}) and (\ref{42}) are difficult to obtain, it is quite easy to solve the instanton equation (\ref{41}) numerically. We do this now in our dimensionless units, i.e. setting \( \eta=\mu^2=\sigma=1\). On figure \ref{f:Z-diff-mass}, we show \(Z(\lambda)\) for different masses.  For real $\lambda$, \(Z(\lambda)\) diverges at two singularities, one at $\lambda=\lambda_c^+>0$, the other at $\lambda=\lambda_c^-<0$. Their values, as a function of \(m\) is plotted on figure \ref{f:lambdacofm}.
Examination of $P_v(\lambda)$ for complex $\lambda$ show that they are branch-cut singularities.
The branch-cut singularities determine the behavior of the function $P^{v}_{\mathrm{approx}}(\dot u)$ defined in (\ref{approxP})
at large $\dot u$, up to possible power-law pre-exponential factors as
\begin{equation}
P_{\mathrm{approx}}(\dot u) \sim \left\{ {\rme^{-\dot u \lambda_c^+}\mbox{ for }\dot u \to + \infty \atop
 \rme^{\dot u \lambda_c^-}\mbox{ ~~for }\dot u \to - \infty
}\right. \ .
\end{equation}
It is important to note that this asymptotic behavior holds at {\em any} driving velocity: Were $P^{v}_{\mathrm{approx}}(\dot u) $ to decay faster (e.g. with a larger constant in front of, or a higher power of $\dot u$ in the exponential), then $\hat P(\lambda_{c})=\rme^{v Z(\lambda_{c})}$ would be finite, which it is not. Similarly, if it would decay slower, then $\hat P(\lambda_{c})=\rme^{v Z(\lambda_{c})}$ would not exist up to $\lambda=\lambda_{c}$. Note that (\ref{41}) also implies that $Z (\lambda)$ is real for real
$\lambda$, at least as long as the solution $\tilde u (t)$ decays to zero at large times, which is the case for
$\lambda_{c}^{-}<\lambda <\lambda_c^+$ (see below).

As shown in Fig.~\ref{f:lambdacofm}, increasing the mass beyond the special value $m^*=3.95402$, the value of \(\lambda_c^-\) decreases again, i.e.\ the tail for negative $\dot u$ becomes again shorter.  For $m\to\infty$ both $\lambda_c^+$ and $\lambda_c^-$ become infinitely large, meaning that $Z(\lambda)$ becomes an analytic function for all $\lambda$. This in agrement with the result found in Sec.~\ref{sec:largemass} where we demonstrate that in the limit $m\to \infty$  the distribution function becomes Gaussian (\ref{eq:GauusianM}).


Beyond allowing to define the function $P_{\mathrm{approx}}(\dot u)$, $Z(\lambda)$ contains
information on the probability distribution  $P(\dot{u}_{x},\dot{u}_{u} )$ of the process $\dot{u} (t)=\dot{u}_{x}+i
\dot{u}_{y} $ in the complex plane (see e.g. Fig. \ref{f:8}) through
\begin{equation}\label{x2}
\hat P(\lambda) = \int_{-\infty}^{\infty} \rmd \dot{u}_{x}
\int_{-\infty}^{\infty} \rmd \dot{u}_{y}  \, \rme^{\lambda[ \dot{u}_{x}+i
\dot{u}_{y} ]} P(\dot{u}_{x},\dot{u}{_y} ) \ .
\end{equation}
It does not allow however to determine $P(\dot{u}_{x},\dot{u}{_y} )$
since we would need a more general generating function for the moments
of $\dot{u}^{*} =  \dot{u}_{x}-i \dot{u}_{y} $. A question is what information
can be extracted, and how does it relate to $P_{\mathrm{approx}}(\dot u)$.
For illustration we can consider a simple toy example such that the probability
factorizes $P(\dot{u}_{x},\dot{u}{_y} ) = P_{x} (\dot{u}_{x}) P_{y} (\dot{u}_{y})$
so that $\hat{P} (\lambda) $ can be written as $\hat{P} (\lambda) = \hat{P}_{x} (\lambda) \hat{P}_{y} (i\lambda )$
where $ \hat{P}_{x} (\lambda)$ and $ \hat{P}_{y} (\lambda)$ are the
Laplace-transforms of $P_{x} (\dot{u})$ and  $P_{y} (\dot{u})$
respectively. From our numerical studies we find that $P
(\dot{u}_{x},\dot{u}{_y} )$ is narrowly distributed in the
$y$-direction, with a width which decreases with increasing
$v$. For large $v$,
$P (\dot{u}_{x}, \dot{u}_{y} )$ will converge against $P_{\mathrm{approx}}
(\dot{u}_{x}) \delta (\dot{u}_{y} ) $, at least for $\dot{u}_{x}>0$.
For finite $v$ consider the simplest example, i.e. a Gaussian for $P_{y} (\dot{u}_{y})$
which leads to
\begin{equation}\label{x6}
\hat{P}_{v} (\lambda) = \hat{P}_{x} (\lambda)  e^{\frac{-\alpha  \lambda ^2}{2}}   \ .
\end{equation}
As discussed above, we found that $\hat{P} (\lambda)$ has two branch-cut
singularities starting at a finite real $\lambda_{c}^{+}$ and
$\lambda_{c}^{-}$. Since the latter control the behavior at large
$\dot{u}$ and $\dot{u}_{x}$, both $P_{\mathrm{approx}} (\dot{u})$ and
$P_{x} (\dot{u}_{x})$ have the same behavior at large (positive
or negative) arguments for this toy example.
We will not pursue further here the study of
$P(\dot{u}_{x},\dot{u}{_y})$.

\subsection{Connection between instanton approach and Fokker-Planck approach\label{sec:instanton}}

We saw in Section \ref{sec:ABBM} that the ABBM model {\em without} inertia can be solved either using the Fokker-Planck equation,
or  the MSR method based on the  non-linear saddle-point  equation (the instanton). We recalled
their equivalence based on the method of characteristics. The same property holds for the $\sqrt{\dot u}$ model,
with a small caveat. The caveat
is that the correspondence  is simple only between the instanton equation and the
{\it Fokker-Planck equation in its Laplace transformed version} Eq.\ (\ref{eq:laplacetransformed}) below.
In real space it corresponds formally to the FP equation (\ref{eq:Fokkerdimnesionless})
where the replacement $|\dot u| \to \dot u$ is made in the diffusion term. This would indeed be
the FP equation associated to the Langevin process (\ref{35}), (\ref{36}) if one forgets that $\dot u$ can become
negative, and consequently complex. Writing an adequate FP equation for such a process requires an
extension of $P(\dot u, a,t)$ with arguments in the complex plane, a route which we do not follow here.

Instead, consider the evolution equation for the following average over trajectories
\bea \label{observ}
\hat{P}(\lambda,\kappa,t)= \overline{ e^{\lambda \dot u(t) + \kappa a(t) } }\ .
\eea
It is in principle an integral $\int \rmd\dot u\,\rmd a\,  e^{\lambda \dot{u}+\kappa a}P(\dot{u},a,t)$ over  the complex \(\dot u\) and \(a\) plane. The average (\ref{observ}) satisfies the Laplace version of the FP equation,
\begin{align}\label{eq:laplacetransformed}
&\frac{\partial \hat{P}}{\partial t}-\frac{\partial \hat{P}}{\partial \kappa}\left(\lambda- \frac{\kappa}{m}\right)-\frac{\partial \hat{P}}{\partial \lambda}\left(- \frac{\kappa}{m}+\frac{\kappa^2}{m^2}\right)= \frac{\kappa}{m}v\hat{P},
\end{align}
in dimensionless variables. As mentioned above it is {\it formally} the Laplace transform (over the real axis)
of the FP equation with a $\dot u$ diffusion term. Eq.\ (\ref{eq:laplacetransformed}) is easily derived from the
Langevin equation (\ref{35}), (\ref{36}) by considering the variation of the observable in an infinitesimal
time interval $\rmd t$. The only subtlety arises when expanding the variation
$\overline{ e^{\frac{\kappa} m \sqrt{\dot u(t)} \rmd\xi(t)} } =1 + \frac{1}{2} (\frac{\kappa} m)^2 \dot u(t) \overline{\rmd\xi(t)^2} + O(\rmd t^2)$
to second order, with $\overline{\rmd\xi(t)^2}=2 \sigma \rmd t$, using stochastic calculus (It\^o). This leads to the
term $(\kappa^2/m^2) \partial_\lambda \hat P$ in (\ref{eq:laplacetransformed}).

We now show the connection to the MSR method and the instanton equation. The solution of (\ref{eq:laplacetransformed}) can be written in the form
\be\label{eq:defforZ}
\hat{P}(\lambda,\kappa,t)=e^{v Z(\lambda,\kappa,t) }\ ,
\ee
where $Z$ is independent of $v$ and satisfies
\begin{align}
\label{eq:partialeqZM}
\frac{\partial Z}{\partial t}-\frac{\partial Z}{\partial \kappa}\left(\lambda- \frac{\kappa}{m}\right)-\frac{\partial Z}{\partial \lambda}\left(- \frac{\kappa}{m}+\frac{\kappa^2}{m^2}\right)= \frac{\kappa}{m}.
\end{align}
To solve this equation we again apply the method of characteristics,
and reduce the partial differential equation to a family of ordinary differential equations,
\begin{eqnarray}
\label{eq:kappaM}
\dot{\kappa}(t)&=&-\lambda(t) +\frac{\kappa(t)}{m},\\
\label{eq:lambdaM}
\dot{\lambda}(t)&=&\frac{\kappa(t)}{m} -\frac{\kappa(t)^2}{m^2},\\
\label{eq:ZMM}
\mathrm{d}Z(t)&=&\frac{\kappa(t)}{m}\mathrm{d}t.
\end{eqnarray}
We have defined $Z(t) = Z(\lambda(t),\kappa(t),t)$, as well as
$\dot{\kappa}(t)=\dif \kappa(t)/\dif t$ and $\dot{\lambda}(t)=\dif \lambda(t)/\dif t$.

Let us now introduce
\bea
\tilde{u}(t)=\kappa(t)/m
\eea
Eliminating $\lambda(t)$ in (\ref{eq:kappaM}) and
 (\ref{eq:lambdaM}) we find that it satisfies
\bea
\label{eq:kappamotionM}
{m\ddot{\tilde{u}}(t)}- \dot{\tilde{u}}(t)+\tilde{u}(t)-\tilde{u}^2(t)=0.
\eea
If we impose $\tilde{u}(t)=0$ and $\dot{\tilde{u}}(t)=0$ for $t>t_0$ and assume $\lambda(t_0)=\lambda$ and $\kappa(t_0)=0$, this equation becomes the instanton equation (\ref{41}) in dimensionless units. To obtain the stationary solution $Z(\lambda,\kappa)$
of (\ref{eq:partialeqZM}) we can solve (\ref{eq:kappamotionM}) on the interval $ t \in ]-\infty,t^*]$,
with boundary conditions at $t=t^*$:
\begin{align}
\label{eq:kappa}
& \tilde{u}(t^*)=\frac{\kappa}{m},\\
\label{eq:lambda}
& \dot{\tilde{u}}(t^*)=-\frac{\lambda}{m} +\frac{\kappa}{m^2},
\end{align}
and $\tilde{u}(-\infty)=0$, $\dot{\tilde{u}}(-\infty)=0$. Then we compute\begin{align}
\label{eq:ZM}
Z(t^*) = \int^{t^*}_{-\infty}\tilde{u}(t)\,\mathrm{d}t := Z(\lambda,\kappa)\ ,
\end{align}
which is precisely $Z(\lambda,\kappa)$ if expressed as a function of
the boundary condition. Since it does not depend on $t^*$,  we  have found the stationary solution.

Hence we have shown, via Eq.~(\ref{eq:laplacetransformed}), that the observable (\ref{observ}) can be obtained from the solution of the instanton equation, although with a slightly more general boundary condition than in (\ref{41}). This is because we now want the
joint distribution of velocity and acceleration (at a given time $t^*$). We can rewrite the observable (\ref{observ}) as\be \hat{P}(\lambda,\kappa,t) =
\overline{ \rme^{\int \rmd t [\lambda \delta(t-t^*) - \kappa \delta'(t-t^*) ] \dot u(t)} } \ .
\ee
Performing the same manipulations using the dynamical MSR action
as in Refs. \cite{LeDoussalWiese2011a,LeDoussalWieseinprep2012,DobrinevskiLeDoussalWiese2011}
as sketched above in Section \ref{sec:defcomplex}, we arrive at the instanton equation
with a source on the right-hand-side $\lambda \delta(t-t^*) - \kappa \delta'(t-t^*)$
which is equivalent to the boundary conditions (\ref{eq:kappa}), (\ref{eq:lambda}).

To summarize, for the $\sqrt{\dot u}$ model the equation (\ref{eq:laplacetransformed}) describes  the time evolution of the observable (\ref{observ}) under the Langevin equation (\ref{35})--(\ref{37}), even though $\dot u$ and $a$ take values in the complex plane. Solving it is equivalent to solving the instanton equation as a function of $\lambda$ and $\kappa$ that determine its boundary conditions (\ref{eq:kappa}) and (\ref{eq:lambda}). Neither equation
admits solutions in closed analytical form for generic values of $m$, but each
has his advantages for numerical or perturbative studies.
For instance, from Eq.~(\ref{eq:laplacetransformed}) we can  easily obtain the moments for this model, as we now show.
Also, note that the Laplace transform of the Fokker-Planck equation for the tree model can also be studied although it is more involved, see App.~\ref{app:deviation}.

\subsection{Moments of the distribution function\label{sec:mainmoments}}

In the stationary case, taking the derivatives  $\partial_{\lambda}^n\partial_{\kappa}^m$ of Eq.~(\ref{eq:laplacetransformed}) where $m$ and $n$ are positive integers and afterwards setting $\kappa$ and $\lambda$ to zero, one obtains a set of linear equations for $\partial_{\lambda}^n\partial_{\kappa}^m\hat{P}(\lambda,\kappa)|_{(0,0)}$. By solving them and using $\partial_{\lambda}^n\partial_{\kappa}^m\hat{P}(\lambda,\kappa)|_{(0,0)}
=\overline{\dot{u}^n a^m}$ and $\hat{P}(0,0)=1$ one finds   \footnote{The
result is exact in the region of parameters discussed at the very beginning
of this section.} the moments characterizing the distribution function for the $\sqrt{\dot u}$ model:
\begin{align}
&\overline{\dot{u}^0 a^1}=0,\\
&\overline{\dot{u}^0 a^2}=\frac{v}{m},\\
&\overline{\dot{u}^0 a^3}=-\frac{2 v}{m (m+2)},\\
&\overline{\dot{u}^0 a^4}=\frac{3 v^2}{m^2}+\frac{6 (5 m+3) v}{m^2 (m+2) (4 m+3)}\\
&\overline{\dot{u}^1 a^0}=v,\\
&\overline{\dot{u}^1 a^1}=0,\\
&\overline{\dot{u}^1 a^2}=\frac{v^2}{m}+\frac{2 v}{m (m+2)},\\
&\overline{\dot{u}^1 a^3}=-\frac{2 v^2}{m (m+2)}-\frac{6 v}{m (m+2) (4 m+3)},\\
&\overline{\dot{u}^2 a^0}= v (v+1),\\
&\overline{\dot{u}^2 a^1}=0,\\
&\overline{\dot{u}^2 a^2}=\frac{1}{m (m+2) (4 m+3)}(4 m^2 v^3+4 m^2 v^2\notag\\&+11 m v^3+27 m v^2+10 m v+6 v^3+18 v^2+12 v),\\
&\overline{\dot{u}^3 a^0}=\frac{4 v }{m+2}+v^3+3v^2,\\
&\overline{\dot{u}^3 a^1}=0.
\end{align}
For brevity we stated only the first few moments, but the procedure can be easily extended to higher moments.
Some of them are given in App.~\ref{app:moments}.

\begin{figure}
\includegraphics[width=\linewidth]{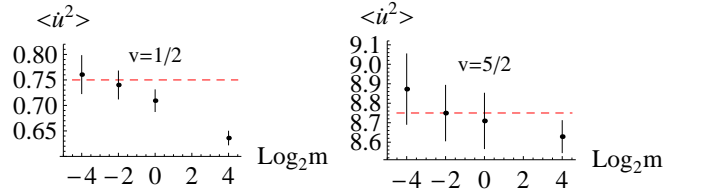}\\
\caption{Average value of $\dot{u}^2$ for the ABBM model with inertia as a function of mass $m$ is shown for two driving velocities $v=1/2$ and $v=5/2$. Dashed line represents the value $v(v+1)$ while the points are the numerical results.
}\label{fig:moments}
\end{figure}

While these results are exact for the $\sqrt{\dot u}$ model we can compare them with the ABBM model with inertia and for the particle on the tree model. In Fig.~\ref{fig:moments} we show $\langle \dot{u}^2\rangle$ as a function of mass for two different driving velocities
for the ABBM model with inertia. For $v=1/2$ we see disagreement between the above obtained $v(v+1)$ value and the numerical result for $m=1$, while for the larger velocity the deviation appears at larger values of the mass. Similar behavior is observed for the tree model.

In the App.~{\ref{app:moments}} the moments of the velocity are computed using an iterative solution of the instanton equation instead of Fokker-Planck equation.

\subsection{Perturbation at small \(m\) for \(P(\dot u)\) \label{sec:perturbation2}}

In this section we obtain the perturbative expansion in small $m$ of the generating function $Z(\lambda)=Z(\lambda,\kappa=0)$ for the steady state
velocity distribution function of the $\sqrt{\dot u}$ model. To this aim we will solve perturbatively in $m$ the non linear instanton equation (\ref{eq:kappamotionM}) and use (\ref{eq:ZM}) to obtain $Z(\lambda)$ from the solution.

The solution $\tilde u(t)$ of (\ref{eq:kappamotionM}) at $m=0$ is given in (\ref{soluabbm}).
It is non-zero for $t<0$, zero for $t>0$ with a jump at $t=0$. In presence of inertia $m>0$
we are now looking for a solution non-zero for $t<0$, zero for $t>0$, but which vanishes at
$t=0$, i.e. $\tilde{u}(0)=0$ since we have set $\kappa=0$, and with $\dot{\tilde{u}}(0)=-{\lambda}/{m}$.
For small $m$ it is then clear that there is a boundary layer for $t = O(m)<0$ where the function varies rapidly
from values of order $0$ to values of order $\lambda$. In addition there is a bulk region $t = O(1) < 0$.
We now study both regions and how they match.

As will become clear below, in the boundary layer region $0<-t\lesssim m$ we can write
\begin{align}\label{eq:1}
\tilde{u}(t)=
\sum_{n=0}^{\infty}m^n f_n\left(\frac{t}{m}\right)
\end{align}
with $\tilde{u}(0)=0$ and $\dot{\tilde{u}}(0)=-{\lambda}/{m}$.
Inserting into the instanton equation (\ref{eq:kappamotionM}) we then find the following
recursion:
\begin{align}
&f_0(t)=\lambda (1-e^{t}),\\\label{eq:recursionf}
&\ddot{f}_{k+1}(t)-\dot{f}_{k+1}(t)+f_{k}(t)-\sum_{\ell=0}^{k}f_{\ell}(t)f_{k-\ell}(t)=0,
\end{align}
 for $k\geq0$. Note that the term $f_0(t)$ already exists in absence of disorder and
 describes the rounding of the disorder-free response function by the mass.
We assume $f_n(0)=0$ and $\dot{f}_n(0)=0$ for $n>0$. Then we find
\begin{align}
f_1(t)=&\frac{\lambda}{2} \Big[4 - 2 t (\lambda-1) - 5 \lambda \Big]  +
\frac{\lambda^2}{2} e^{2 t} \notag\\& +
e^{t} \lambda \Big[2 \lambda-2 -t( 2 \lambda-1)\Big].
\end{align}
Higher-order terms can be  found, but for brevity  are not state  here. In the App.~{\ref{app:perturbationinstanton}} we give $f_2$. In
general  $f_n$ is a sum of exponentials and a polynomial, of the form
$f_n(t)=\sum_{k=1}^{n+1}e^{kt}\mathcal{C}_k^{(n)}(t,\lambda) +\sum_{k=0}^nA_{k}^{(n)}(\lambda)(-t)^k$
further discussed in App.~{\ref{app:perturbationinstanton}}.

In the region $-t = O(1) \gg m$ the solution can be written as
\begin{align}\label{eq:2}
&\tilde{u}(t)=\sum_{n=0}^{\infty}m^n y_n(t)
\end{align}
where $y_n$ satisfy the following differential equations
\begin{align}
\label{eq:yo}
&-\dot{y}_0+y_0-y_0^2=0,\\\label{eq:yn}
&\ddot{y}_{k-1}(t)-\dot{y}_k(t)+y_k(t)-\sum_{\ell=0}^{k}y_{\ell}(t)y_{k-\ell}(t)=0.
\end{align}
The last line is for $k>0$. The boundary condition is $y_n(-\infty)=0$. Eqs.~({\ref{eq:yo}) and (\ref{eq:yn}}) can be solved recursively.  This gives for each $y_n(t)$ a first-order differential equation, for which we need to fix $y_n(0)$ by  the condition that solutions in different regions (\ref{eq:1},\ref{eq:2}) match at $-t$ small but $- t/m$ large. We discuss this in detail in App.~\ref{app:perturbationinstanton}. Here we state the solution
\begin{align}\label{eq:y0}
y_0(t)=&\frac{\lambda}{\lambda+(1-\lambda)e^{-t}},\\ \label{eq:y1} y_1(t)=&-\frac{e^{-t} \lambda}{2 (-e^{-t} (-1 + \lambda)+\lambda)^2} \big[-4 - 2 t (-1+\lambda)\notag\\ &+ 5 \lambda-4 (-1 + \lambda) \log{(e^{-t} + \lambda - e^{-t} \lambda)}\big].
\end{align}
Additionally, in  App.~\ref{app:perturbationinstanton} we give $y_2(t)$.
Now we can determine
\begin{align}\label{eq:Zexpansion}
Z(\lambda,0)=\sum_n m^n Z_n(\lambda)
\end{align}
given by Eq.~(\ref{eq:ZM}) with $t^*=0$. We find $Z_0=\int^{0}_{-\infty}\mathrm{d} t\, y_0(t)$ and $
Z_n=\int^{0}_{-\infty}\mathrm{d} t \sum_{k=1}^{n}e^{kt}\mathcal{C}^{(n-1)}_{k}(t,\lambda)+\int^{0}_{-\infty}\mathrm{d} t y_n(t)$.
Hence there is a contribution from both regions. The contribution of the $A_k^{(n)}$ terms is already
taken into account through the contribution of the $y_n(t)$ at small $t$. Calculating the integrals we obtain:
\begin{align}\label{eq:Z0}
Z_0&=-\log{(1-\lambda)},\\\label{eq:Z1}
Z_1&=-\lambda -
 \frac{1}{2} \lambda \left(\frac{\lambda}{1 - \lambda} + \frac{2 \log{\left(1 - \lambda\right)}}{\lambda}\right),\\
Z_2&=\frac{\lambda  (\lambda  (5 \lambda
   (\lambda +2)-36)+24)}{24 (\lambda
   -1)^2}+\ln(1-\lambda).
\end{align}
Now  $\hat{P}(\lambda,0)=\overline{e^{\lambda\dot{u}}}=e^{vZ(\lambda,0)}$ and from that follows the perturbative expansion of the moments $\overline{\dot{u}^k}$ in agreement with the  results in Sec.~\ref{sec:mainmoments}. Assuming that for small mass we can ignore contribution in $\hat{P}$ coming from complex velocities we perform the inverse Fourier transform of $\hat{P}$ and find the distribution function
\begin{align}\label{eq:pu}
 & P_{approx}(\dot{u})=\frac{e^{-\dot{u}}
   \dot{u}^{v-1}}{\Gamma (v)}\theta(\dot{u})-\frac{m e^{-\dot{u}}
   \dot{u}^{v-2}}{
   2 \Gamma (v)}\notag\\&
   \times \left(-2 \dot{u} v
   \log \left(\dot{u}\right)+2
   \dot{u} v \psi
   (v)+\dot{u}^2-v^2+v\right)\theta(\dot{u}), \nn \\
   & + O(m^2)
\end{align}
where $\theta(x)$ is the Heaviside theta function, and $\psi(v)$ is the digamma function. The result to this order is thus a probability  normalized to unity (to this order) for $v>1$. The correction $O(m^2)$ is given in appendix \ref{app:perturbationtheory} by Eq.~(\ref{eq:distributionmm}) and is a bona-fide distribution
for $v>2$. It also gives nonzero contribution for $\dot{u}>0$ only and it is normalized to zero. Note that there is no a priori reason why at a fixed $m>0$ the inverse Laplace transform of $Z(\lambda)$ should exist, and furthermore with support on $\dot u>0$ since the present model leads to complex velocities. However, it appears that (i) for large enough driving velocity $v>n$ the perturbative result can be trusted up to order
$O(m^n)$, i.e. the negative velocity events are sufficiently rare not to spoil the result to that order; (ii) the perturbative result seems to be correct for any $v$ provided $\dot{u}\gtrsim m$. These findings are consistent with the analysis for the tree model, for which
we found the same result (\ref{eq:distributionm}) in the region $\dot{u}\gtrsim m$ although with unknown constants
$c_3$ and $c_5$ while for the $\sqrt{\dot u}$ model they are fixed by Eqs. (\ref{eq:constants1},\ref{eq:constants2}).

\subsection{Critical case $m=1/4$}\label{sec:mcritical}

Here we analyze the specific case of \(m=1/4\), and work in our dimensionless units.
For the system {\it without disorder} this is the critical case which separates the over-damped to under-damped regime where
the particle starts to oscillate. It is also the mass which brings the particle back to the origin of the parabola in the shortest time.
In presence of disorder there is a priori nothing special to this case. Since, however, perturbative formulas simplify for
$m=1/4$ our aim in this Section is to obtain to high numerical accuracy the behaviour of $Z(\lambda)$
and of $P_{\mathrm{approx}}(\dot u)$. While the results are specific to $m=1/4$, they are representative of other masses.

Since neither $Z(\lambda)$ nor $P_{\mathrm{approx}}(\dot u)$ can be obtained exactly, we consider
several schemes to compute them to excellent precision (i) a perturbative expansion in \(\lambda\); (ii) an analysis of $Z$ close to the branch-cuts, and (iii) the behavior of \(Z\) for $\lambda\to \infty$. Each of these schemes gives an estimates of $P_{\mathrm{approx}}(\dot u)$,
(the latter two are are denoted $P_{\mathrm{BC}}(\dot u)$ and $P_{\mathrm{asymp}}(\dot u)$ below) depending on which region
of the complex plane dominates in the Laplace inversion. They are valid for different values of $v$ and $\dot u$, as discussed below.
In addition we use numerical solution of the instanton equation. At the end we compare our results to the numerical simulations on the
ABBM model.

\subsubsection{Perturbative expansion for \(Z(\lambda)\)}
\begin{figure}
\Fig{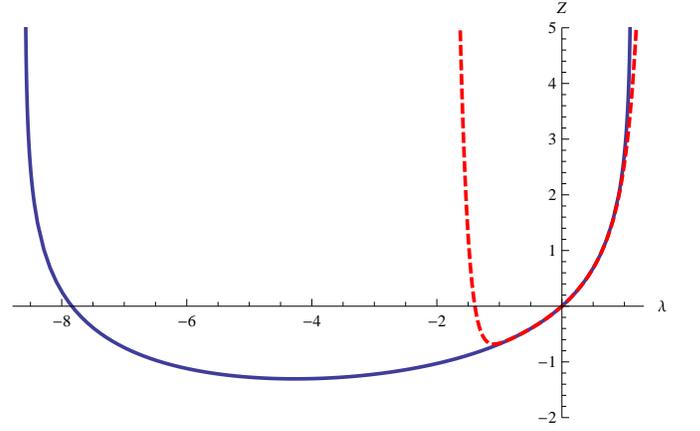}
\caption{\(Z(\lambda)\) in the critical case \(m=1/4\), together with its Taylor expansion (\ref{48}).}
\end{figure}

The response function is \begin{equation}
R(t)=4 t \rme^{-2 t} \Theta(t)\ , \qquad \left( \frac{\partial_t}{2}+1 \right)^{\!2} R(t) = \delta(t)
\label{e:46}\end{equation}
The instanton equation (\ref{41}) becomes
\begin{equation}\label{inst-eq-crit}
\left(1-\frac{\partial_{t}}2 \right)^{\!2} \tilde u_{t} -\tilde u_{t}^{2} = \lambda \delta (t)
\end{equation}
The boundary conditions are given by Eq.~(\ref{prop1}). Eq.\ (\ref{inst-eq-crit}) can be solved iteratively in \(\lambda\), as shown in App.~\ref{app:moments}. Integrating over time yields a perturbative expansion of $Z(\lambda)$,
\begin{eqnarray}
Z(\lambda) &=&\lambda +\frac{\lambda ^2}{2}+\frac{8 \lambda ^3}{27}+\frac{29 \lambda
   ^4}{144}+\frac{4094 \lambda ^5}{28125}+\frac{599431 \lambda
   ^6}{5467500}\nn\\&&+\frac{650366396 \lambda ^7}{7657689375}+\frac{4815122286049
   \lambda ^8}{71693475840000}\nn\\&&+\frac{289088854220889511 \lambda
   ^9}{5357932381952640000}\nn\\&&+\frac{16329405133190713144717 \lambda
   ^{10}}{372078637635600000000000}\nn\\&&+\frac{47848267999001244408297501187913
   \lambda
   ^{11}}{1326979721280974188091895000000000}\nn\\&&+\frac{89670625665999315214636277707
   2877141 \lambda ^{12}}{30017728909617033561025674240000000000}\nn\\&&+O\left(\lambda
   ^{13}\right)\label{48}
\end{eqnarray}
The branch cuts described in Section \ref{sec:defcomplex} are at \begin{equation}\label{49}
\lambda_c^+=1.10647\ , \qquad \lambda_c^-=-8.58563\ .
\end{equation}
\begin{figure}
\Fig{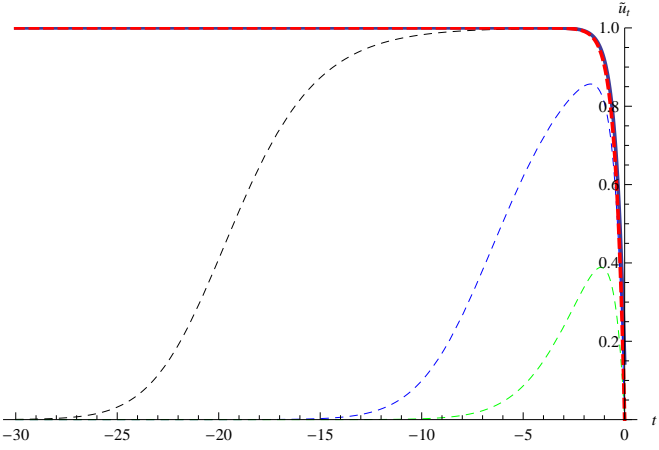}
\caption{The function $\tilde u_{t}$ is increasing with increasing $\lambda$, as seen by numerically integrating the instanton solution:
$\lambda =1/2$ (green, dashed, smallest curve),  $\lambda =1$
(blue, dashed), $\lambda =1.106$ (black, dashed), $\lambda
=\lambda_{c}^{+}$ (red/blue, thick). The last curve shows superimposed the numerical solution of the instanton equation, and of the expansion (\ref{147}).   }
\label{f:ut-crit}
\end{figure}
\subsubsection{Perturbative expansion for the critical instanton and \(\lambda_c^+\)}
The value of the branch cuts can either be obtained from the numerical solution of the instanton equation, or via the following observation: The critical instanton, \(\tilde u^+_c(t):=\tilde u_{\lambda_c^+}(t)\) converges for $t\to -\infty$ rapidly against 1, see figure \ref{f:ut-crit}. Thus while \(Z(\lambda_c^+)=\infty\), the solution of the instanton equation is still defined at this point. This allows for a series expansion, making  the ansatz
\begin{equation}\label{a2}
\tilde u_c^+ ({t+t^{*}}) = 1-\rme^{\alpha t} + \sum_{n=2}^{\infty} b_{n}
\rme^{n \alpha t}\ .
\end{equation}
The parameters \(\alpha>0\) and $b_n$ have to be determined, such that (\ref{a2}) satisfies the instanton equation (\ref{41}), and finally $t^*$ is chosen such that \(u_c^+ ({t^{*}}) =0.\)
To find the parameters, expand (\ref{a2}) inserted into the instanton equation (\ref{41}) in a series in $y:=\rme^{\alpha t}$. At first
order in $y$, the condition is $4+4\alpha -\alpha^{2}=0$, leading to
$\alpha =2(1+\sqrt{2})$. Then solve order by order in $y$, determining the $b_n$. Finally, find
$t^{*}$, such that $\tilde u_c^+ (t=0)=0$. (This is necessary since the
higher orders shift the origin.) The result for the critical instanton is
\begin{eqnarray}\tilde u_c^+ (t)&=&1-1.08835 y+0.0935856 y^2-0.00550998 y^3
\nn\\&& +0.000284501 y^4-0.00001369
   y^5\nn\\&&+6.30567\times 10^{-7}
   y^6-2.81909\times 10^{-8}
   y^7\nn\\&&+1.23335\times 10^{-9}
   y^8-5.30802\times 10^{-11}
   y^9\nn\\&&+2.25516\times 10^{-12}
   y^{10}-9.48217\times 10^{-14}
   y^{11}\nn\\&&+3.95296\times 10^{-15} y^{12}+O(y^{13})\ \ \ \ \ \ \label{147}
~~\end{eqnarray}
This determines \(\lambda_c^+=1.10647\).
Of course, all coefficients could be given explicitly. Also note that already for \(n=2\) one  gets \(\lambda_c^+\) within an error of \(0.01\).
Each further order in \(n\) improves the accuracy of \(\lambda_c^+\) by about one order of magnitude.
Thus one could use this algorithm as an efficient estimator for \(\lambda_c^+\), as a function of \(m\).

A similar solution could be constructed for  \(\lambda_c^-\).

\subsubsection{A good approximation for \(Z(\lambda)\) from the branch cuts}
There is an astonishingly good approximation for \(Z(\lambda)\), which is dominated by the branch cuts:
\begin{equation}\label{a10}
Z_{\mathrm{BC}} (\lambda) = -1.20711  \log
\left(\frac{(\lambda-\lambda_{c}^{-} ) (\lambda - \lambda_{c}^{+})}{\lambda_{c}^{-}\lambda_{c}^{+}} \right)
\end{equation}It is obtained by  starting from the massless case, \(Z_{m=0}(\lambda) =-\ln(1-\lambda/\lambda_{c}^+)\) with $\lambda_{c}^+|_{m=0}=1$, and adding a similar  term \(-\ln(1-\lambda/\lambda_c^-)\) for the negative tail.
The prefactor was then determined by asking that the highest known moment,
in (\ref{48}) of order $\lambda^{12}$ be given exactly. Then
\begin{eqnarray}&&Z(\lambda)-Z_{\mathrm{BC}} (\lambda) = 0.0496465 \lambda -0.00117287 \lambda ^2
\nn\\&&
-0.0000986924 \lambda
   ^3-2.769730882007382\times 10^{-6} \lambda
   ^4\nn\\&&+2.48119\times 10^{-8} \lambda
   ^5-5.57763\times 10^{-11} \lambda
   ^6\nn\\&&-2.96177\times 10^{-9} \lambda
   ^7-1.82040\times 10^{-9} \lambda
   ^8\nn\\&&-1.04881\times 10^{-9} \lambda
   ^9-5.56368\times 10^{-10} \lambda^{10}
   \nn\\&&-2.25950\times 10^{-10} \lambda^{11}
   +O\left(\lambda ^{13}\right)
\end{eqnarray}
What is remarkable about this approximation, besides its accuracy in reproducing higher moments, is that both expected terms, \(\ln (1-\lambda/\lambda_c^+)\) and \(\ln (1-\lambda/\lambda_c^-)\) appear with the {\em same} amplitude. Performing the inverse-Laplace transform
for $v>0$ one finds, defining $\tilde v = 1.20711 v$:
 \begin{eqnarray}
P_{\mathrm{BC}}(\dot u)&=&\frac{e^{-\frac{1}{2} \dot u (\lambda
   _c^-+\lambda _c^+)}
   (-\lambda _c^- \lambda_c^+)^{\tilde v}
   \left(\frac{\dot u}{\lambda_c^+- \lambda_c^-}\right)^{\tilde v-\frac{1}{2}}
  }{\sqrt{\pi } \Gamma
   ({\tilde v})} \nn\\
 &&\times K_{{\tilde v}-\frac{1}{2}}\left(\frac{1}{2} \dot u[\lambda _c^+-\lambda
   _c^-]\right)\theta(\dot u>0)\nn\\
   &&+\frac{e^{-\frac{1}{2} \dot u (\lambda
   _c^-+\lambda _c^+)}
   (-\lambda _c^- \lambda_c^+)^{\tilde v}
   \left(\frac{\dot u}{\lambda_c^-- \lambda_c^+}\right)^{{\tilde v}-\frac{1}{2}}
  }{\sqrt{\pi } \Gamma
   ({\tilde v})} \nn\\
 &&\times K_{{\tilde v}-\frac{1}{2}}\left(\frac{1}{2} \dot u[\lambda _c^--\lambda
   _c^+]\right)\theta(\dot u<0)\ . \label{153}
\end{eqnarray}
 \(K\) is the Bessel-$K$ function. The function \(P_{\mathrm{BC}}(\dot u)\) decays as $\rme^{-\lambda_c^+ \dot u}$ for $\dot u$ to $\infty$, and as $\rme^{-\lambda_c^- \dot u}$ for $\dot u$ to
$-\infty$.
When compared to the numerical inverse-Laplace transform, we find that this is a good approximation for all but
small $\dot u$.

In the limit of $v=0^+$, we know (see  \cite{LeDoussalWiese2011a} for more details)
that one can expand $\hat P(\lambda) = e^{v Z(\lambda)}=1 + v Z(\lambda)+ O(v^2),$ and
that upon Laplace inversion one obtains
\be
P(\dot u) = (1- v p') \delta(\dot u) + v p'  \tilde P^{}(\dot u) + O(v^2)\ .
\ee
The first term represents events when the particle is pinned, and the second one yields the velocity distribution in an avalanche $\tilde P(\dot u)$
via $Z(\lambda) = v p' \int \rmd  \dot u (e^{\lambda \dot u}-1) \tilde P(\dot u)$ where
$v p'$ is the probability that $t=t_0$ belongs to an avalanche. From the form (\ref{a10}) for  \(Z_{{\mathrm{BC}}}(\lambda)\)
we  get
\begin{equation}
p' \tilde P_{\mathrm{BC}} (\dot u)=  1.20711\left[\frac{\rme^{-\dot u\lambda_c^+}}{\dot u} \theta(\dot u>0)+ \frac{\rme^{-\dot u\lambda_c^-}}{|\dot
u|} \theta(\dot u<0) \right]
\end{equation}
Note that for $\dot u>0$ it is very
similar to the result for the ABBM model with $m=0$ \cite{LeDoussalWiese2011a}
up to the different
value for $\lambda_c^+$. More interestingly, it also gives a non-trivial
prediction for the tail on the negative-velocity side
in the avalanche regime.

\subsubsection{Large \(\lambda\)-behavior}
\begin{figure}\Fig{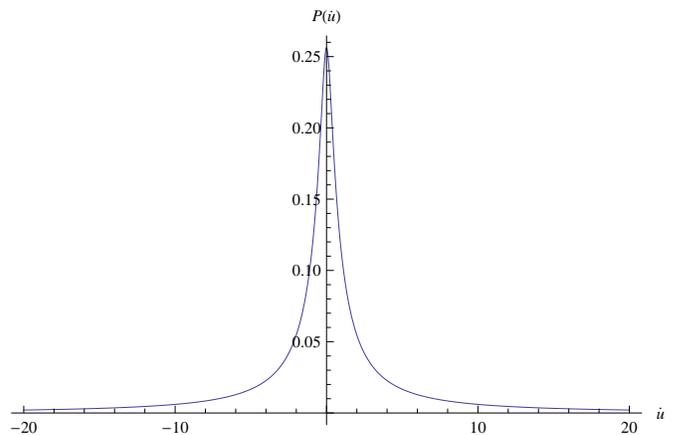}\caption{The contribution of the large-\(\lambda\)
behavior (\ref{b7+}) to $P_{\rm asymp}(\dot u)$ at \(v=2\), \(m=1/4\), as given in Eq.~(\ref{Pv}).}
\label{f:Pv}
\end{figure}

In order to obtain the small-\(\dot u\) behavior of \(P_{\mathrm{approx}}(\dot u)\), one has to analyze \(Z(\lambda)\) for $|\lambda| \to \infty$. For real \(\lambda\) this is impossible since branch-cuts appear at \(\lambda=\lambda_c^\pm\). As we show in appendix \ref{app:Z-large-lambda}, the behavior for $\lambda \to \pm i \infty$ {\em on the imaginary axis} can be calculated analytically, and is given by
\begin{eqnarray}\label{b7+}
Z_{\mathrm{asymp}} (\lambda) &=&- \frac{15 \sqrt[3]{6} \sqrt{\pi } \,\Gamma
\big(\frac 5{6}\big)}{\Gamma \big(\frac{1}{6}\big)^{\! 2}}
   \left({|\lambda| m} \right)^{2/3 }\nn\\
&=&-1.76006   \left({|\lambda| m} \right)^{2/3 }  \ .
\end{eqnarray}
The contribution of expression (\ref{b7+}) to \(P_{\mathrm{asymp}}(\dot u) \) at finite driving velocity \(v\) is

\begin{eqnarray}\label{Pv}
P_{\mathrm{asymp}}(\dot{u}) &=&  \int_{0 }^{\infty} \frac{\rmd \lambda}{ \pi} \cos( \lambda
|\dot u|) \rme^{v Z (\lambda i)}\nn \\
& \approx & \int_{0}^{\infty} \frac{\rmd \lambda}{ \pi} \cos( \lambda |\dot
u |)  \rme^{-1.76006  \lambda^{2/3} vm^{2/3}} \nn \\
&=&-\frac{m^{2/3} v \exp\left({\frac{0.403877 m^2 v^3}{\dot u^2}}\right)
   }{|\dot u|^{7/3}}\nn\\
&&\times \bigg[1.43193 m^{2/3} v \text{Ai}\left(\frac{0.715967
   m^{4/3} v^2}{|\dot u|^{4/3}}\right)\nn\\
&&\quad +1.6923 |\dot u|^{2/3}
   \text{Ai}'\left(\frac{0.715967 m^{4/3}
   v^2}{|\dot u|^{4/3}}\right)\bigg]\nn\\
\end{eqnarray}%
Ai is the Airy function, and the result (\ref{Pv}) a positive
function, peaked around zero, even for relatively large driving velocities,
see figure \ref{f:Pv}. Its value at $\dot u=0$ is \begin{equation}\label{P0}
P_{\mathrm{asymp}}(0)= \frac{0.181215}{m v^{3/2}}\ .
\end{equation}
Note that while (\ref{Pv}) is relevant at finite $v$, it does not contribute to the large-deviation function discussed below.
We will discuss below its domain of applicability.

\subsubsection{From $Z(\lambda)$ to $P_{\mathrm{approx}}(\dot u)$}

The instanton equation can be solved numerically for any complex $\lambda$
away from the branch cuts on the real axis at $\lambda>\lambda_c^+$ or $\lambda<\lambda_c^-$.
This yields $Z(\lambda)$ for complex $\lambda$. We have performed the
numerical inverse Laplace transform for $Z(\lambda)$. A convenient choice of the contour
is $\lambda = \alpha (1\pm i)$, $\alpha >0$ for $\dot u>0$ and $\lambda = - \alpha (1\pm i)$, $\alpha >0$ for $\dot u<0$.
This gives an excellent numerical accuracy, except when $\dot u$ and $v$ are both very small ($<0.1$).

We can now compare with the asymptotic estimates of $P_{\mathrm{approx}}(\dot u)$ discussed above (see Figure \ref{f:asdfasdf}).
Why the different approximations work or fail, can be traced to an analysis of the inverse Laplace transform. Depending on $\dot u$ and $v$, it is dominated by one of the three special points: $\lambda=0$ for the perturbative approximation, valid for $v\to \infty$, $\lambda=\lambda_c^\pm$ for the tails, and $\lambda\to \pm i \infty$, for small $\dot u$, as long as $v$ is not large enough s.t.\ $\lambda=0$ dominates.

First the perturbative expansion in $\lambda$ works well for large $v$. This will be quantified in section \ref{sec:largedeviation}, where we discuss the large-deviation function.

Second, the branch-cut approximation (\ref{153}) is a reasonable approximation to \(P_{\rm approx}(\dot u)\) for all $v$ and all $\dot u$,
and becomes a good approximation in the tails.
The latter can be expected since it uses the knowledge about the branch-cut singularities at $\lambda_c^+$  and  $\lambda_c^-$.

Third, the approximation $P_{\mathrm{asymp}}(\dot{u})$, given in (\ref{Pv}). It is a reasonable approximation at small $\dot u$, as long as $v$ is small enough. For $v=1/2$ e.g.\  it predicts, with relative precision of $10^{-3}$, the value of $ [P_{\mathrm{approx}}(
0^+)+ P_{\mathrm{approx}}(0^-)]/2$, where we note that \( P_{\mathrm{approx}}(\dot
u)\) jumps at $0$, from $1.68172$ for $\dot u=0^-$ to
$2.42249$ at $\dot u=0^+$. The worse disagreement at $v=0.1$ is probably due to problems in the numerical inverse Laplace-transform ($\hat P(\lambda)$ oscillates  strongly on the chosen contour). For $v=5$, it does not work.

\subsubsection{The \(\sqrt{\dot u}\)-model as an approximation for
ABBM}
\label{s:sudvABBM}
\begin{figure}
\Fig{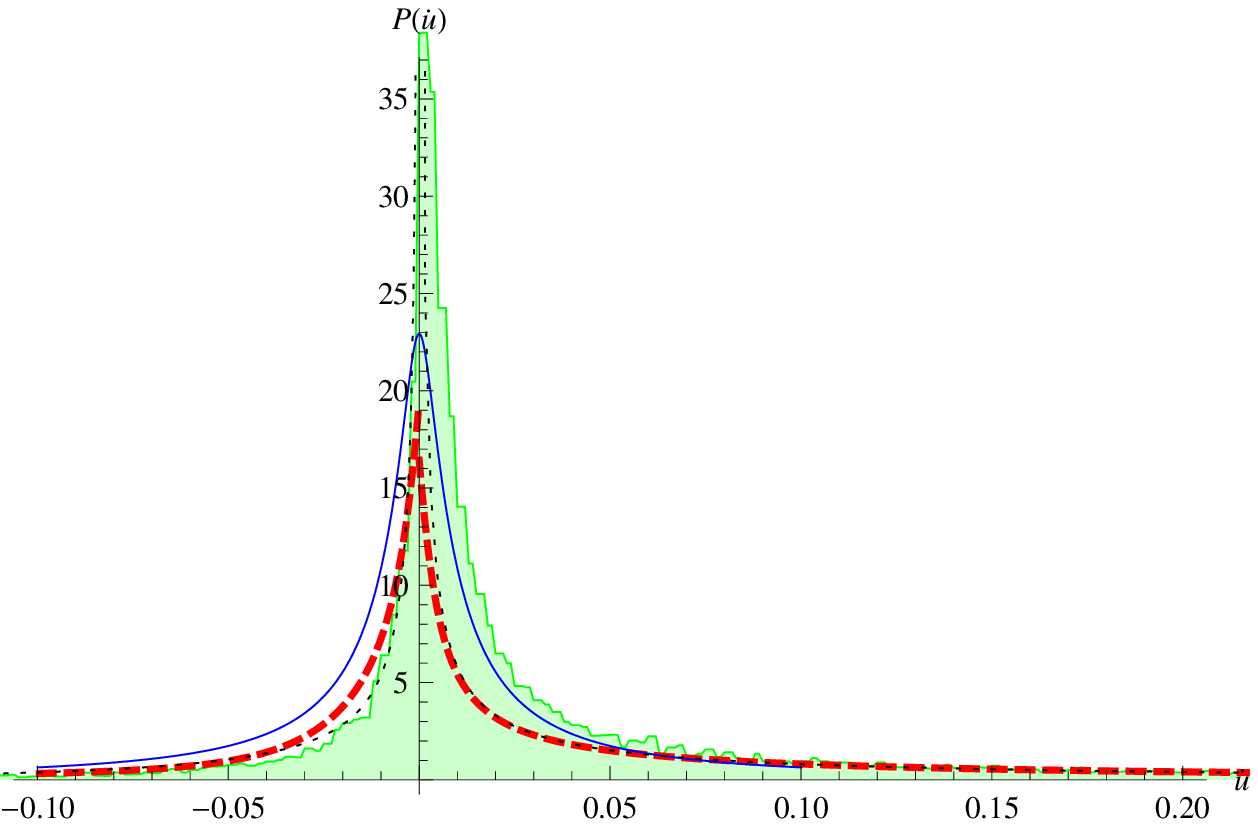}
\Fig{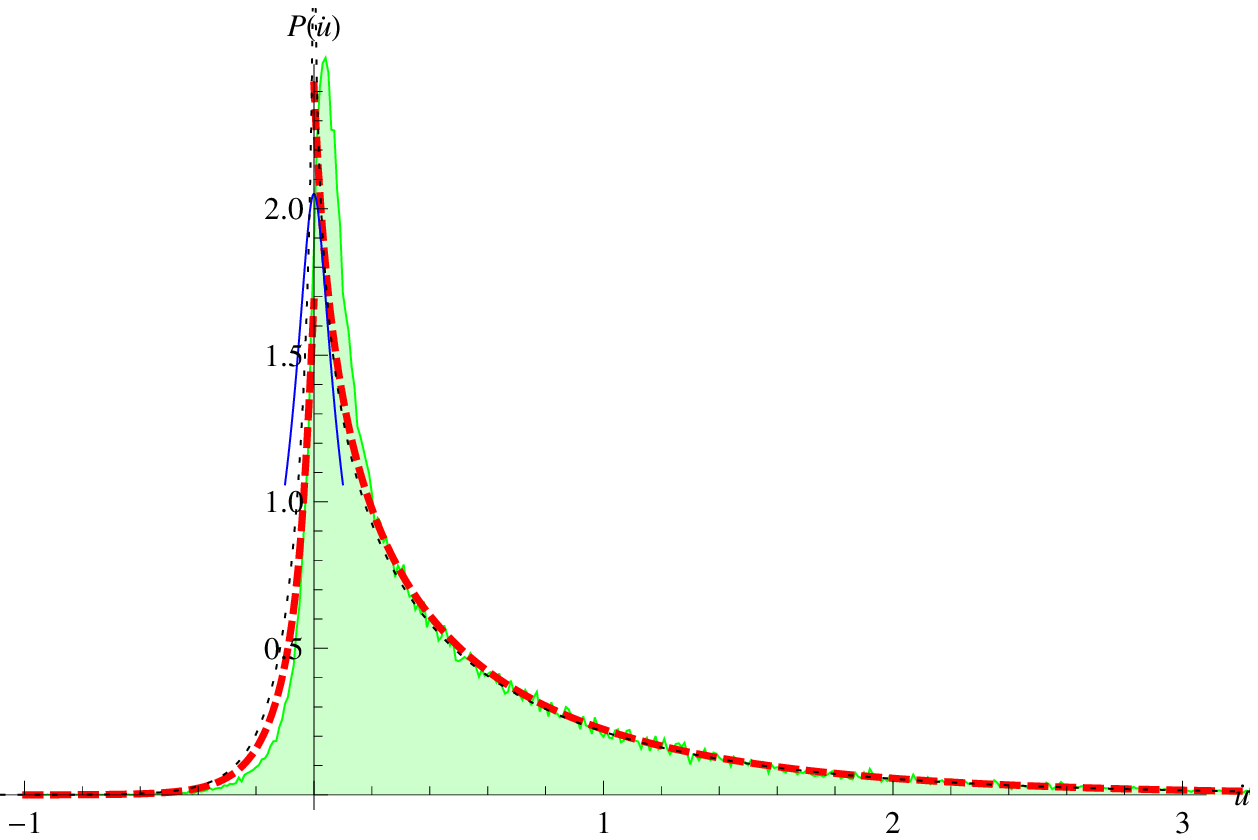}
\Fig{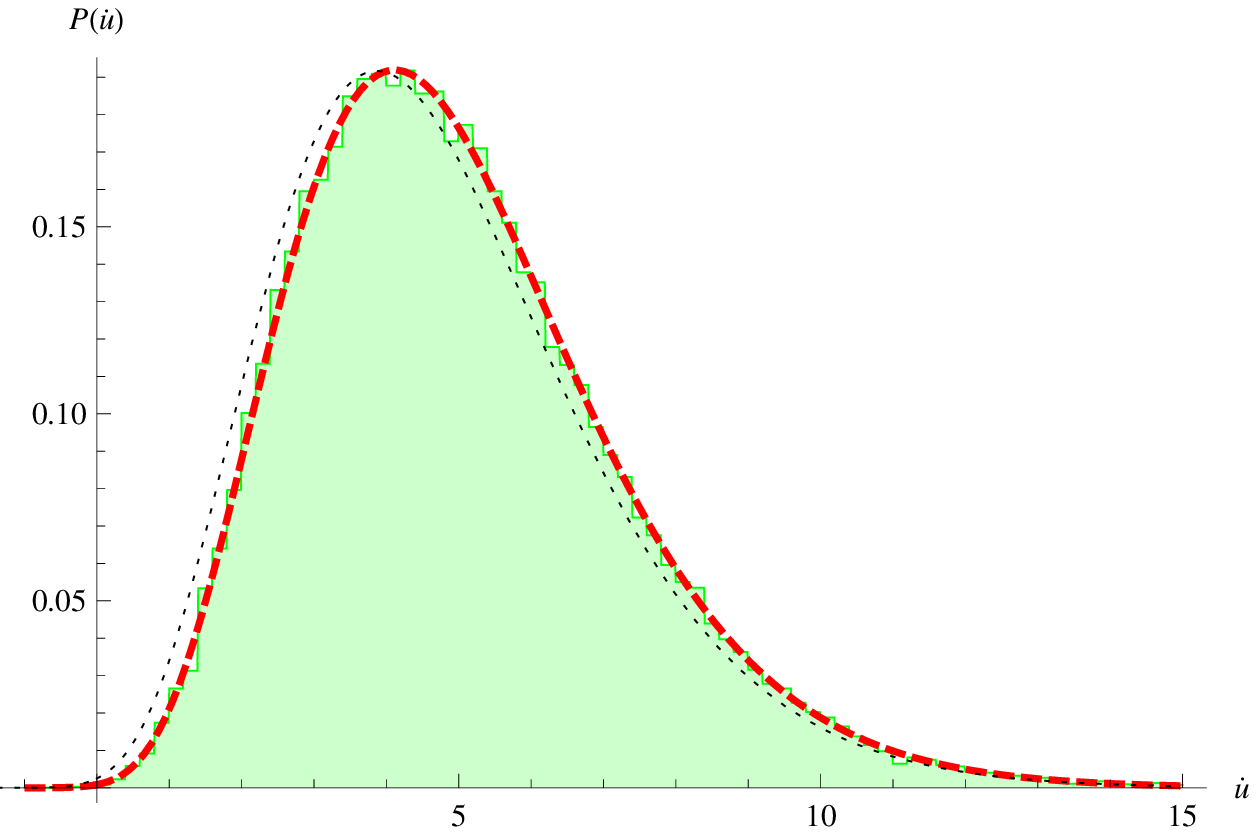}
\caption{Comparison of $P_{\mathrm{approx}}(\dot u)$ obtained by numerically inverse-Laplace transforming $\rme^{vZ(\lambda)}$ (thick red dashed lines) with a simulation of the ABBM model (green shaded histogram plot),  the approximation (\ref{153}) (black dotted lines), and \(P_{\rm asymp}(\dot u)\) given by . The mass is $m=1/4$, and the driving velocities are from top to bottom: $v=1/10$, $v=1/2$, and $v=5$. }
\label{f:asdfasdf}
\end{figure}
On Figure \ref{f:asdfasdf}, we show the data obtained for the probability distribution \(P(\dot u)\) from a numerical simulation of the ABBM model at $m=1/4$ (green shaded area), compared to results obtained  in this section. The driving velocities are $v=1/10$, \(v=1/2\) and $v=5$ (from top to bottom).  First compare with  the result for  \(P_{\mathrm{approx}}(\dot
u) \), obtained by numerically inverse-Laplace transforming $\hat P(\lambda)$ (thick red dashed line). The agreement is excellent for $v=5$,  and reasonable for $v=1/2$ (it does not  give well the peak for $\dot u$ close to $0$, but quite well the tails, even on the negative side.
For $v=0.1$ only the tails are reasonably well approximated.
(We cannot however exclude numerical problems in the inverse-Laplace transform for $\dot u<0.01$.)

\subsection{\label{sec:exactm625}Exact results for $m=6/25$}

While Eq.~(\ref{eq:kappamotionM}) cannot be solved in closed form for generic $m$, and one had to
resort to expansions or numerical solutions, there exists a magic value of the mass $m=6/25$ for which
there exist analytic solutions in closed form. These are known in the context of the Fisher-Kolmogorov equation
and Abel equations, which is intimately linked to our problem as we now explain.

\subsubsection{Fisher-Kolmogorov equation}\label{sec:kolmogorov}

Eq.~(\ref{eq:kappamotionM}) is a second order differential equation and it has many solutions that are characterized by different set of values $(\kappa_0,\lambda_0)$ define by Eqs.~(\ref{eq:kappa},\ref{eq:lambda}) where $t^*$ is an arbitrary time.
One solution for $m=6/25$ reads as\cite{fishers}:
\begin{align}\label{eq:solution}
\tilde{u}(t)=\frac{1}{(1+D e^{-5t/6})^2},
\end{align}
where $D$ is an arbitrary parameter. Here we used that traveling wave solutions of the Fisher-Kolmogorov equation satisfy  Eq.~(\ref{eq:kappamotionM}), and that for a special value of the speed of the propagating wave it has the analytic solution Eq.~(\ref{eq:solution}) that corresponds to the special value $m=6/25$ in our case.

For $D>0$ it follows that $0<\tilde{u}<1$ and using Eq.~(\ref{eq:ZM}) we get
\begin{align}\label{eq:z1}
Z(\lambda(\kappa),\kappa)=
-\frac{6}{5}\left(\log{(1-\sqrt{25\kappa/6})}+\sqrt{25\kappa/6} \right),
\end{align}
we determine below which values $\lambda$ takes as a function of $\kappa$.
For $D<0$ solution $\tilde{u}(t)$ (given by Eq.~(\ref{eq:solution})) takes some value that is greater than $1$ at two different times. If one chooses
$t^*=-{6}\log{\left({\left(1+\sqrt{{6}/({25\kappa})}\right)}/({-D})\right)}/5$
then
\begin{align}\label{eq:z2}
Z(\lambda(\kappa),\kappa)=-\frac{6}{5}\left(\log{(1+\sqrt{25\kappa/6})}-\sqrt{25\kappa/6} \right).
\end{align}
Here $\tilde{u}$ takes all values greater than zero.
If one takes
$t^*=-{6}\log{\left(\left({1-\sqrt{{6}/{25\kappa}}}\right)({-D})\right)}/5$, then $Z(\lambda(\kappa),\kappa)=\infty$ because the integration in Eq.~(\ref{eq:ZM}) is over the divergence that happens at $t=6\log{(|D|)}/5$.
Taking into account these results one finds the following correlation functions
\begin{align}\label{eq:correlation}
&\overline{e^{\lambda(\kappa)\dot{u}+\kappa a}}=e^{v Z(\lambda(\kappa),\kappa)},\\\label{eq:lambdaofkappa}
&\lambda(\kappa)=\begin{cases}
5\kappa/2+50\kappa^{3/2}/6^{3/2}& D>0 \;\text{and}\; 0\leq\kappa<6/25 \\
5\kappa/2-50\kappa^{3/2}/6^{3/2}& D<0\;\text{and} \; \kappa\geq0.
 \end{cases}.
\end{align}
In Eq.~(\ref{eq:correlation}), the first and the second line of Eq.~(\ref{eq:lambdaofkappa}) correspond to $Z(\kappa)$
given by Eq.~(\ref{eq:z1}) and (\ref{eq:z2}), respectively. Hence the Laplace transform of the joint distribution
of velocities and acceleration is known exactly on the curve $\lambda=\lambda(\kappa)$, which, upon expanding in
$\kappa$ can be translated into non-trivial relations between moments of this distribution.

\subsubsection{Abel differential equation of the second kind\label{sec:abel}}

Next, we find more general solution of Eq.~(\ref{eq:kappamotionM}) than in the previous section. Then using it we obtain an exact result for $Z(\lambda,0)$ for some range of $\lambda$ values. Additionally, we obtain an analytic expression for $\lambda_c^+$ that tell us about behavior of the tails of $P(\dot{u})$ for positive velocities, see Eq.~(\ref{observ}).

Introducing $A={\lambda}/{m}-{\kappa}/{m^2}$ and $B=\kappa/m$, the set of Eqs.~(\ref{eq:kappaM},\ref{eq:lambdaM},\ref{eq:ZMM}) can be rewritten as:
\begin{align}\label{eq:Abel}
\frac{\rmd A}{\rmd B}&=\frac{B^2-B-A}{m A},\\
\frac{\rmd Z}{\rmd B}&=-\frac{B}{A}.
\end{align}
Eq.~(\ref{eq:Abel}) is the so-called Abel differential equation of the second kind. Its parametric solution reads as \cite{polyanin}
\begin{align}\label{eq:parametric}
B_C(s)&=\frac{1}{4}E^2(s,C)s,\\\label{eq:parametric1}
A_C(s)&=-\frac{5}{24}E^2(s,C)\left(\sqrt{1+s^3}E(s,C)+2s\right),\\\label{eq:parametric2}
E(s,C)&=\int_{0}^s d\tau (1+\tau^3)^{-1/2}+C,
\end{align}
where $C$ is an arbitrary constant. We see that $B_C(0)=0$ while $A_C(0)=- C^3 5/24$ can be arbitrary, meaning that we have found the family of solutions determined by parameter $C$. Then
\begin{align}
Z(\lambda_C(s),\kappa_C(s))=-\int \rmd s \frac{B_C(s)}{A_C(s)}\frac{dB_C(s)}{ds}+const,
\end{align}
where the constant is to be determined by knowing that $Z(0,0)=0$.

For example, if we want to calculate
\begin{align}\label{eq:correlation625}
\overline{\exp{\left(-C^3\dot{u}/20\right)}}=\exp{\left(vZ(-C^3/20,0)\right)},
\end{align}
we need to find first
\begin{align}
Z(-\frac{1}{20}C^3,0)&=-\int_{t_{C0}}^{0}  \rmd s \frac{B_C(s)}{A_C(s)}\frac{dB_C(s)}{ds}\notag\\
&=\frac{3}{10}\int_{t_{C0}}^{0} \rmd s \frac{s}{\sqrt{1+s^3}}E(s,C),
\end{align}
where $t_{C0}$ is defined as $E(t_{C0},C)=0$.
We find analytic expression for
\begin{align}\label{eq:zm6over25}
Z(-\frac{1}{20}C^3,0)=&\frac{1}{15}
   \left(t_{C0}^3+1\right)\;_3F_2\left(\frac{5}{6},1,1;\frac{3}{2},2;t_{C0}^3 +1\right)\notag\\
   &-\frac{3}{5\sqrt{\pi}} \Gamma \left(\frac{7}{6}\right)
   \Gamma \left(\frac{4}{3}\right) \notag \\&\times \sqrt{t_{C0}^3+1}
    \,_2F_1\left(\frac{1}{3},\frac{1}{2}
   ;\frac{3}{2};t_{C0}^3+1\right)\notag\\
   &-\frac{3}{20} C
   t_{C0}^{2} \,
   _2F_1\left(\frac{1}{2},\frac{2}{3};\frac{5}{3};-t_{C0}^3\right)\notag\\
   &-\frac{1}{30}\left(-\sqrt{3} \pi -3 \log
   \left(\frac{27}{16}\right)\right).
\end{align}
Here $\;_2F_1$ is the hypergeometric function and $\;_3F_2$ is generalized hypergeometric function. $t_{C0}$ is implicitly given by  $E(t_{C0},C)=0$ that could be rewritten as
\begin{align}\label{eq:tco}
E(t_{C0},0)&=-C.
\end{align}
We find that $E(s,C)=C+s \,_2F_1\left({1}/{3},{1}/{2};{4}/{3};-s^3\right)$.
The function $E(s,0)$ becomes complex for $s<-1$. For $s>-1$ it is increasing function of $s$, and satisfies:
\begin{align}
-\frac{\Gamma{(\frac{1}{6})}\Gamma{(\frac{4}{3})}}{2\sqrt{\pi}}\leq E(s,0)\leq\frac{\Gamma{\left(\frac{1}{6}\right)}
\Gamma{\left(\frac{4}{3}\right)}}{\sqrt{\pi}}.
\end{align}

\begin{figure}
\includegraphics[width=0.9\linewidth]{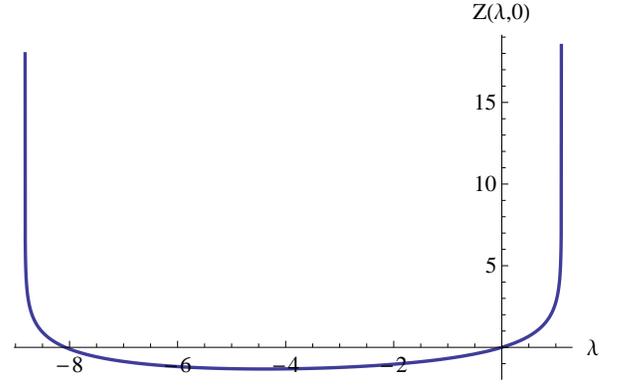}\\
\caption{Numerical result for $Z(\lambda,0)$ is shown for $m=6/25$.}\label{fig:zoflambda}
\end{figure}

We see that for $\lambda_{\mathrm{min}}\leq\lambda=-C^3/{20}<\lambda_{\mathrm{max}}$, where
\begin{align}
\lambda_{\mathrm{min}}=-\frac{1}{20}\left[\frac{\Gamma{\left(\frac{1}{6}\right)}
\Gamma{\left(\frac{4}{3}\right)}}{2\sqrt{\pi}}\right]^3=-0.137843,\\
\lambda_{\mathrm{max}}=
\frac{1}{20}\left[\frac{\Gamma{\left(\frac{1}{6}\right)}
\Gamma{\left(\frac{4}{3}\right)}}{\sqrt{\pi}}\right]^3=1.10274,
\end{align}
there exist a solution of Eq.~(\ref{eq:tco}) and one can find $t_{C0}$. Since at $s=t_{C0}$ both $\lambda(s)$ and $\kappa(s)$ are zero, this constraint on $\lambda$ might mean that a solution of Eq.~(\ref{eq:kappamotionM}) with boundary conditions $\kappa(t=0)=0$ and $\dot{\kappa}(t=0)=\lambda$ reaches $(\dot{\kappa}(t),\kappa(t))=(0,0)$, i.e. it is convergent meaning that at large times it goes to zero.
By considering numerically this instanton solution of Eq.~(\ref{eq:kappamotionM}) with given boundary conditions, one indeed sees that it starts to diverge for this value of $\lambda$ and $\lambda_c^+=\lambda_{\mathrm{max}}$. However, for $\lambda_{\mathrm{min}}$ the instanton solution is still convergent. The numerics gives $\lambda_c^- = -8.8219< \lambda_{\mathrm{min}}=-0.1378$. In Fig.~\ref{fig:zoflambda} we show the numerical result for $Z(\lambda,0)$. We conclude that the parametric solution (\ref{eq:parametric},\ref{eq:parametric1},\ref{eq:parametric2}) does not "catch" whole $\kappa(t)$ dependence on time with given boundary conditions $\kappa(t=0)=0$ and $\dot{\kappa}(t=0)=\lambda$ for $\lambda<\lambda_{\mathrm{min}}$.

Next, we solve $E(t_{C0},C)=0$ perturbatively  in $C$ and  find
\begin{align}
t_{C0}&=-C+\frac{C^4}{8} -\frac{C^7}{112}+\frac{C^{10}}{1792}-\frac{3
   C^{13}}{93184}\notag\\&
   +\frac{37C^{16}}{20873216}-\frac{75 C^{19}}{793182208}+
  \mathcal{O}(C^{20}),\\
Z(\lambda,0)&=\lambda+\frac{\lambda
   ^2}{2}+\frac{25 \lambda
   ^3}{84}+\frac{125 \lambda
   ^4}{616}\notag\\&
   +\frac{375 \lambda
   ^5}{2548}+\frac{115625 \lambda
   ^6}{1039584}+\mathcal{O}(\lambda^7).
\end{align}
Now, expanding Eq.~(\ref{eq:correlation625}) in $\lambda=-C^3/20$ one can find the moments $\overline{\dot{u}^k}$ in agreement with Sec.~\ref{sec:mainmoments} and App.~\ref{app:moments}.

\section{Large-deviation function}\label{sec:largedeviation}

\subsection{Definitions and numerical determination}
\label{sec:dev}
It is suggestive that we ``do a good job for large driving
velocities'', since when the particle does not move backward, all
three models considered here are indistinguishable. We actually
show below that the  so-called  {\em large-deviation} function
coincides  for all three models for $x>0$.  The large-deviation function is defined as
the leading behaviour for large driving velocity $v$ of the
distribution of instantaneous velocity $P(\dot u)$ as follows:
\begin{eqnarray}
F_{v} (x)&:=&-\frac{\log{\left[P(x v)\right]}}{v}\\\label{eq:Fdefinition}
F(x)&:=&\lim_{v\to \infty}F_{v} (x)
.
\end{eqnarray}
Analogously, one defines, if that limit exists,
\begin{eqnarray}\label{p1}
Z_{v} (\lambda)&:=&\frac{ \ln \overline{\rme^{\lambda \dot{u} }}}v\\
Z (\lambda)&:=& \lim_{v\to \infty}Z_{v} (\lambda )\ .
\end{eqnarray}
If the limits exist, then for large $v$ the Laplace-transform
\begin{equation}
\overline{\rme^{\lambda \dot{u}}} = \rme^{v Z_v (\lambda)} = v\int \rmd
x\, \rme^{-vF_v (x)} \rme^{\lambda x v}
\end{equation}
can be approximated by its saddle point, and $Z (\lambda)$ and $F (x)$
are related via a Legendre transform:
\begin{eqnarray}\label{p2}
Z (\lambda)+F (x) &=& \lambda x\\
\frac{\rmd}{\rmd x}F (x) &=& \lambda\label{p2b}  \\
\label{eq:xoflambda}
\frac{\rmd}{\rmd \lambda}Z (\lambda) &=& x\ .
\end{eqnarray}
It is assumed that $F''(x)>0$. Note that $Z (\lambda)$ is easier to measure numerically than $F
(x)$, since the former  does not need binning.

Let us now review our numerical results and how they are consistent
with the following scenario:

(i)  $Z_{v} (\lambda )$  becomes $v$-independant at large $v$
for each model in some range of $\lambda$.

(ii) the asymptotic curves
coincide for $\lambda > \lambda^*$,\bea\label{eq:ZlargeV}
Z(\lambda) = Z_{\sqrt{\dot u}}(\lambda) = Z_{\mathrm{ABBM}}(\lambda) = Z_{\mathrm{tree}}(\lambda)\ ,
\eea
 where
\bea
Z'(\lambda^*)=0
\eea
i.e.\ the minimum of $Z(\lambda)$ is at $\lambda^*$. This corresponds to the point of zero velocity since Eq.~(\ref{eq:xoflambda}) implies $x(\lambda^*)=0$. Another way to state Eq.~(\ref{eq:ZlargeV}) is to say that for $x>0$
\begin{align} \label{idd}
F(x) = F_{\sqrt{\dot u}}(x) = F_{\mathrm{ABBM}}(x) = F_{\mathrm{tree}}(x)\ .
\end{align}
We assumed that $Z'(\lambda)>0$ for $\lambda>\lambda^*$.
A simple argument, which shows that (\ref{eq:ZlargeV}) and (\ref{idd})
holds, is given below.
For $\lambda<\lambda^*$ the $Z(\lambda)$ for each model is dominated by negative velocities. Since the models differ for these velocities, there is no reason why their $Z(\lambda)$ should be the same, and consequently $F(x)$ for $x<0$ is expected to be different for the different models.

\begin{figure*}[t]
\fig{10.5cm}{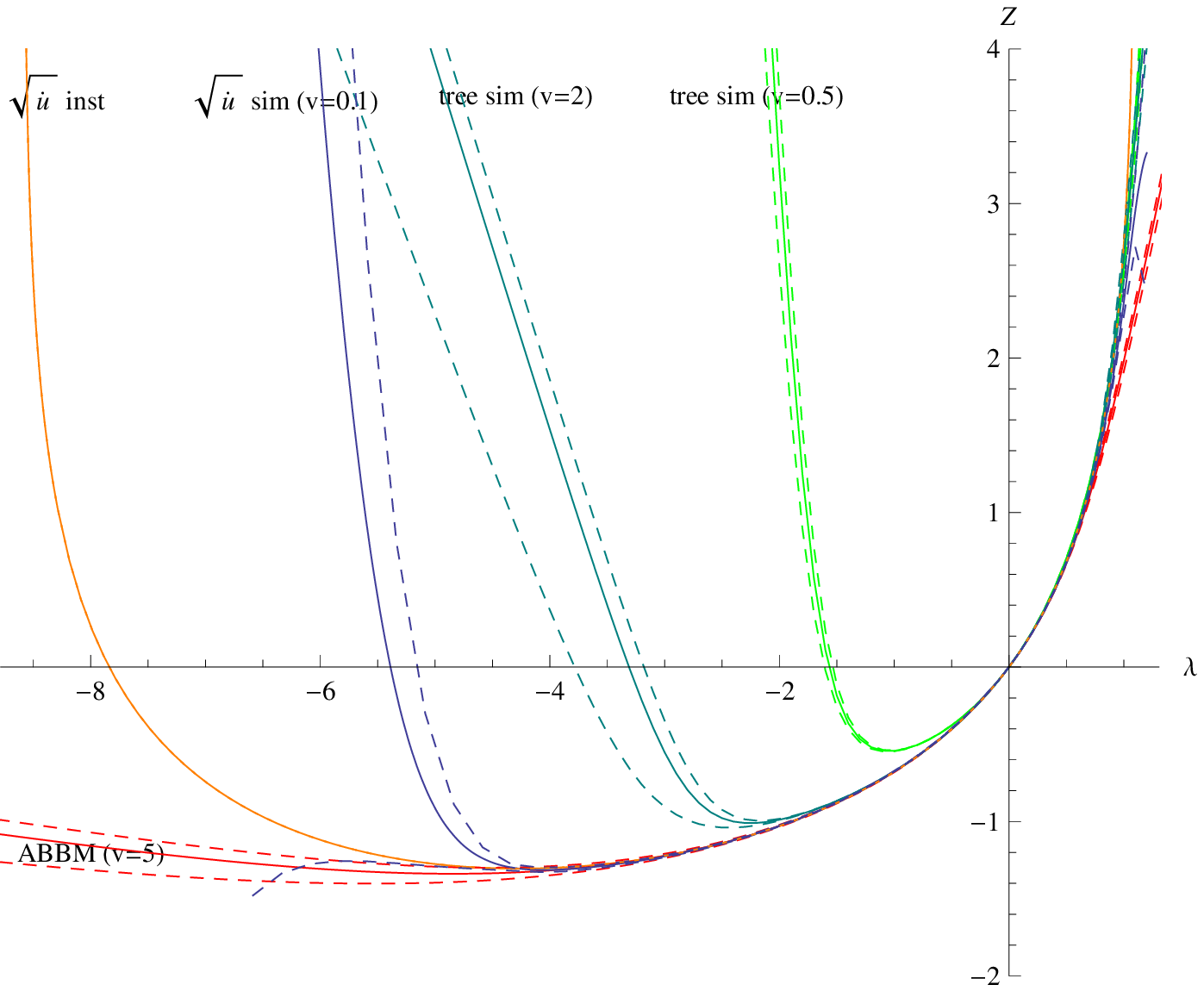}\
\fig{7cm}{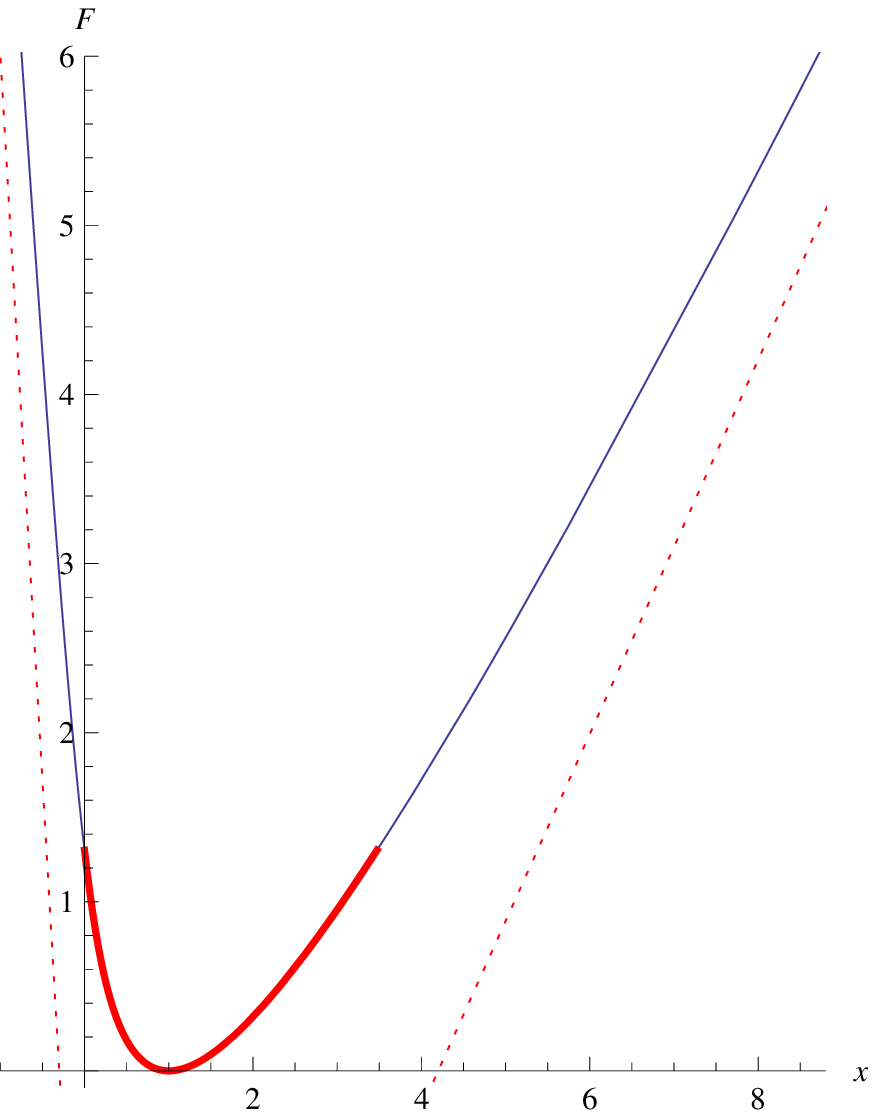}
\caption{Left: Different curves for $Z (\lambda)$, $m=1/4$: The numerical solution of the instanton equation (\ref{41}) is used to obtain $Z$ (see (\ref{42})) denoted as $\sqrt{\dot u}~ \textrm{inst}$ and shown by a thick orange line. Numerical simulation of this $\sqrt{\dot{u}}$-model for
driving velocity $v=0.1$ (grey-blue line), with an estimate of the numerical error bars,
(one-$\sigma$ error, dashed curves of the same color).  Within error-bars, the simulation
has converged to the orange curve $\sqrt{\dot u}~ \textrm{inst}$; errors are small for $-4\le \lambda\le1$ .
The simulations for the tree
model at $v=2$ (dark green) and $v=0.5$ (bright green), both with
error bars (dashed) show clear deviations from instanton solution at
small $v$, but get closer to the latter for larger $v$. The last curve
is a simulation of the ABBM model at $v=5$ (red). It has also
converged to the orange curve within error bars. This is not the
case for smaller driving velocities (not shown).  All simulations are for $5 \times
10^{7}$ data points, apart from the one for ABBM, which has 93.000
data points. Due to the small number, the error bars in that case
are  underestimated.\smallskip\\
Right: The large-deviation function $F (x)$, obtained by performing a numerical Legendre transform of
$Z (\lambda)$ given by the numerical solution of the instanton equation (\ref{41}), and
(\ref{42}). We note $F (0)=F (3.47268)= 1.30459$. The thick red part is the
domain for which $F (x)\le F (0)$, which can be obtained with arbitrary precision
from a simulation with negligible probability of negative velocity as explained in the text, hence must coincide
for the three models. The dashed curves show the asymptotic behavior $\lambda_c^\pm x +\mbox{const}$. }
\label{f:compare-Zs}
\end{figure*}

Our numerical data for $Z_{v} (\lambda)$ are presented on the left of Fig.~\ref{f:compare-Zs} for $m=1/4$. First of all, we have checked through
large-scale simulations ($5\times 10^{7}$ data points) that for the
$\sqrt{\dot{u}}$-model $Z_v(\lambda)=Z(\lambda)$ is velocity-independent, and
given by the numerical solution of the instanton equation (\ref{41}), using
(\ref{42}) (thick orange line). Only data for $v=0.1$ are presented
on the plot (dark grey-blue line, with error bars given by the dashed
lines of the same color). Relative errors are  \(\le 10\% \) for $-4.2<\lambda<1$.

We have then checked that $Z_v(\lambda)$ for the tree-model converges, for \(v\to \infty\), towards the numerical solution of the instanton equation (\ref{42}) from above, while the ABBM model converges from below. (This fact is consistent with Figs.~\ref{fig:numerics2} and \ref{fig:numerics2v}, where one  observes that the probability distribution for the tree model has larger tails than for the ABBM model with inertia.) On Fig.~\ref{f:compare-Zs}, data for $v=0.5$ and $v=2$ are shown for
the tree model, and for $v=5$ for the ABBM model. Simulations for other
velocities (not shown) confirm this picture. Note, however, that we
have found  convergence of the simulations of \(Z_v(\lambda)\)  for both ABBM and the tree model to \(Z(\lambda)\) of the \(\sqrt{\dot u}\)-model  only in some domain $\lambda > \lambda^*$
and certainly not
for \(\lambda <\lambda^*\), corresponding to negative
velocity,  consistent with the above scenario. For $m=1/4$ we find $\lambda^*=- 4.26953$.

Once the $Z(\lambda)$ curves have been measured for each model,
one can obtain the corresponding large-deviation functions via
a numerical Legendre transform. On figure \ref{f:compare-Zs}
we have plotted the large-deviation function, Legendre-transform of $Z (\lambda)$
  for the $\sqrt{\dot u}$ model. Its minimum is at
$x=1$, and for large positive $x$ it grows  like $F (x)\approx \lambda_{c}^{+}x+\mbox{const}$, with
$\lambda_{c}^{+}$ given by equation (\ref{49}). For large negative
$x$, the growth is $ F (x)\approx \lambda_{c}^{-}x+\mbox{const} $  with
$\lambda_{c}^{-}$ also given by equation (\ref{49}).
These slopes are indicated by the dashed curves in figure \ref{f:compare-Zs} (right).

\subsection{Convergence of the large-deviation functions}\label{p3}

We now show the main result of this section, namely that the three models have the same large-deviation function in the  region \( 0<x<x_{+}\):
\bea \label{123}
 F(x)&=&F_{\sqrt{\dot{u} }} (x) = F_{\rm tree} (x) = F_{\mathrm{ABBM}} (x)
\nn\\ &&\mbox{ for all } 0<x<x_{+}\ , \\
 F(x_{+}) &=& F(0)
\eea
The argument is  simple and in fact much more
general: Two models, which have the exact same dynamics for positive velocities,
 have the same large-deviation function in that interval.
The idea
is the following. Consider a simulation with a set of $N_v=e^{d v}$ data points.
If $d > F(0)$ there will be negative velocities in a typical such set (with
probability one at large $v$). But if $d< F(0)$ there will be none, again with
probability one at large $v$. Consider any $x$ such that $F(x)<F(0)$.
We can then measure the value of $F(x)$ with arbitrary  accuracy
(as $v$ becomes large) if we use $e^{d v}$ data points with $F(x) < d < F(0)$
(in fact $d=F(x)^+$ is sufficient). As stated above, this set almost surely does not
contain negative velocities. Since the dynamics for the three models
exactly coincides for trajectories with positive velocities, this shows the
above property (\ref{123}), provided at least for one of the models the large-deviation function {\em exists}. However the latter is true for the \(\sqrt{\dot u}\)-model, since there \(Z(\lambda) \) is \(v\)-independent, which completes the argument.

The argument is based on considerations of under-sampling of rare events, and is
reminiscent of similar considerations used for the multi-fractal spectrum of wave-functions,
e.g.\ when comparing the size-dependence-exponents of participation ratios for a typical sample
or disorder-averaged ones \cite{multifractal1,multifractal2}. Here, the additional input is the identification
of the dynamics for positive velocities. Note the restriction that $x < x_{+}$,
which means that  rare events with positive velocities but as rare as
the negative velocity can not be controlled neither. In the present case we
do not see a reason why the various functions   $F(x)$ would not  coincide  for $x>x_{+}$.
The restriction  comes from the  generality of the argument: One
could for instance imagine a dynamics such that the particle jumps discontinuously from
large positive velocities to negative ones.
The above estimates are made more precise in appendix \ref{s:convergenceF}.

\subsection{Large deviation function in perturbation theory for small $m$}\label{sec:smallm}

\subsubsection{Expression obtained from previous results}

In this section we compute the large-deviation function for the $\sqrt{\dot u}$ model in a perturbative expansion at small $m$. For $x>0$, as argued above, it  gives the result for all three models. A further discussion of this equivalence is given in App.~{\ref{app:deviation}}.

We slightly generalize the discussion of the previous Section by considering
the large-deviation function for both velocity and acceleration defined as
\begin{align}\label{eq:Fdefinitionxy}
F\left(x,y\right)=-\lim_{v\to \infty}\frac{\log{\left[P(x v,y v)\right]}}{v},
\end{align}
where $x=\dot{u}/v$ and $y=a/v$. The connection to the
generating function introduced in (\ref{observ}, \ref{eq:defforZ}) is through a
Legendre transform in both variables, namely
\begin{align}\label{eq:195}
Z(\lambda,\kappa)=\max_{x,y}\{-F(x,y)+\lambda x+\kappa y\}\ .
\end{align}
To compute $F(x,y)$ from $Z(\lambda,\kappa)$, one looks for $(x_0,y_0) $ that satisfy $\partial_x F|_{x_0,y_0}=\lambda$ and $\partial_y F|_{x_0,y_0}=\kappa$, to get
\be
F(x_0,y_0)=-Z(\lambda,\kappa)+\lambda x_0+\kappa y_0,
\ee
These formula assume somehow the convexity of $F(x,y)$ and $Z(\lambda,\kappa)$
which we did not attempt to prove but seems to hold.

Now let us use our results from Sec.\ref{sec:perturbation} to obtain $F(x,y)$ perturbatively in $m$. To this aim we use Eq.~(\ref{eq:Fdefinitionxy}) with $P(\dot{u},a)$ given by Eqs.~(\ref{eq:distributionperturbative}) and (\ref{eq:F0text})-(\ref{eq:F2text}) where $c_3$ and $c_5$ are determined by Eqs.\ (\ref{eq:constants1}) and (\ref{eq:constants2}), and $\dot{u}$ is restricted to positive values. We find
\begin{align}
F(x,y)=&
   \left(\frac{\tilde{y}^2}{2
   x}+x-\log (x)-1\right)+\frac{\sqrt{m} \tilde{y}^3}{6
   x^2}\notag\\
   &+m \left(\frac{\tilde{y}^4}{48
   x^3}-\frac{\tilde{y}^2}{4
   x^2}+\frac{\tilde{y}^2}{4
   x}+\frac{x}{2}-\frac{1}{2 x}-\log
   (x)\right)\notag\\
   &+m^{3/2}
   \left(-\frac{\tilde{y}^5}{240
   x^4}-\frac{\tilde{y}^3}{36
   x^3}-\frac{\tilde{y}^3}{18
   x^2}\right)+\mathcal{O}(m^2), \label{resFxy}
\end{align}
where $\tilde{y}=\sqrt{m} y$.

Similarly we  obtain the large-deviation function for the velocity only, defined in Eq.\ (\ref{eq:Fdefinition}).
Legendre
transforming the results of Sec.~\ref{sec:perturbation2} for $Z(\lambda)$ for small $m$, using (\ref{p2}) , we get
\begin{align}\label{eq:Fpert}
F(x)=&\  x-\log (x)-1+m
   \left(\frac{x}{2}-\frac{1}{2x}-\log (x)\right)\notag\\&
   +m^2 \left(-\frac{1}{12 x^2}-\frac{5x}{12}+\frac{3}{4 x}+\log  (x)-\frac{1}{4}\right)\nn\\&+\mathcal{O}(m^3).
\end{align}
We note that $F(x)=F(x,y=0),$ which presumably holds
to all orders.

One can check that convexity $F''(x)>0$ holds for $x>m/(1+m)$ for the two first orders in $m$.
This indicates that the expansion is valid at small $m$ only for $x \gg O(m)$. We will
see below, on the example of $m=1/4$, that indeed the small-$m$ expansion is accurate for large enough $x$ and
breaks down for small $x$, see figure \ref{f:F-compare}.

\subsubsection{Equation from Fokker-Planck}

From the Fokker-Planck equation it is possible to obtain a differential equation for
the large-deviation function. Inserting the form $P(\dot u, a) = e^{- v F_v(x=\frac{\dot u}{v},y=\frac{a}{v})}$
in the Fokker-Planck equation (\ref{eq:Fokkerdimnesionless}) we find that $F(x,y)$ satisfies, to dominant order at large $v$,
\be \label{F}
m^2 y \partial_x F - m (x+y-1) \partial_y F + D(x) (\partial_y F)^2 = 0
\ .\ee
Here $D(x)=x$ for the $\sqrt{\dot u}$ model and $D(x)=|x|$ for the
tree model. If we study this equation for $x>0,$ the two equations are the
same. While the equation for the tree model can  be studied
for all $x$, the meaning of the one for the $\sqrt{\dot u}$ model for $x<0$
 requires further analysis due to possible complex velocities.

We now use  equation (\ref{F}), and the emerging structure of the above perturbative results (\ref{resFxy}) and (\ref{eq:Fpert}), to construct the expansion in
$m$ to higher orders. One way to analyze (\ref{F}) is to perform a Taylor expansion around $y=0$, \bea
F(x,y) = F_0(x) + \sum_{n \geq 2} F_n(x) y^n,
\eea
where we assume $F(x)=F(x,y=0)$.
Then the $F_n(x)$ obey the recursion relations
 $F_1(x)=0$, $F_2(x) = F'_0(x)/[2 (x-1)]$, and for $k\ge2$
\bea
F_{k+1}(x) = \frac1{\sqrt{m} (x-1) (k+1) } \bigg[-k F_k(x) + \sqrt{m} F'_{k-1}(x)\nn\\
+ x \sum_{p=1}^{k-1} (p+1) (k-p+1) F_{p+1}(x) F_{k-p+1}(x) \bigg].\nn\\
\eea
If we assume that the
structure of (\ref{resFxy}) holds to higher orders, in particular that there are no poles in $1/(x-1)$,
we find that the conditions for their cancelations order by order in $m$ give enough conditions  to determine  $F_0(x)$ entirely. The
form which we find by inspection is\be
 F_k(x) = \sum_{n=0}^\infty m^{n + \frac{k-2}{2}} \sum_{p=k-1}^{p=n+k-1} \frac{a^k_{np}}{x^p}
\ee
for $k>0$.
One checks that the previous result (\ref{resFxy}) satisfies  equation (\ref{F}).
One finally obtains
\begin{align}
 &F(x) = F_0(x)=
(x-\log (x) -1)\nn\\&+\frac{1}{2} m \left(x-\frac{1}{x}-2 \log (x)\right)\nn\\
&+m^2
   \left(\log (x)-\frac{5 x^3-9 x+1}{12 x^2} - \frac{1}{4} \right)\nn\\
&+m^3 \left(\frac{x
   \left(107 x^3-240 x+22\right)+5}{144 x^3}-2 \log (x) + \frac{53}{72} \right) \nn \\
& +m^4
   \Big(\frac{669-x (x (14809 x^3-38610
   x+2090)+2905)}{8640 x^4}\nn\\&\qquad\qquad +5 \log
   (x) - \frac{3895}{1728} \Big)+O(m^5)\ .\label{202}
\end{align}
We have fixed the integration constant by requiring that $F(1)=0$, a consequence
of $\overline{\dot u}=v$. Note that the above result satisfies $F'(1)=0$  and $F''(1)=1$, which is
consistent with our analysis in Section \ref{sec:largev}, namely that the bulk of the distribution is the
Gaussian (\ref{eq:GaussianV}) in the scaling region $|\dot u - v | \sim \sqrt{v}$.
The first corrections to the Gaussian arise from $F'''(1)=-4/(2+m)$
and may already be visible in the tail $| \dot u - v | \sim v^{2/3}$. This can be compared to the perturbative expansion of $F(x)$ around $x=1$, given in Eq.\ (\ref{205}) below. The complete large-deviation
function describes the far tails $| \dot u - v | \sim v$. Another interesting feature is that
one obtains an expansion for $\lambda_c^+(m)$ as
\bea
\lambda_c^+(m) &=& \lim_{x \to \infty} \frac{F(x)}{x} \\ &=&1+\frac{m}{2} -\frac{5
   m^2}{12}+\frac{107
   m^3}{144}-\frac{14809
   m^4}{8640}+O(m^5). \nn
\eea
Finally note that for large $x,y$ the large deviation function takes the form:
\bea
F(x,y) \approx x g(y/x)
\eea
and that an ordinary differential equation can be written from (\ref{F}) for $g(z)$.

\subsection{Large deviation function for large $m$ and
matching of small and large $m$\label{sec:largem}}

In App.~C  it is explained how to calculate the perturbative solution of $Z(\lambda)$ in powers of $\lambda$.
Examining the results in powers of $\lambda$ keeping the complete $m$ dependence
of the coefficients shows that it actually turns into  a large-$m$ expansion. Indeed one finds\begin{align}
&Z(\lambda)=\lambda +\frac{\lambda ^2}{2}+\frac{2 \lambda ^3}{3
   m+6}+\frac{\lambda ^4 (5 m+6)}{16 m^2+44
   m+24}\nn \\
&+\frac{2 \lambda ^5 (103 m^2+198 m+72)}{5
   (m+2) (m+6) (4 m+3) (9 m+4)} \nn\\
&+\frac{\lambda ^6 (695 m^4+4396 m^3+7666 m^2+4284 m+720)}{3 (m+2)^2
   (m+6) (4 m+3) (9 m+4) (16 m+5)}\nn\\
&+O(\lambda^7)\ , \label{201}
  \end{align}
and each two orders more in $\lambda$ come with an additional factor of $1/m$ at large \(m\). We have obtained two more orders, which are not displayed here due to their length. Legendre transforming yields
\begin{align}\label{205}
&F(x)=\frac{1}{2} (x-1)^2-\frac{2 (x-1)^3}{3 m+6}\nn\\&+\frac{(12+16m-5m^{2}) (x-1)^4}{4 (m+2)^2 (4 m+3)}\nn\\&+\frac{4 (61 m^4+420 m^3-338 m^2-540 m-144) (x-1)^5}{5 (m+2)^3 (m+6) (4 m+3) (9 m+4)}\nn\\&+\frac{ (x-1)^6}{6 (m+2)^4 (m+6) (4 m+3)^2 (9 m+4)
   (16 m+5)}\times \nn\\ &~~~\times \Big[(17280 + 143136
m + 386448 m^2 + 239312 m^3 \nn\\&\qquad~~- 488936 m^4 -
 556346 m^5 - 3195 m^6 + 5240 m^7)\Big]\nn\\&+O(x-1)^7
\end{align}
which is actually an expansion in $x-1$ at fixed $m$, in other words
deviations from the Gaussian solution of Sec.~\ref{sec:largev} at large velocity.
Again, we have obtained two more orders, which are not displayed
here due to their length.  Since it yields the
derivatives $F^{(n)}(1)$ for any $m$, one can check that at small
$m$ they  match the result obtained from (\ref{202}) above.

From the above expansion one can now obtain a good approximation
for $F(0)$, which gives an estimate of the probability
\bea
p \sim e^{- v F(0)}
\eea
for negative velocities at large $v$, and improves on the estimate of Sec.~\ref{sec:largev}.

\begin{figure}
\Fig{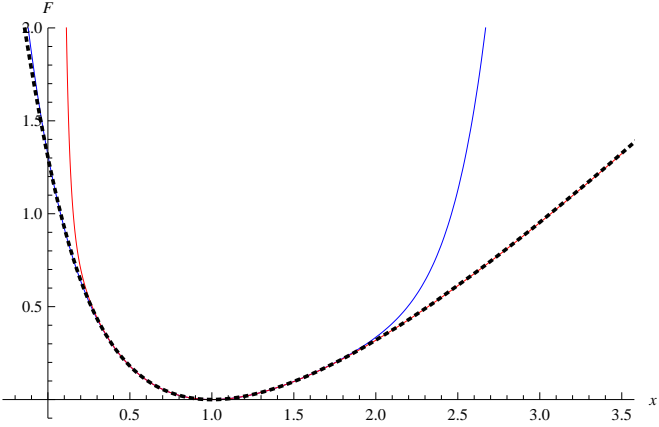}
\caption{$F(x)$ for $m=1/4$ as obtained by the numerical solution of the instanton equation (dashed line). Large-$m$ expansion (\ref{205}) to order $1/m^{3}$, thus two more orders in $(1-x)$ as given in Eq.~(\ref{205}) (blue solid line); this approximation substantially deviates for $x>2$. Small $m$ expansion (\ref{202}), which works well at large $x$, but breaks down at about $x=1/4$ (red solid line).}
\label{f:F-compare}
\end{figure}

As a test, we can  compare  the small-\(m\) expansion (\ref{202}) and the large-\(m\) expansion (\ref{205}) with the numerical solution of the instanton equation, followed by a  Legendre transform (in the form of a parametric representation).  The result for $m=1/4$ is shown on figure \ref{f:F-compare}. One sees that the small-\(m\) expansion works well for large $x$, but breaks down for $x\to 0$, while the large-\(m\) expansion converges for $x=0$, but may have a finite radius of convergence in $x-1$. Taking both expansions together, we have an analytical approximation for the range \(0<x<x_{+}\  \) drawn in red in Fig.~\ref{f:compare-Zs} in its right part of Sec.~\ref{sec:dev} where the large-deviation functions for the three models have been argued to coincide.

\subsection{Large-deviation function for $m=6/25$\label{sec:exact} }

Here we determine exactly, in a parametric form, the large-deviation function for the special value of the mass $m=6/25$ and for a certain range of values for $x$, using the results obtained in Sec.~\ref{sec:abel}.

One finds $F(x>0),$ using Eq.~(\ref{p2}), where $Z(\lambda=-C^3/20)$ is determined by Eq.~(\ref{eq:zm6over25}).  In Eq.\ (\ref{p2b}),  $x$ takes the values
\begin{align}
x\left(-\frac{C^3}{20}\right)=&\frac{1}{3 C^2
   t_{\text{C0}}}\Big[4
   \sqrt{t_{\text{C0}}^3+1}
   t_{\text{C0}}^3  \,_2F_1\left(\frac{5}{6},1;\frac{3}{
   2};t_{\text{C0}}^3+1\right)\notag\\
   & -6 C
   t_{\text{C0}}^2+3 t_{\text{C0}}^3
   \,
   _2F_1\left(\frac{1}{2},\frac{2}{3}
   ;\frac{5}{3};-t_{\text{C0}}^3\right)\notag\\
   &+\frac{18 \Gamma
   \left(\frac{7}{6}\right) \Gamma
   \left(\frac{4}{3}\right)
   \left(-t_{\text{C0}}^3\right){}^{2
   /3}}{\sqrt{\pi }}\Big]
\end{align}
where $t_{\mathrm{C0}}$ is given by Eq.~(\ref{eq:tco}). This expression is real as long as $t_{\mathrm{C0}}>-1$.

\section{Thermal and quantum fluctuations}

As long as we neglect backward trajectories, it is possible to include thermal and quantum fluctuations. Here we
obtain some new results within that approximation. We then discuss its expected range of validity.

\subsection{Classical systems: Thermal fluctuations}

For a classical system in presence of a thermal noise, we can generalize the equation of
motion in the laboratory frame (\ref{12}), (\ref{13}) as
\be
m \ddot u(t) + {\eta} \dot u (t) =\mu^2[v-\dot{u}(t)] + \partial_t F\big(u(t)\big) + \xi_T(t)\ .
\ee
To remain slightly more general, and cover the case of a colored noise, we define the correlations of the noise as
\bea \label{thermal}
\langle \xi_T(t) \xi_T(t') \rangle = B(t-t')\ .
\eea
The standard thermal white noise is
\be
B(t) = B_T(t) = 2 \eta T \delta(t)\ .
\ee
Hence when we can neglect the backward motion, and for $F(u)$ a Brownian landscape, the problem
reduces again to a Langevin equation for the velocity,
\be \label{newlangevin}
m \dddot u(t) + {\eta} \ddot u (t) =\mu^2[v-\dot{u}(t)] + \sqrt{\dot u} \xi(t) + \dot \xi_T(t)\
\ee
in presence of both a thermal noise and a noise generated by the quenched disorder. It still describes the equation of motion for the center-of-mass velocity of an elastic
manifold moving in a Brownian force landscape in presence of a thermal noise, under the
same (but more stringent) approximation of neglecting any backward movement of the interface (i.e. even of a piece
of it). Indeed for  forward motion, the general argument given in Refs.\ \cite{LeDoussalWiese2011a,LeDoussalWieseinprep2012}
and  \cite{DobrinevskiLeDoussalWiese2011} (section IV.A) still applies.

Also note that the Langevin equation (\ref{newlangevin}) defines a noisy version of the $\sqrt{\dot u}$ model,
as a legitimate model provided one accepts complex velocities which arise from backward trajectories.

It turns out that (\ref{newlangevin}) can be solved exactly for an arbitrary noise correlator $B(t)$. This is remarkable, since
except when $\dot \xi(t)$ is a white noise itself (see below) no simple Fokker-Planck version seems to exist. However, within
the MSR formalism the problem is much simpler. The Laplace transform of the velocity distribution can be written as in Eq.\ (\ref{39}), where the dynamical action $S[\dot u, \tilde u]$ contains the additional term\bea
&& - \frac{1}{2} \int_{tt'} \tilde u(t) \tilde u(t') \partial_{t} \partial_{t'} B(t-t') \\
&& = - \frac{1}{2} \int_{tt'} \partial_{t} \tilde u(t) \partial_{t'} \tilde u(t') B(t-t')\ .
\eea
The derivation follows the same steps as in Section \ref{sec:defcomplex}. The action is still linear in $\dot u$, hence the instanton equation (\ref{41}) is unchanged and inserting this equation in the action gives
\be \label{res1}
\overline{ \langle e^{\int \lambda \dot u(t_0)} \rangle } = e^{ v Z(\lambda) + \frac{1}{2} \int_{tt'} \partial_{t} \tilde u(t) \partial_{t'} \tilde u(t') B(t-t')  }
\ee
using that $\tilde u(t)$ vanishes at $t=\pm \infty$, with the same $Z(\lambda)=\mu^2 \int_t \tilde u(t)$ as in (\ref{42}). For a thermal noise it gives
\bea \label{res2}
\overline{ \langle e^{\int \lambda \dot u(t_0)} \rangle } = e^{ v Z(\lambda) + \eta T \int_{t} [\partial_{t} \tilde u(t)]^2   } \ .
\eea
Note that a double average is performed over thermal fluctuations and disorder realizations \footnote{Note that here the natural unit of temperature is $m_\mu v_\mu^2 = \mu^6/\sigma^2$ as can be seen from (\ref{res2}) and using that the dimensionfull instanton solution is $\tilde u^{\rm dimfull}_{\lambda,m}(t) = \frac{\mu^2}{\sigma} \tilde u^{\rm dimless}_{\lambda v_\mu,m \mu^2/\eta}(t/\tau_\mu)$. The
model now depends on three dimensionless parameters, $m$,$v$ and $T$ in the units defined here.}. Note also that these results can  be obtained from the general expression given in Ref.\ \cite{DobrinevskiLeDoussalWiese2011} (see Eq.\ (6) there), valid for an arbitrary forward driving and a forward motion
$w(t),$ by substituting $w(t) \to v t + \xi_T(t)/\mu^2$ and performing the Gaussian average over $\xi_T$
\footnote{$\mu$ here is noted $m$ there}.

The calculation of the contribution of the noise requires a small-time cutoff. For standard thermal white noise with no intrinsic cutoff, (\ref{res2}) can be evaluated only for a non-zero inertial mass, which provides the small-time cutoff. Its evaluation turns out to be remarkably simple, as we now show: Multiplying the instanton equation (\ref{41}) by $\partial_t \tilde u$, it can be rewritten, for $t<t_0$, as
\bea
&& - \partial_t \left[ \frac{m}{2} (\partial_t \tilde u(t))^2 + W(\tilde u(t))\right] = - \eta [\partial_t \tilde u(t)]^2 \\
&& W(\tilde u) = \frac{\mu^2}{2} \tilde u^2 - \frac{\sigma}{3} \tilde u^3\ .
\eea
This can be interpreted as a classical particle of position variable $\tilde u(t),$ undergoing (backward in time) a damped motion in
a potential $W(\tilde u)$. Integrating over time we find
\bea
 \eta \int_{-\infty}^{t_0} \rmd t (\partial_t \tilde u)^2 &=& \left[  \frac{m}{2} (\partial_t \tilde u(t))^2 + W(\tilde u(t)) \right]^0_{-\infty} \nn\\
& =& \frac{\lambda^2}{2 m} \ ,
\eea
using that $\tilde u(t_0)=0$ and $\tilde u'(t_0)=\lambda/m$, as well as that $\tilde u(t)$ decays to zero at $t=-\infty$. In other words, the
total dissipated energy is the total initial energy since the fictitious particle settles to rest at $t=-\infty$. Hence, we find  the exact result  for any $m$ and $v$, \bea \label{res3}
\overline{ \langle e^{\int \lambda \dot u(t_0)} \rangle } = e^{ v Z(\lambda) + \frac{T}{2 m} \lambda^2 } \ .
\eea
 The new term corresponds to the thermal broadening (i.e.\ $\dot u \to \dot u + \delta \dot u$) of the
$T=0$ velocity distribution by a Gaussian of variance $\langle \delta \dot u^2 \rangle^c  = \frac{T}{2 m}$. It can be interpreted
as the system reaching  kinetic-energy equipartition, as   in the system without disorder. Keeping only the
terms of $O(\lambda)$ and $O(\lambda^2)$ in (\ref{res3}) leads to the large-$v$ Gaussian distribution (\ref{eq:distributionm}) with $T_{\mathrm{eff}} \to T_{\mathrm{eff}} + T$,
and indicates that at large $v$ negative velocities are negligible when $T + T_{\mathrm{eff}} \ll m v^2$. More precisely, a new large-deviation function
at large $v$ can be defined if one scales $T \sim v$, i.e in the high-temperature, high-driving-velocity limit.
The Legendre transform of $Z(\lambda)+\theta \lambda^2/2$ yields to $F_{\theta}(x)$. Defining $\theta = T/(m v v_\mu^2)$ we have
\bea \label{res4}
&& P(\dot u) \sim  e^{ - v F_\theta(x=\frac{\dot u}{v}) } \\\label{eq:Fff}
&& F'(x) = F'_\theta(x + \theta F'(x)) .\
\eea
For the case $m=0$ (ABBM model without inertia) we find using $F_\theta(1)=0$
\bea \label{newF}
 F_\theta(x) &= &(x-1)\left [ \frac{1}{2} + \frac{x-\theta}{x+\theta + \sqrt{4 \theta + (x-\theta)^2}}\right] \nn \\
&& - \ln\left ( \frac{1}{2}\Big(x-\theta + \sqrt{4 \theta + (x-\theta)^2} \Big)\right)
\ .\eea
The second solution that follows from (\ref{eq:Fff}) is not applicable since it does not give Gaussian distribution expected at large $v$, Sec.~\ref{sec:largev}.
This formula should, following the general argument given in Section {\ref{sec:largedeviation}} also hold   in the positive velocity region $x>0$
for the original ABBM model (without inertia) and with  temperature, even though it cannot be solved exactly because of possible backward motion due to thermal fluctuations. Extension of (\ref{newF}) in presence of inertia can be studied using the array of methods introduced in this paper, but we refrain from doing so here.

The result (\ref{res3}) is remarkable since the effect of thermal fluctuations is exactly Gaussian, despite the presence of quenched randomness which is highly non-linear. This property can
be generalized and traced to the fact that the noise-dissipation satisfies the fluctuation dissipation relation
(FDR), i.e. for {\it an equilibrium} thermal bath. To see that consider a slightly more general bath
and response function, in frequency space,
\bea
&& R^{-1}(\omega) = - m \omega^2 + \mu^2 + \eta(\omega) i \omega\ ,
\eea
where $\eta(\omega)$ is an even function in $\omega$. The classical FDR reads\
\bea
2 T \eta(\omega) = B(\omega)\ .
\eea
Since the response function is changed, the instanton equation becomes  different. It involves
the transpose of the inverse response and reads
\be
(m \partial_t^2 + \mu) \tilde u - \sigma \tilde u^2 + \int_\omega e^{i \omega t} \eta(\omega) i \omega \tilde u(\omega)=0\ .
\ee
Again, multiplying with $\partial_t \tilde u(t)$ and integrating over time yields
\bea
\frac{\lambda^2}{2 m} = \int_\omega \eta(\omega) \omega^2 \tilde u(\omega) \tilde u(-\omega)\ .
\eea
Hence
\bea
 \frac{1}{2} \int_{tt'} \partial_{t} \tilde u(t) \partial_{t'} \tilde u(t') B(t-t') = \frac{\lambda^2}{2 m}
\eea
if the FDR holds. Eq.\ (\ref{res1}) again implies Eq.\ (\ref{res2}).

Formula (\ref{res1}) also allows to compute $\overline{e^{\lambda \dot u + \kappa a}}$, and the joint distribution of
velocity and acceleration, replacing $Z(\lambda)$ by $Z(\lambda,\kappa)$ and $\tilde u(t)$ by the
solution of the instanton equation with boundary conditions ({\ref{eq:kappa},\ref{eq:lambda}}). The formula (\ref{res2}) however cannot be used,
as it contains a divergent integral since $\tilde u(t)$ has a jump of $\kappa/m$ at $t_0$. This is because the
distribution of acceleration is not well defined, unless we add an intrinsic small-time cutoff to the thermal bath,
i.e. the mass is not sufficient to act as a cutoff. It can be seen  within the large-$ v$ analysis of Section {\ref{sec:largev}}
since $\langle a^2 \rangle$ is defined by the same integral with an additional $\omega^2$ in the numerator.

A non-Gaussian contribution to the velocity distribution can arise from a Gaussian noise {\it if it is a non-equilibrium one}, such as, for instance colored
noise with a frequency-independent constant-friction dissipation term $\eta$. We  mention here one simple example, when
$\dot \xi(t)$ is a white noise, $\langle \dot \xi(t) \dot \xi(t) \rangle = 2 D \delta(t-t')$, i.e. $\xi(t)$ is a Brownian. Then it is easy to see that for
$m=0$ one recovers  the result for the ABBM model without inertia (\ref{eq:distribution}) for $P(\dot u)$ with the replacement
\bea
P_{v,D}(\dot u) = A P_{v+D,D=0}(\dot u+D)
\eea
This is the only case which is amenable to a Fokker-Planck equation (with $\dot u \to \dot u+D$ in the diffusion term).
Not of course, that negative velocities  are simply neglected, hence the normalization factor $A$.

\subsection{Quantum system}

The extension to the quantum system in presence of a bath is straightforward. It can be done by generalizing the
MSR methods of Ref.~\cite{LeDoussalWiese2011a,DobrinevskiLeDoussalWiese2011} to the Keldysh path integral. Let us write the Keldysh action for a quantum particle in a
Brownian-force landscape. The Keldysh path-integral over the fields $u(t), \hat u(t)$ between an initial and a final time involves $e^{-S_K}$ with Keldysh action
$S_K = S_K^0 + S_K^1$ and
\begin{eqnarray}
 S_K^0 &=& \int_{tt'} \int_{t'} i \hat{u}(t) G_0^{-1}(t-t') u(t')
\nn\\&&-  \frac{1}{2} \int_{tt'} i \hat{u}(t) B(t-t') i \hat{u}(t')  \\
  S_K^1 &=& \frac{i}{\hbar}  \int_{t} \sum_{\epsilon=\pm 1} \epsilon V(u(t) + \epsilon \frac{\hbar}{2} \hat{u}(t)) \\
 S_K^2 &=& \frac{i \mu^2}{2 \hbar}  \int_{t} \sum_{\epsilon=\pm 1} \epsilon (u(t) + \epsilon \frac{\hbar}{2} \hat{u}(t) - w(t) )^2 \qquad \\
& =&   \int_{t} i \hat u(t) \mu^2 (u(t) - w(t))
\end{eqnarray}
The path integral can be expressed alternatively in terms of the upper and lower Keldysh fields
$u^{\pm}(t) = u(t) \pm \frac{\hbar}{2} \hat u(t)$. The form of the functions
$G_0^{-1}(t)$ and $B(t)$ depends on the details of the bath.
One convenient choice is\bea
 S^0_K + S^2_K &=& \int_{t} i \hat{u}_{t} ( m \partial_t^2 +
\eta \partial_t - \mu^2 ) u_{t}  \\
&& -  \frac{1}{2}  \int_{tt'} i \hat{u}_{t} B(t-t') i \hat{u}_{t'}  + \int_{t} i \hat{u}_{t} \mu^2 w_t \nn
\eea
with
\be
B(\omega)= \eta \hbar \omega coth(\beta \hbar \omega/2)\ ,
\label{235}\ee
which for $\eta>0$ represents an Ohmic bath. A realistic bath has
a large-frequency cutoff $\omega_c$.

After averaging over disorder the system is described by the same Keldysh action
with $S^1_K$ replaced by
\be
 S^1_K = \frac{1}{2 \hbar^2} \int_{tt'} \sum_{\epsilon,\epsilon'=\pm1} \epsilon \epsilon'
R\Big(u_{t} - u_{t'} + \frac{\hbar}{2} (\epsilon \hat{u}_{t} - \epsilon' \hat{u}_{t'})\Big)\ . \label{K1}
\ee
This is because the Keldysh path integral is normalized to unity.

In the classical limit $\hbar \to 0,$ one recovers the classical
MSR functional with the thermal white noise $B(\omega)= 2 \eta T$ and the usual
disorder part
\be
S_1^{\mathrm{classical}} = - \frac{1}{2} \int_{tt'} i \hat{u}_{t} i \hat{u}_{t'}
\Delta(u_{xt} - u_{xt'})\ .
\ee
 $\Delta(u)=- R''(u)$ is the correlator of the pinning force. Note that
 $i \hat u$ is sometimes denoted by $\hat u$ in the MSR formalism.

Now we can treat the case of a Brownian force landscape choosing
\be
R(u)  = R(0) - \Delta(0) \frac{u^2}{2}  + \sigma \frac{|u|^3}{6}\ .
\ee
This corresponds  to $\Delta(u)= \Delta(0) - \sigma u$. Inserting this into Eq.\ (\ref{K1}),
we obtain a complicated expression. However, if we make the replacement
\bea \label{approx}
{\rm sgn}(u^\epsilon(t) - u^{\epsilon'}(t')) \to {\rm sgn}(t-t')
\eea
for all four couples $(\epsilon,\epsilon')=(\pm1,\pm1)$,
then it simplifies into
\bea
S^1_K = \frac{1}{2} \int_{tt'} \hat u_t \hat u_{t'} \Big[ \Delta(0) - \sigma (u_t - u_{t'})  {\rm sign}(t-t')\Big]\ . \nn
\eea
The observable we are computing is the following average over the Keldysh action
\bea
\hat P[\lambda] = \langle e^{\int_t \lambda_t \dot u_t} \rangle_{S_K}\ .
\eea
The study of more general observable is left for the future. To recover the velocity theory we define
\bea
i \hat u_t = - \partial_t \tilde u_t\ ,
\eea
and consider $\tilde u_t$ vanishing at $t=\pm \infty$. It is very similar to what is
done in \cite{DobrinevskiLeDoussalWiese2011}, and to which we refer for details. It yields, after integration
by parts
\bea
S^1_K &=& - \sigma \int_{tt'} \tilde u_t \tilde u_{t'} \dot u_t \\
 S^0_K + S^2_K &=& \int_{t} \tilde u_t ( m \partial_t^2 +
\eta \partial_t - \mu^2 ) \dot u_{t}  \\
&& -  \frac{1}{2}  \int_{tt'} \partial_t \tilde u_t B(t-t') \partial_{t'} \tilde u_{t'} - \int_{t} i \tilde{u}_{t} \dot w_t  \ .\nn
\eea
Integration over $\dot u_t$ leads to the instanton equation (\ref{41}), and inserting its solution into the action,
we find again for $w(t)=v t$ the same result (\ref{res1}). Note however that for any non zero $\hbar$ the
bath cutoff time is needed to get a finite result - the mass only cutoff leads to a logarithmic divergence when inserting into
(\ref{res1}). Since they require the corresponding dissipation related by FRD, we leave explicit
calculations to future studies. To summarize however, one can say that
everything works as if the quantum system is described by a semi-classical equation of motion
with the noise correlator (\ref{235}).

Of course, there are two crucial ingredients here: (i)~the Brownian disorder-force landscape; (ii) the
approximation (\ref{approx}). For $\epsilon'=\epsilon$ it amounts to neglecting any trajectory with backward motion.
We see however that for $\epsilon'=-\epsilon$ the approximation cannot be correct at short time differences
$t-t'$, even for forward-only trajectories. This should be valid if this time scale is much smaller than the
other ones considered here. A more detailed discussion of the validity of this approach is left for the future.

\section{Conclusions and discussion}

In this paper we studied an extension of the ABBM model including inertia, i.e.\ the motion of a particle of inertial mass $m$ driven at externally imposed average velocity $v>0$ in a 1D Brownian random-force landscape in presence of damping. Its main interest, besides modeling a particle, is that in some cases it describes the center-of-mass dynamics of an interface in a Brownian correlated disorder. When all the segments of the interface move forward it is certainly true, and in more general situations it remains to be understood. For any $m$ this model can also be derived for the center of mass of a manifold with inertia in a short-range disorder potential, in the limit of a fully connected model (infinite-ranged interactions). In that sense it is
a mean-field approximation. Whether that property also extends to finite range interactions in high enough internal space dimension $d$, as is
the case for $m=0$ for $d>d_c$, remains to be understood. Our aim here was to calculate exactly, or using approximate methods, the distribution of the instantaneous velocity, and of the acceleration for this particle model.

We started by recalling the $m=0$ limit which is exactly solvable (standard ABBM model). It is characterized by a relaxation time scale $\tau_\mu$, a spatial scale
$S_\mu$ and a velocity scale $v_\mu=S_\mu/\tau_\mu$, see Sec.~\ref{sec:ABBM}. For $v<v_\mu$ the motion proceeds via avalanches, while for $v>v_\mu$ the motion becomes smoother. In presence of inertia $m>0$, the ABBM model cannot easily be solved  because
backward motion occurs and induces memory effects. Numerical simulations and qualitative arguments
showed new characteristic time scales for oscillations $\tau_0$, and damping $\tau_m$, and new velocity scales $v_0=S_\mu/\tau_0$ and $v_m=S_\mu/\tau_m$ (Sec.~\ref{sec:model}). An avalanche regime survives for $v<\min(v_m,v_\mu)$ similar to the one for $m=0$ except that
the smallest avalanches have merged into bigger ones. As the mass increases, overshoots and oscillations become more pronounced before the damping allows relaxation into a metastable state. As $v$ increases further the motion becomes smoother, but  oscillations persist. As a general rule the inertia tends to make the motion less jerky and to smoothen the abrupt jumps of a center-of-mass position in time.
At the same time, the distribution of velocities evolves from being strictly positive but with a
divergent limit $P(\dot{u}=0^+)= \infty$ for $m=0$ and $v<1$, corresponding to intermittent motion
where the particle is part of the time at rest, to developing a finite weight for negative velocities and a
finite $P(\dot{u}=0)$, corresponding to oscillatory motion.

To make quantitative progress we introduced two variants of the model which share the exact same dynamics with the ABBM model for all forward trajectories, and are analytically more tractable. The analysis of the first, the tree model, is based on a Fokker-Planck equation. The study of the second one, the $\sqrt{\dot u}$ model, is based on a saddle-point equation of the dynamical action. For both of them we calculated the joint distribution of acceleration and velocity perturbatively in small  and large mass. The $\sqrt{\dot u}$ model could also be solved exactly for a magic value of the mass in terms of hypergeometric functions, and was studied with very high precision for other values of the mass. From these variant models we obtained two sets of results for the ABBM model:

(i) at large driving velocity: Since by increasing $v$ the probability distribution for negative velocities decreases, all the considered models become more and more similar. The bulk of the velocity distribution, i.e.~for velocities $|\dot{u}-v| = O(v)$, then tends to the same Gaussian.
To characterize more accurately the
tails we defined for each model the large-deviation function which describes the rare events when the instantaneous velocity $\dot u$ deviates from the average $v$. We proved that these large-deviation functions become identical for positive $\dot{u}$ at large $v$ for the three models, and obtained analytical expressions at small and large $m$ and for the magic value of the mass.

(ii) at any driving velocity: We compared the three models and discuss differences and similarities. Although agreement is not
exact anymore, some features of the ABBM model are, in some cases, quite well reproduced.

Finally we showed how thermal and quantum fluctuations can also be treated within the approximation of
neglecting backward trajectories. For thermal fluctuations it is expected to be a reasonable approximation at fixed
$v$ only for small $T$ or for any $T$ at large $v$. For quantum fluctuations, the discussion is more
subtle but basically it should hold within a semi-classical approximation.

In conclusion this paper proposes a first step to the description of classical and quantum avalanches of pinned elastic systems in presence
of inertia, and the effect of driving. A more elaborate theory should incorporate barrier crossing by thermal and
quantum fluctuations, and a treatment of memory and oscillation effects. However we believe that we have introduced a
useful framework. For instance a key observable is $Z(\lambda) = \frac{1}{v} \ln \overline{e^{\lambda \dot u}}$
which in the $\sqrt{\dot u}$ model is independent of $v$, and is well characterized by branch-cut singularities
describing the tails of the velocity distribution. At the same time, as $v \to 0^+$ it describes the avalanche
dynamics. Hence, numerical or experimental determination of this quantity in realistic systems could
help developing  further understanding.

It would be interesting to extend the current analysis to different time-dependent and space-dependent driving, as well as to analyze the spatial correlations of the probability distribution taking into account the spatial extension of the interface and not just its center-of-mass position.
Additionally, due to retardation effects appearing in soft magnets \cite{review} there is a need to generalize the current approaches and to use a more general equation of motion, with a more general response function. Finally in quantum systems where velocity translates into current,
developing a more general connection with the full counting-statistics problem would be very interesting.

\section{Acknowledgments}
We are grateful to A. Dobrinevski for numerous helpful remarks. We thank Z. Ristivojevic for suggesting the method of numerical solving of the Fokker-Planck equation (\ref{eq:Fokkerdimnesionless}), pointing out Ref.~\cite{discretization} and useful discussions. We
thank D. Bernard, Y. Fyodorov and S. Majumdar for helpful discussions. This work was supported by ANR grant 09-BLAN-0097-01/2. We are
grateful to KITP for hospitality and partial support through NSF grant PHY05-51164.

\appendix

\bigskip

\appendix

\section{\label{app:perturbationtheory}Perturbation theory at small $m$ for \(P(\dot u)\) for the tree model}

In this appendix we give details on the perturbation theory discussed in the Sec.~\ref{sec:perturbation}.
We start with the differential equation (\ref{eq:recursion}) with $n=0$. It has  the solution
\begin{align}
F_0(a,u)=\sqrt{\frac{\pi }{2}} \sqrt{\dot{u}} c_1(\dot{u})
   \text{erfi}\left(\frac{a}{\sqrt{2}
   \sqrt{\dot{u}}}\right)+c_2(\dot{u}),
\end{align}
where $\text{erfi}$ denotes the imaginary error function $\mathrm{erfi}(z)=\text{erf}(iz)/i$. In order to have properly defined moments $\overline{a^i\dot{u}^j}$, the distribution function has to decay exponentially
fast at large $a$ and $\dot{u}$. This implies that $c_1(\dot{u})=0$. Next, \begin{align}\label{eq:F1}
F_1(a,\dot{u})&=c_4(\dot{u})\nn\\&+
   \frac{1}{2} \left[-\frac{1}{3} a^3 c_2(\dot{u})+ a \dot{u}c_2(\dot{u})-
      2 a \dot{u}^2 c_2'(\dot{u})\right],\\\label{eq:F2}
F_2(a,\dot{u})&= \frac{1}{2} a^2 \dot{u}^3
\left[-v c_2'(\dot{u})+\dot{u} c_2'(\dot{u})-\dot{u}
c_2''(\dot{u})\right] \notag\\& \times\,
_2F_2\left(1,1;\frac{3}{2},2;\frac
{a^2}{2 \dot{u}}\right)\notag\\&
+\frac{1}{72} a^6
c_2(\dot{u})+\frac{1}{48} a^4 \dot{u} \left[8
\dot{u} c_2'(\dot{u})-5
c_2(\dot{u})\right]\notag\\&
-\frac{1}{6} a^3
c_4(\dot{u})\nn\\ &+\frac{1}{4} a^2 \dot{u}^2 \left[2
\dot{u}^2 c_2''(\dot{u})+v c_2(\dot{u})-\dot{u}
c_2(\dot{u})\right]\notag\\&
+\sqrt{\frac{\pi
}{2}} \sqrt{\dot{u}} c_5(\dot{u})
\text{erfi}\left(\frac{a}{\sqrt{2}
\sqrt{\dot{u}}}\right)
\notag\\&-\frac{1}{2} a \dot{u}
\left[2 \dot{u} c_4'(\dot{u})-5
c_4(\dot{u})\right]+c_6(\dot{u}),
\end{align}
where $\,_2F_2(a;b;z)$ is a generalized hypergeometric function.
The first line of Eq.~(\ref{eq:F2}) has to vanish due to its large-$a$ behavior, which  gives us a differential equation for $c_2(\dot{u})$.
Solving it, we obtain\begin{align}
c_2(\dot{u})= c_3 (-1)^v  \Gamma
   (1-v,-\dot{u})+c_4\ .
\end{align}
Analyzing it, we conclude that $c_3=0$. Then, finally we obtain that
$
F_0(a,\dot{u})=c_4.
$
Since the distribution has to be normalized by $\int\dif a\,\dif\dot{u\,}P(\dot u,a)=1$ for every $m$, we conclude that $F_0(a,\dot{u})=1$.

Also, $c_5(\dot{u})$ appearing in $F_2$, has to be zero. We proceed further in a similar way. In order to find $c_4(\dot{u})$ entering $F_1(a,\dot{u})$, we have to solve Eq.~(\ref{eq:recursion}) for $n=3$, and the procedure continues for higher-order terms in $m$. To be able to find $F_i$ we have to solve the differential equations (\ref{eq:recursion}) for all $n\leq i+2$.

Apart from the already stated Eqs.~(\ref{eq:F1text}) and (\ref{eq:F2text}) we give  the final expressions for
\begin{align}
&F_3(a,\dot{u})=\frac{1}{6} \dot{u}^4 v \log (\dot{u})
   \left(-\tilde{a}^3+3 \tilde{a} \dot{u}+6 c_3
   \dot{u}^2\right)\notag\\&
   +\frac{-10 \tilde{a}^9+135 \tilde{a}^7
   \dot{u}+180 \tilde{a}^6 c_3 \dot{u}^2}{12960}\notag\\&
   +\frac{27 \tilde{a}^5 \dot{u}^2 (20
   \dot{u}-20 v-3)-1350 \tilde{a}^4 c_3 \dot{u}^3}{12960}\notag\\
   &+\frac{45 \tilde{a}^3
   \dot{u}^3 \left(-4 \left(12 c_5+5\right)
   \dot{u}+24 \dot{u}^2-24 v^2+80
   v-17\right)}{12960}\notag\\
   &-\frac{3240 \tilde{a}^2 c_3 \dot{u}^4
   (\dot{u}-v)}{12960}\notag\\
   &-\frac{135 \tilde{a} \dot{u}^4 \left(-8 \left(6
   c_5-5\right) \dot{u}+24 \dot{u}^2-24 v^2+20
   v+9\right)}{12960}\notag\\
   &-\frac{270 \dot{u}^5 \left(c_3
   \left(24 \dot{u}^2-24 v^2+36
   v-5\right)-48 c_7 \dot{u}\right)}{12960}.
   \end{align}
Higher orders can be calculated in the same manner.

As an approximation for the tree model that seems justified for small enough mass, and large enough driving velocity, one can write $1=\int_{-\infty}^{\infty} \rmd a \int_{-\infty}^{\infty}\rmd u {P}\approx\int_{-\infty}^{\infty} \rmd a \int_{0}^{\infty}\rmd u {P}^{(1)}$. From this condition follow Eqs.~(\ref{eq:constants1}), (\ref{eq:constants2}), and we find
\begin{align}\label{eq:distributionmm}
&P_{\rm approx}(\dot{u})=\frac{e^{-\dot{u}}
   \dot{u}^{v-1}}{\Gamma (v)}-\frac{m e^{-\dot{u}}
   \dot{u}^{v-2}}{
   2 \Gamma (v)}\notag\\&
   \times \left(-2 \dot{u} v
   \log \left(\dot{u}\right)+2
   \dot{u} v \psi
   ^{(0)}(v)+\dot{u}^2-v^2+v\right)\notag \\ &
   +\frac{m^2 e^{-\dot{u}} \dot{u}^{v-3}}{24 \Gamma(v)}\notag\\&
\times\Big(
-12 \dot{u} v \log
\left(\dot{u}\right)
\left(-\dot{u} v \log
\left(\dot{u}\right)+\dot{u}
\left(\dot{u}+2\right)-v^2+v\right
)\notag\\&
+12 \dot{u} v \Big(\psi
   (v)\notag\\&
\times   (-2 \dot{u} v \log
   \left(\dot{u}\right)+\dot{u} v
   \psi ^{}(v)+\dot{u}
   \left(\dot{u}+2\right)-v^2+v
   )\notag\\&
   -\dot{u} v \psi
   ^{(1)}(v)\Big)\notag\\ &-6 \dot{u}^2 (v-4)
   v+6 \dot{u} (5-3 v) v+3
   \dot{u}^4+10 \dot{u}^3
   \notag\\&+(v-2) (v-1)
   v (3 v+5))\Big)
\end{align}

A similar analysis can be done for the $\sqrt{\dot{u}}$ model, and there Eq.~(\ref{eq:distributionmm}) is obtained for small mass after neglecting complex velocities, as shown using the instanton solution in Sec.~\ref{sec:perturbation2}.

\section{\label{app:matching}Matching of $\dot u \sim m$ and $\dot u \gg m$ at small mass}

Next we discuss some remaining details of the perturbation theory presented in Sec.~\ref{sec:perturbation}. Although we could not solve analytically the equations that determine the solution in  region 2, in this appendix we demonstrate that we properly organized the perturbation theory in  region 2. We will prove the matching between the distribution function in  regions 1 and 2, without explicitly solving the equations in  region 2.

The matching condition at the boundary reads  $P^{(1)}({\dot{u}},a)\approx P^{(2)}(\dot{u},a)$. From that follows
\begin{align}\label{eq:P2} P^{(2)}(\dot{u},a)=&m^{v-1}\sum_{n=0}^{\infty}
\tilde{P}_n^{(2)}(a,\tilde{\dot{u}})m^n\notag\\
&+\log{(m)}m^{v-1}\sum_{n=0}^{\infty}
\tilde{\mathcal{P}}_n^{(2)}(a,\tilde{\dot{u}})m^n+\ldots,
\end{align}
where $\ldots$ denotes that there are other terms $\sim(\log{m})^i$ where $i>1$, as well as $\sim m^{v-1/2}$. Note that $\log{(m)}$ terms come from the $\log(u)$ dependence in $P^{(1)}$, while $\sim m^{v-1/2}$ comes from terms like $\sim c_3$ in $P^{(1)}$. Note, that $\tilde{\mathcal{P}}_n^{(2)}$ satisfies the same equations as $\tilde{P}_n^{(2)}$, but with different boundary conditions, i.e.~it contains only some terms from $\tilde{P}_n^{(2)}$ (see below). The same holds for the omitted terms which  can be analyzed in the same manner.

For large enough $a,\tilde{\dot{u}}$
\begin{align}\label{eq:largeexpansion}
\tilde{P}_0^{(2)}(a,\tilde{\dot{u}})&\approx e^{-\frac{a^2}{2\tilde{\dot{u}}}}\sum_{n=0}^{\infty}
\frac{\tilde{\dot{u}}^{v-\frac{3}{2}-2n}}{\Gamma(v)\sqrt{2\pi}}
\tilde{F}_n(a,\tilde{\dot{u}}),\\
\tilde{F}_n(a,\tilde{\dot{u}})&=\mathrm{lim}_{m\to 0}\frac{F_n(a\sqrt{m},\tilde{\dot{u}}m)}
{m^{\frac{3n}{2}}},
\end{align}
where $F_n$ are determined by Eq.~(\ref{eq:recursion}).
Analyzing expressions for $F_n$, we find\begin{align}\label{eq:F}
\tilde{F}_n(a,\tilde{\dot{u}})={\sum_{i=0}^{[3n/2]}a^{3n-2i}\;\tilde{\dot{u}}^i c^n_i}\ ,
\end{align}
where $[x]$ rounds $x$ to an integer such that $[x]\leq x$ and $c^n_i$ are numbers. Strictly speaking Eq.~(\ref{eq:F}) holds for $n<7$ and for $n\geq 7$ there might be some additional $(\log u)^i$ terms, but in that case the discussion would be  similar.

By plugging Eq.~(\ref{eq:P2}) into Fokker-Planck Eq.~(\ref{eq:Fokkerdimnesionless}) one finds
\begin{align}\label{eq:P02}
a\frac{\partial \tilde{P}^{(2)}_0}{\partial \tilde{\dot{u}}}+\frac{\partial}{\partial a}\{(-a+v)\tilde{P}^{(2)}_0\}
-\frac{\partial^2}{\partial a^2}\left({ \tilde{\dot{u}}\tilde{P}^{(2)}_0}\right)=0\ .
\end{align}
In order that both Eqs.~(\ref{eq:largeexpansion}) and (\ref{eq:P02}) hold,\begin{align}
\label{eq:condition}
&c^n_i(3n-2i)-c^n_{i-1}(3n-2i+2)(3n-2i+1)\notag\\&
+c^{n-1}_{i-1}(-\frac{1}{2}-2n+i)
+\frac{1}{2}c^{n-1}_i\notag\\
&+c^{n-1}_{i-2}v(3n-2i+1)=0\ ,
\end{align}
where $c^n_i=0$ for $i<0$ as well as for $i>[3n/2]$.
 Indeed, our results satisfy this structure.

Note that for $i=0$ we obtain $c^n_0 3n+c^{n-1}_0/2=0$. To conclude, using Eq.~(\ref{eq:condition}) and knowing $c_0^0=1$ we can find all  others coefficients $c^n_i$. Then the distribution of velocities has for large enough velocities the form
\begin{align}
\tilde{P}^{(2)}_0(\tilde{\dot{u}})=\sum_{k=0}^{\infty}
\frac{\tilde{\dot{u}}^{v-k-1}}{\Gamma[v]\sqrt{2\pi}}\sum_{i=0}^{3k}2^{3k-i+1/2}c^{2k}_i
\Gamma(3k-i+1/2).
\end{align}
Now we consider the next term in the expansion $\tilde{P}_1^{(2)}$. Knowing $P^{(1)}$, from the matching condition follows that
\begin{align}\label{eq:expansionP1}
\tilde{P}_1^{(2)}(a,\tilde{\dot{u}})\approx & e^{-\frac{a^2}{2\tilde{\dot{u}}}}\sum_{n=0}^{\infty}
\frac{\tilde{\dot{u}}^{v-\frac{3}{2}-2n}}{\Gamma(v)\sqrt{2\pi}}
\tilde{Q}_n(a,\tilde{\dot{u}}),\\
\tilde{Q}_n(a,\tilde{\dot{u}})=&-\tilde{\dot{u}}\tilde{F}_n(a,\tilde{\dot{u}})
+\tilde{\dot{u}}\sum_{i=2}^{[3n/2]}a^{3n-2i}\tilde{\dot{u}}^i p^n_i
\notag\\
&+
\tilde{\dot{u}}\log{\tilde{\dot{u}}}\sum_{i=3}^{[3n/2]}a^{3n-2i}\tilde{\dot{u}}^i d^n_i,
\end{align}
for  large enough velocities. On the other hand, the equation for $\tilde{P}^{(2)}_1(a,\tilde{\dot{u}})$ reads\begin{align}\label{eq:P11}
&a\frac{\partial \tilde{P}^{(2)}_1}{\partial \tilde{\dot{u}}}+\frac{\partial}{\partial a}\left[(v-a)\tilde{P}^{(2)}_1\right]
\notag\\
&-\frac{\partial^2}{\partial a^2}\left({ \tilde{\dot{u}}\tilde{P}^{(2)}_1}\right)
-\tilde{\dot{u}}\frac{\partial}{\partial a}\tilde{P}^{(2)}_0=0.
\end{align}
From Eqs.~(\ref{eq:expansionP1}) and (\ref{eq:P11}) one obtains
\begin{align}\label{eq:condition2}
&p^n_i(3n-2i)-p^n_{i-1}(3n-2i+2)(3n-2i+1)
\notag\\&
+p^{n-1}_{i-1}(\frac{1}{2}-2n+i)
+\frac{1}{2}p^{n-1}_i+p^{n-1}_{i-2}v(3n-2i+1)\notag\\&-c^{n-1}_{i-2}(3n-2i+1)+d^{n-1}_{i-1}=0,\\
\label{eq:condition3}
&d^n_i(3n-2i)-d^n_{i-1}(3n-2i+2)(3n-2i+1)\notag\\&
+d^{n-1}_{i-1}(\frac{1}{2}-2n+i)
+\frac{1}{2}d^{n-1}_i
+d^{n-1}_{i-2}v(3n-2i+1)=0,
\end{align}
where $p^n_i=0$ for $i<2$ or $i>[3n/2]$, and $d^n_i=0$ for $i<3$ or $i>[3n/2]$. Note that knowing $d^{2}_3=v$ we can find all  other coefficients. By examining the expressions for $\tilde{Q}_n$ that follow from $F_n$, we found that both Eqs.~(\ref{eq:condition2}) and (\ref{eq:condition3})  hold. One can repeat the procedure for higher-order terms in the same way.

 Similarly, we find that in  region 1
\begin{align}
\mathcal{P}_0^{(2)}&=0,\\
\mathcal{P}_1^{(2)}&=e^{-\frac{a^2}{2\tilde{\dot{u}}}}
\frac{\tilde{\dot{u}}^{v-\frac{1}{2}-2n}}{\Gamma(v)\sqrt{2\pi}}
\sum_{i=3}^{[3n/2]}a^{3n-2i}\tilde{\dot{u}}^i d^n_i.
\end{align}
The analysis presented in this appendix confirms that perturbation theory in  region 2 is properly organized and matching between the distributions in  regions 1 and 2 holds.

\section{Large-deviation function\label{app:deviation}}

Here we comment shortly on the Laplace transformed Fokker-Planck equation for the tree model and then present a calculation
of its large deviation function. It allows to show that the equivalence of the large-deviation function for the tree and $\sqrt{\dot{u}}$ model for positive velocities holds self-consistently. It may allow to obtain a complete proof of the equivalence if one subcase (see below) could be ruled out,
but we have not succeeded at this stage in doing so.

The Laplace-transformed Fokker-Planck equation reads
\begin{align}\label{eq:hatPplus}
& \frac{\partial \hat{P}_{+}}{\partial t}-\frac{\partial \hat{P}_{+}}{\partial \kappa}\left(\lambda -\frac{\kappa}{m}\right)-\frac{\partial \hat{P}_{+}}{\partial \lambda}\left(-\frac{\kappa}{m}+\frac{\kappa^2}{m^2}\right)\notag\\ &=\frac{\kappa}{m} v \hat{P}_{+}+ \Phi(\kappa),\\
& \frac{\partial \hat{P}_{-}}{\partial t}-\frac{\partial \hat{P}_{-}}{\partial \kappa}\left(\lambda -\frac{\kappa}{m}\right)-\frac{\partial \hat{P}_{-}}{\partial \lambda}\left(-\frac{\kappa}{m}-\frac{\kappa^2}{m^2}\right)\notag\\ &=\frac{\kappa}{m} v \hat{P}_{-}- \Phi(\kappa),
\end{align}
where
\begin{align}\label{eq:Phi}
\Phi(\kappa)=\frac{\partial}{\partial \kappa}\int_{-\infty}^{\infty}\mathrm{d}a
e^{\kappa a}P(\dot{u}=0,a)
\end{align}

Here we kept only the one boundary term at $\dot{u}=0$, assuming that other terms at ``infinity" vanish. Here $\hat{P}_{+}(\lambda,\kappa)=\int_{0}^{\infty}\mathrm{d}\dot{u}
\int_{-\infty}^{\infty}\mathrm{d}a
e^{\lambda \dot{u}+\kappa a}P(\dot{u},a)$. We define $\hat P_-$ similarly, with the difference that the integration is over negative velocities.

Eq.~(\ref{eq:hatPplus}) (which  describes the contribution coming from positive velocities) can be compared with the corresponding equation for the $\sqrt{\dot{u}}$ model (\ref{eq:laplacetransformed}). One sees that the main difference is the additional boundary term $\Phi(\kappa)$. When this boundary term vanishes, the two models become equivalent.

In the stationary case, introducing the time parameterization as in Eqs.~(\ref{eq:kappaM}) and (\ref{eq:lambdaM}) we obtain
\begin{align}
\frac{\dif \hat{P}_+}{\dif t}-\frac{\kappa(t)}{m}v\hat{P}_+-\Phi(\kappa(t))=0.
\end{align}
The solution of this equation is
\begin{align}\label{eq:app}
\hat{P}_+(\lambda_0,\kappa_0)=&\hat{P}_+(0,0)e^{v Z(\lambda_0,\kappa_0)}\notag\\&+
\int_{-\infty}^{t}
e^{v\left[Z(\lambda_0,\kappa_0)-Z(\lambda(s),\kappa(s)) \right]}\Phi(\kappa(s))\dif s,
\end{align}
where $Z$ is given by Eq.~(\ref{eq:ZM}) with $\kappa(t)=\kappa_0$ and $\lambda(t)=\lambda_0$.

Introducing $F(x,y)$ as in Eq.~(\ref{eq:Fdefinitionxy}), we find $\Phi(\kappa)=a_0e^{v[G(\kappa)-f_0]}$, where $\hat{P}_+(0,0)=e^{-v f_0}=e^{-v\mathrm{min}_{x,y}F(x,y)}$. Here
\begin{align}\label{eq:G}
G(\kappa)-f_0=-\mathrm{min}_{y}\{F(0,y)-\kappa y\},
\end{align}
and $a_0$ is the corresponding value of the acceleration when the minimum is reached. If we assume that the first term in Eq.~(\ref{eq:app}) gives the main contribution in the limit $v\to \infty$, we obtain
\begin{align}\label{eq:ZZ}
-f_0+Z(\lambda,\kappa)=-\mathrm{min}_{x,y}\{ F(x,y)-\kappa y-\lambda x\}.
\end{align}
Then  indeed   the second term is smaller than the first one in Eq.~(\ref{eq:app}), since it reads
\begin{align}\label{eq:more}
&e^{-v f_0}e^{v Z(\lambda,\kappa)}\int_{-\infty}^t e^{-v\left[ Z(\lambda(s),\kappa(s))-G(\kappa(s))\right]} \rmd s=\notag\\
&e^{-v f_0}e^{v Z(\lambda,\kappa)} e^{-v\,\mathrm{min}_{s\in (-\infty,t)}\{Z(\lambda(s),\kappa(s))-G(\kappa(s))\}},
\end{align}
and using (\ref{eq:G}) and (\ref{eq:ZZ}) it follows that $\phi(\lambda_0,\kappa_0)=\mathrm{min}_{s\in (-\infty,t)}\{Z(\lambda(s),\kappa(s))-G(\kappa(s))\}\geq 0$.

However, if we assume that the second term gives the main contribution, then  (\ref{eq:more}) implies that\begin{align}\label{eq:new}
&-f_0+Z(\lambda,\kappa)-\phi(\lambda,\kappa) \nn \\ &\qquad=-\mathrm{min}_{x,y}
\{F(x,y)-\kappa y-\lambda x\}.
\end{align}
From this equation and Eq.~(\ref{eq:G}) follows that $\phi(\lambda_0,\kappa_0)\leq Z(\lambda_0,\kappa_0)-G(\kappa_0)$. This statement is not in contradiction with the assumption, but on the other hand it does not follow from (\ref{eq:new}) that $\phi(\lambda_0,\kappa_0)\leq 0$, as it should be if the second term is the dominant one. It would be nice to show that this possibility is ruled out, which would provide
a proof of the equivalence of large deviation functions, independent of the one given in the main text.

Note, that $f_0$ is expected to be zero and then (\ref{eq:ZZ}) is equivalent to Eq.~(\ref{eq:195}).
\section{Moments of the distribution function for $\sqrt {\dot u}$ model}\label{app:moments}

Using the equations derived in Sec.~{\ref{sec:instanton}} we find  the moments characterizing the  distribution function of the $\sqrt{\dot{u}}$ model  discussed in Sec.~\ref{sec:instanton}.
Choosing $t^*=0$ with a constraint $\tilde{u}(t\geq 0)=0$ and $\dot{\tilde{u}}(0)=-\lambda/m$, we can rewrite  Eq.~(\ref{eq:kappamotionM}) as
\begin{align}\label{eq:dimensionleseq}
(m \partial_t^2-\partial_t+1)\tilde{u}-\tilde{u}^2=\lambda\delta(t).
\end{align}
Then  $Z(0,\lambda)=\int^0_{-\infty}\tilde{u}(t)\rmd t$.

One can write  $\tilde{u}=\sum_{n=0}^{\infty}\tilde{u}_n\lambda^n$, where
\begin{align}
(m \partial_t^2-\partial_t+1)\tilde{u}_n(t)-\sum_{\ell=0}^{n}
\tilde{u}_{\ell}(t)\tilde{u}_{n-\ell}(t)=\delta_{n,1}\delta(t).
\end{align}
A solution of this set of equations is\begin{align}
\tilde{u}_0&=0,\\
\tilde{u}_1(t)&=R(t),\\
\tilde{u}_n(t)&=\int\mathrm{d}\tau R(t-\tau)\sum_{\ell=1}^{n}
\tilde{u}_{\ell}(\tau)\tilde{u}_{n-\ell}(\tau)\quad\quad\text{for}\;\;n>1,
\end{align}
with
\begin{align}\label{eq:response}
&(m \partial_t^2-\partial_t+1)R_{t-t'}=\delta(t-t'),\\
&R_t=\begin{cases} \frac{2\theta(t)}{\sqrt{4m-1}}e^{t/(2m)}\sin(\frac{t}{2m}\sqrt{4 m-1}),& m>1/4\\
\frac{2\theta(t)}{\sqrt{1-4m}}e^{ t/(2m)}\sinh{(\frac{t}{2m}\sqrt{1-4m})},& m<1/4.
\end{cases}.
\end{align}
The case \(m=1/4\) was treated in equation (\ref{e:46}).
 Using that $\hat{P}(\lambda,\kappa)=e^{v Z(\lambda,\kappa)}$ and $\partial_{\lambda}^n\partial_{\kappa}^m\hat{P}(\lambda,\kappa)|_{(0,0)}
=\overline{\dot{u}^n a^m}$, we obtain the first few moments exactly:
\begin{widetext}
\begin{align}
\overline{\dot{u}}&=v,\\
\overline{\dot{u}^2}&=v (\sigma+v),\\
\overline{\dot{u}^3}&=v \left(\frac{4\sigma ^2}{m'+2}+3\sigma v+v^2\right),\\
\overline{\dot{u}^4}&=v \left(\frac{6 \sigma^3 (5 m'+6)}{(m'+2) (4 m'+3)}+\frac{\sigma^2 (3 m'+22) v}{m'+2}+6 \sigma v^2+v^3\right),\\
\overline{\dot{u}^5}&=v \left(\frac{48 \sigma^4 (m' (103 m'+198)+72)}{(m'+2) (m'+6) (4 m'+3) (9 m'+4)}+\frac{10 \sigma^3 (31 m'+30) v}{(m'+2) (4 m'+3)}+\frac{5 \sigma^2 (3 m'+14) v^2}{m'+2}+10 \sigma v^3+v^4\right),\\
\overline{\dot{u}^6}&=v \Big(\frac{240 \sigma^5 (m' (m' (m' (695 m'+4396)+7666)+4284)+720)}{(m'+2)^2 (m'+6) (4 m'+3) (9 m'+4) (16 m'+5)}\notag\\
&+\frac{2 \sigma^4 (m' (m' (9 m' (225 m'+4138)+130916)+137592)+39456) v}{(m'+2)^2 (m'+6) (4 m'+3) (9 m'+4)}\notag\\
&+\frac{15 \sigma^3 (m' (4 m'+105)+90) v^2}{(m'+2) (4 m'+3)}+\frac{5 \sigma^2 (9 m'+34) v^3}{m'+2}+15 \sigma v^4+v^5\Big),\\
\overline{\dot{u}^7}&=v \Big(\frac{2880 \sigma^6 (m' (m' (m' (m' (22015 m'+190244)+521534)+594996)+277920)+43200)}{(m'+2)^2 (m'+6) (m'+12) (4 m'+3) (9 m'+4) (9 m'+10) (16 m'+5)}\notag\\
&+\frac{168 \sigma^5 (m' (m' (m' (20438 m'+114479)+179748)+93924)+15120) v}{(m'+2)^2 (m'+6) (4 m'+3) (9 m'+4) (16 m'+5)}\notag\\
&+\frac{14 \sigma^4 \left(m' \left(m' \left(3105 m'^2+39756 m'+125468\right)+122256\right)+33408\right) v^2}{(m'+2)^2 (m'+6) (4 m'+3) (9 m'+4)}+\notag\\
&\frac{105 \sigma^3 (m' (4 m'+53)+42) v^3}{(m'+2) (4 m'+3)}+\frac{35 \sigma^2 (3 m'+10) v^4}{m'+2}+21 \sigma v^5+v^6\Big),
\end{align}
\end{widetext}
where $m'=m\mu
^2/\eta^2$, $\sigma=\frac{|\Delta'(0^+)|}{(\eta\mu^2)}$.
Here we used dimension-full quantities.
These results are in agreement with the results from  Section \ref{sec:mainmoments}.

\section{Perturbation theory for the instanton solution\label{app:perturbationinstanton}}

Here we give  more details on the derivation of the perturbative solution of Eq.~(\ref{eq:kappamotionM}) considered in Sec.~\ref{sec:perturbation2}.
We analyze the matching of the expansions in the different regions (\ref{eq:1}) and (\ref{eq:2}) at $t\sim - m$, in order to find $y_n$ appearing in (\ref{eq:2}).

After determining $f_n$ (see Sec.~\ref{sec:perturbation2}) we notice that for $n\geq0$ it has the structure
\begin{align}
f_n(x)=\sum_{k=1}^{n+1}e^{kx}\mathcal{C}_k^{(n)}(x,\lambda)
+\sum_{k=0}^n \mathcal{A}_{k}^{(n)}(\lambda)(-x)^k.
\end{align}
From Eq.~(\ref{eq:recursionf}), and by comparing the coefficient in front of $x^k$ we obtain
\begin{align}\label{eq:recursionA}
(k+1)\mathcal{A}_{k+1}^{(k+1)}+\mathcal{A}_{k}^{(k)}
-\sum_{\ell=0}^{k}\mathcal{A}_{k-\ell}^{(k-\ell)}\mathcal{A}_{\ell}^{(\ell)}=0.
\end{align}
We find that
\begin{align}\label{eq:solutiona}
\mathcal{A}_{k}^{(k)}(\lambda)=\frac{S_k(\frac{\lambda-1}{\lambda})}{k!},\\
S_k(x)=\sum_{n=0}^{\infty}x^n n^k.
\end{align}
This comes from
\begin{align}\label{eq:proof}
&\sum_{\ell=0}^{k}\frac{k!}{\ell!(k-\ell)!}S_{k-\ell}(x)S_{\ell}(x)=\sum_{\ell=0}^{k} {k \choose \ell}\sum_{a=0}^{\infty}x^a a^{k-\ell}\sum_{b=0}^{\infty}x^b b^{\ell}\notag\\&=\sum_{a,b=0}^{\infty}(a+b)^k x^{a+b}=\sum_{t=0}^{\infty}t^k x^t (t+1)=S_{k+1}(x)+S_k(x)
\end{align}
By multiplying Eq.~(\ref{eq:proof}) with $1/k!$, we obtain Eq.~(\ref{eq:recursionA}), if $\mathcal{A}_{k}^{(k)}(\lambda)={S_k(x)}/{k!}$. Using the ``boundary condition" $S_0(x)=1/(1-x)=\mathcal{A}_0^{(0)}=\lambda$, we find Eq.~(\ref{eq:solutiona}).

It follows that the matching condition holds in zeroth order in $m$ if
\begin{eqnarray}
y_0(x)&=&\sum_{n=0}^{\infty}S_n\left(\frac{\lambda-1}{\lambda}\right)\frac{(-x)^n}{n!}
\notag\\& =&\frac{\lambda}{\lambda+(1-\lambda)e^{-x}}.
\end{eqnarray}
In order to have a smooth function $\tilde{u}(t)$ at $t\sim - m$ in each order $n\geq 0$ in $m$, it should hold \begin{align}\label{eq:matching}
y_n(x)=\sum_{k=0}^{\infty}\mathcal{A}_{k}^{(n+k)}(\lambda)(-x)^k.
\end{align}
Then, $y_{n>0}(0)=\mathcal{A}_{0}^{(n)}$ and we find Eq.~(\ref{eq:y1}). Higher-order terms $y_{n}$ can be found easily. We state here:
\begin{align}
f_2(t)&=-\frac{1}{6} e^{3 t} \lambda^3 \notag\\&+
\frac{1}{12} e^{2 t} \lambda \left(39 \lambda - 12 t \lambda -54 \lambda^2 + 24 t \lambda^2\right) +\notag\\&
\frac{1}{12} e^{t} \lambda \Big(-72 + 36 t - 6 t^2 + 192 \lambda-144 t \lambda+12 t^2 \lambda\notag\\&-126 \lambda^2+132 t \lambda^2-12t^2\lambda^2\Big)\notag\\&+\frac{1}{12}\lambda(72 +36 t + 6 t^2 - 231 \lambda-114 t \lambda-18 t^2\lambda+182 \lambda^2\notag\\&+84 t\lambda^2+12 t^2\lambda^2),\\
y_2(t)&=-\frac{e^{-t} \lambda}{12 (e^{-t} (-1 + \lambda) - \lambda)^3}  \notag\\ &\times\Big(\lambda\big(-96 + 6 t^2 (-1 + \lambda) + 141 \lambda - 25 \lambda^2 \notag\\ &-6 t (-8 + 9 \lambda)\big) \notag\\&+
e^{-t} (72 + 6 t^2 (-1 + \lambda)^2 - 135 \lambda +41 \lambda^2 + 25 \lambda^3 \notag\\ &-6 t (4 - 9 \lambda + 5 \lambda^2))
\notag\\&-12(e^{-t} (-1+\lambda) + \lambda) (-6 - 2 t (-1 + \lambda) + 7 \lambda)\notag\\ &\times \log{[e^{-t} + \lambda - e^{-t} \lambda]} +24 (-1 + \lambda)\times   \notag\\ &  \times(e^{-t} (-1 + \lambda) + \lambda) \log{[e^{-t} + \lambda - e^{-t} \lambda]}^2\Big).
\end{align}

\section{The behavior of $Z(\lambda)$ for $\lambda \to \pm i \infty  $ in the \(\sqrt{\dot u}\)-model}
\label{app:Z-large-lambda}

In this appendix, we derive the large-\(\lambda\) behavior of $Z(\lambda)$ for the  \(\sqrt{\dot u}\)-model.

\begin{figure*}
\parbox{8cm}{\begin{minipage}{8.0cm}{\fig{8cm}{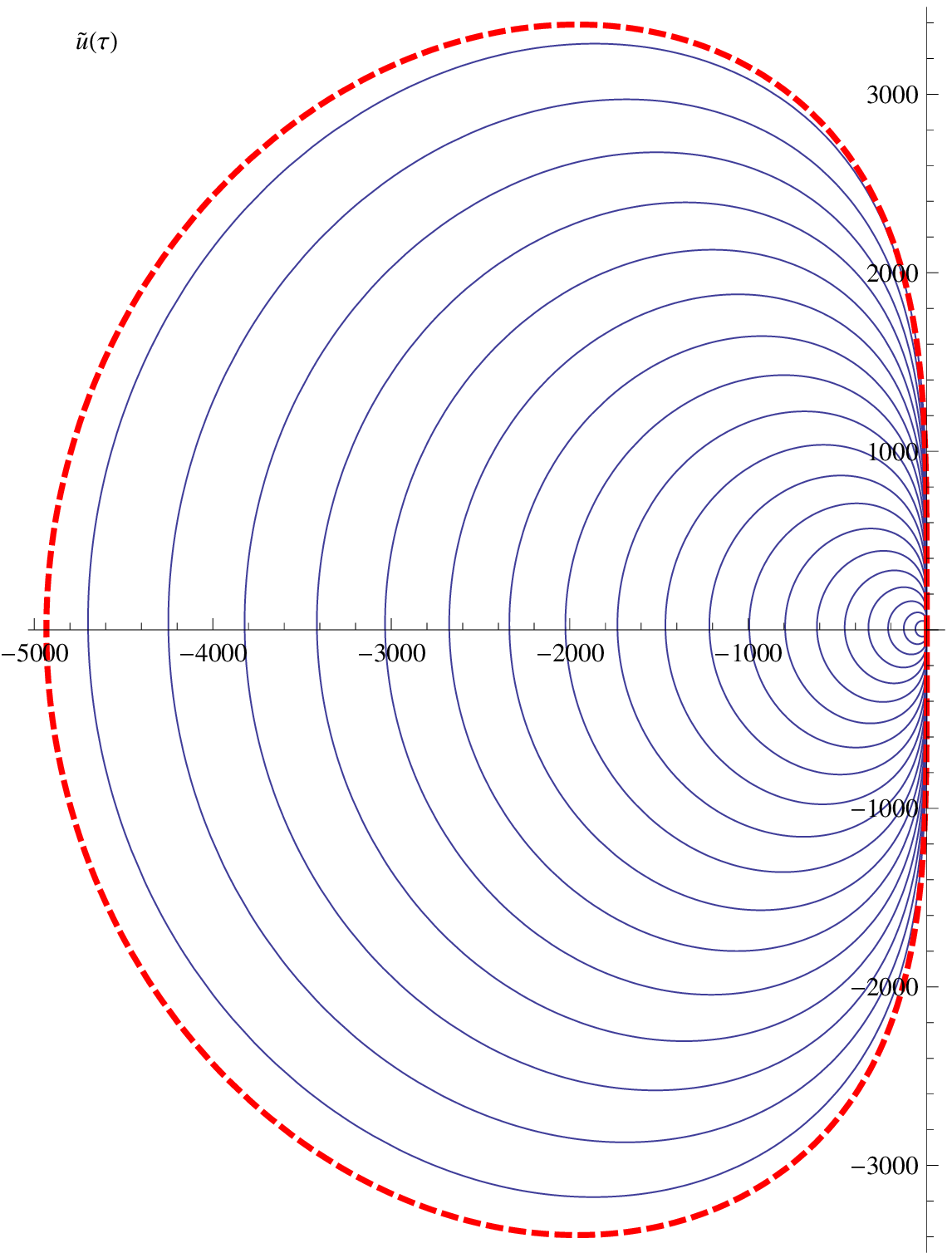}}\\
\fig{8.5cm}{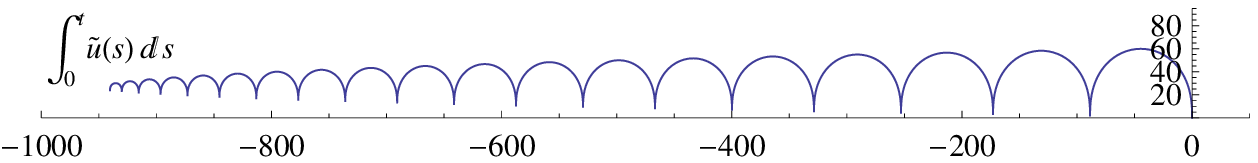}\end{minipage}}\hspace{1cm}
\parbox{8.5cm}{\begin{minipage}{8.5cm}\fig{8.5cm}{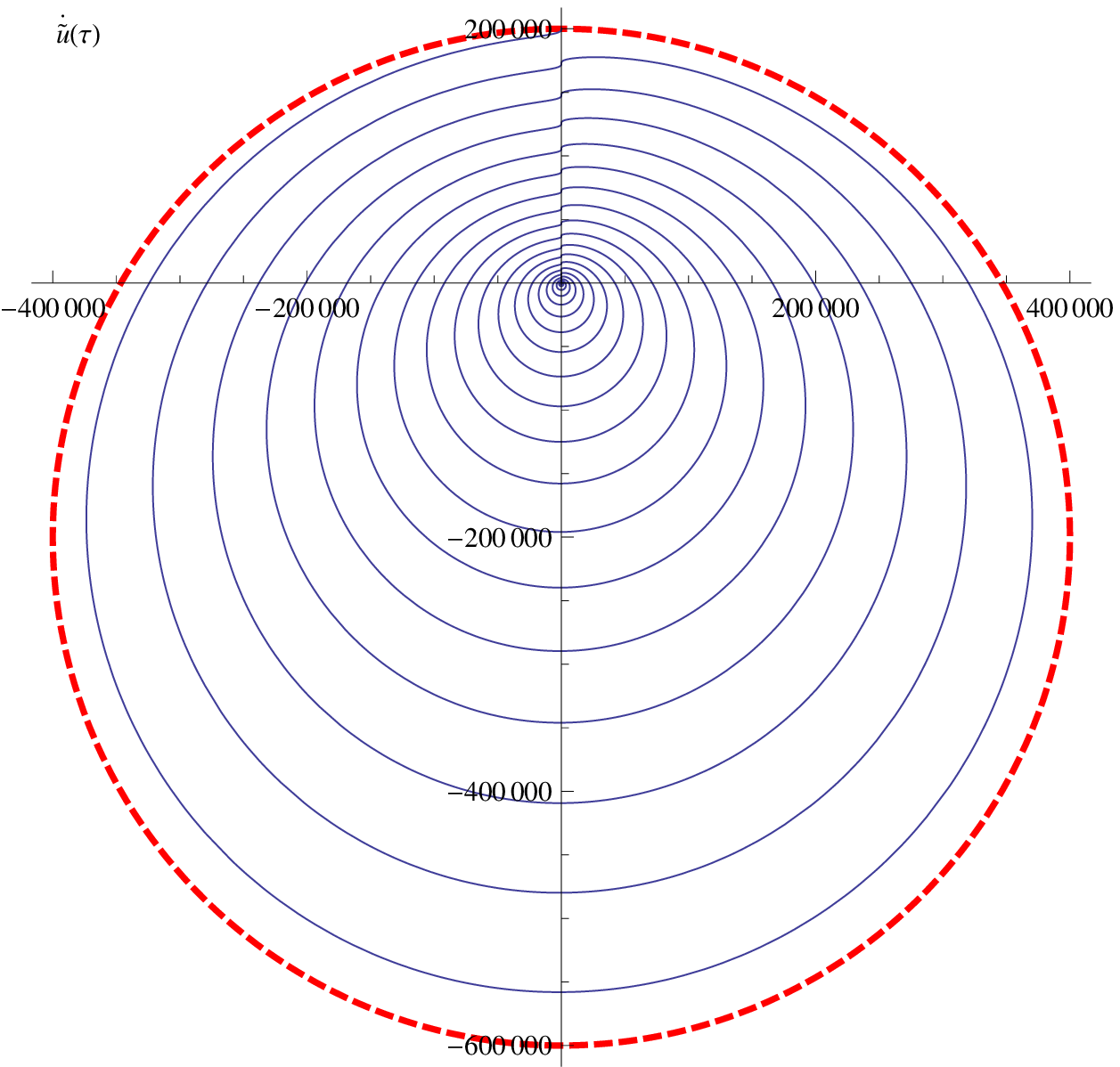}\\
\fig{8.5cm}{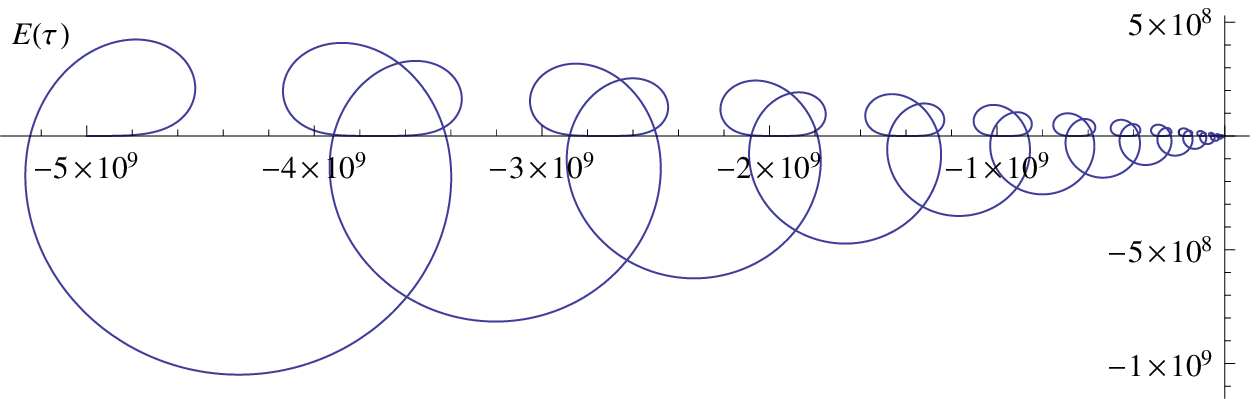} \end{minipage}}
\caption{(Color online) Evolution in the complex plane of $\tilde u(\tau) $ (top left), $\dot {\tilde u}(\tau)$ (top right), $E_{\mathrm{kin}}(\tau)$ (bottom right),   and $\int_0^t \rmd s\, \tilde u(s) \rmd s$ (bottom left), for $\lambda = 50\,000 i  $, \(m=1/4\). The movement of $\tilde u(\tau)$ starts at 0, moving in counterclockwise circles inwards. The kinetic energy evolves from left to right, and $\int_0^\tau \rmd s\, \tilde u(t) \rmd s$ from zero to the left. The dashed  lines are the solution for $\mu=\eta=0$, given in (\ref{a14}) and (\ref{a15}).}
\label{f:largelambda}
\end{figure*}

In order to facilitate our thinking, we denote \(\tau:=-t\), such that the time $\tau$ of the instanton goes from zero to $\infty$.
The behavior for $\lambda \to \infty$ is complicated. Analyzing $Z(\lambda)$ for real $\lambda$, one finds that there is  a branch-cut singularity. On the diagonal in the complex plane, \(\lambda\sim 1+i\), the instanton solution of (\ref{41}) looks rather chaotic. For  complex $\lambda$ with vanishing real part, the behavior is slightly simpler: In the complex plane, $\tilde u(\tau)$, its derivative $\dot {\tilde u}(\tau )$, and the  energy
\begin{equation}
E(\tau) :=\frac m2  {\left[\partial_{\tau} \tilde u(\tau) \right] ^2} +\frac {\mu^2}2     {\tilde u(\tau)^2}-\sigma\frac{\tilde u^3(\tau)}{3}
\end{equation}
are behaving as indicated on Fig.~\ref{f:largelambda} (blue solid lines).
The  energy is dissipated as
\begin{equation}
\frac{\rmd}{\rmd \tau} E(\tau)  = -\eta[\partial_\tau  \tilde  u(\tau)]^ 2\ .
\end{equation}
We are interested in the case when $\lambda$ is equal to the imaginary unit times a large positive number. We start  by neglecting  dissipation and  the term $\sim\mu^2$ . One can check later on the trajectories we find that  this approximation is justified.

We will be working in dimensionless units \(\sigma=\eta=\mu=1\). We have to solve
\begin{eqnarray}\label{a13}
m\,\partial_{\tau}^{2} \tilde u (t) &=& \tilde u (\tau)^{2} ,\\
\partial_{\tau} \tilde u (0) &=& \lambda/m, \\
 \tilde u (0) &=&0.
\end{eqnarray}
The  solution for \(\partial_{\tau} \tilde u \equiv\dot {\tilde u}\) can be parameterized
by  \(\phi,\)\begin{equation}\label{a14}
 \dot{\tilde u} (\phi) = \left(2 \rme^{i \phi}-1 \right)\frac \lambda m
\ .\end{equation}
While this is of course always possible, we claim and check in Eq.\ (\ref{a16}) below, that \(\phi\) is real, and thus has the natural interpretation of an angle.
Using energy conservation, $E (0) = \frac{\lambda^{2}}{2m}$, and $
E(\tau) =\frac m2  {\left[\partial_{\tau} \tilde u(\tau) \right] ^2} -\frac{\tilde u^3(\tau)}{3}
{\ ,}$
this yields
\begin{equation}\label{a15}
 {\tilde u} (\phi) = \sqrt[3]{\frac{-6\lambda ^2}m} e^{\frac{i \phi }{3}+\frac{2 i \pi }{3}} \sqrt[3]{1-e^{i
   \phi }}\ .
\end{equation}
This allows to obtain the ``angular velocity'' (attention: the prime indicates the derivative
w.r.t.\ the argument $\phi $)
\begin{equation}\label{a16}
\dot{\phi} (\phi) =\frac{\dot{\tilde u} (\phi)}{\tilde u' (\phi)}  = \sqrt[3]{\frac{9}{2} \frac{\lambda} {i m^2}} e^{\frac{i(5\pi/2- \phi) }{3}} \left(1- e^{i
   \phi }\right)^{2/3}
\ .\end{equation}
Note that for \(\phi \in [0,2\pi],\) expression  (\ref{a16}) is real, thus the curves are indeed parameterized by a real angle \(\phi\).

Noting the differential $\rmd t $,
\begin{equation}\label{a18}
\rmd t = \frac{\rmd \phi}{\dot{\phi} (\phi)} \ ,
\end{equation}this allows to calculate  the period as
\begin{eqnarray}\label{a17}
T &=& \int_{0}^{2\pi} \frac{1}{\dot \phi (\phi)} \rmd \phi = \frac{2^{2/3} 3^{5/6} \Gamma \left(\frac{1}{3}\right) \Gamma
   \left(\frac{7}{6}\right)}{\sqrt{\pi } }   \sqrt[3]{\frac{i m^2}{\lambda}}  \nn\\&=&   {5.56022} \, \sqrt[3]{\frac{i m^2}{\lambda}}\ .
\end{eqnarray}

Next we switch on  dissipation.  It leads to a change in energy, for a  time $T$ or the
equivalent phase $\Phi$,
\begin{eqnarray}\label{a20}
&&E (\Phi)-E (0)= E( T)-E (0) = \int_{0}^{T} \dot{E} (\tau) \rmd \tau \nn\\&&\qquad = - \int_{0}^{T}\dot{\tilde u} (\tau)^{2}\rmd \tau
= -
\int_{0}^{\Phi}\dot{\tilde u} (\phi)^{2}     \frac{\rmd \phi}{\dot{\phi} (\phi)} \ ,\qquad
\end{eqnarray}
In principle, the exact trajectory has to be put. Using the
dissipation-less one, we find for one period
\begin{eqnarray}\label{a21}
E (2\pi ) -E (0) &\approx &  -\frac{\Gamma \left(-\frac{5}{6}\right) \Gamma \left(\frac{1}{3}\right)}{2\sqrt[3]{2}
   \sqrt[6]{3} \sqrt{\pi } \sqrt{m}} \left(\frac{\lambda}{i\sqrt m} \right)^{5/3}   \nn \\
&=& \frac{3.33613}{\sqrt m}  \left(\frac{\lambda}{ i \sqrt m} \right)^{5/3}\ .\end{eqnarray}
If we call $n$ the number of periods, then
\begin{eqnarray}\label{a22}
-E (0) &=&- \frac{\lambda^{2}}{2m} > 0\\
\frac{\rmd}{\rmd n}\left[- E (2\pi n) \right] &\approx &  \frac{\Gamma \left(-\frac{5}{6}\right) \Gamma \left(\frac{1}{3}\right)}{2\sqrt[3]{2}
   \sqrt[6]{3} \sqrt{\pi} \sqrt m }  \Big[-2 E (2\pi n) \Big]^{5/6} \nn\\&=& \frac{-3.33613}{ \sqrt m} \;\Big[-2 E (2\pi n)\Big]^{5/6}
\end{eqnarray}
where we have re-expressed $\lambda$ by the energy
itself, in order to allow for an iteration. Defining
$\lambda_{\mathrm{eff}}$ by
\begin{equation}\label{a23}
E (2\pi n) =:  \frac{\lambda_{\mathrm{eff}} (2\pi n)^{2}}{2m} \ ,
\end{equation}
we get
\begin{eqnarray}\label{a24}
\frac{\rmd}{\rmd n} \frac{\lambda_{\mathrm{eff}} (2\pi n)}{i} &=&
\frac{\rmd}{\rmd n} \sqrt{-2m E (2\pi n)} \nn \\
&\approx &  \frac{\Gamma \left(-\frac{5}{6}\right) \Gamma \left(\frac{1}{3}\right)}{2\sqrt[3]{2}
   \sqrt[6]{3} \sqrt{\pi }}  \Big[-2 E (2\pi n) \Big]^{1/3} \nn \\
&=&  \frac{\Gamma \left(-\frac{5}{6}\right) \Gamma \left(\frac{1}{3}\right)}{2\sqrt[3]{2}
   \sqrt[6]{3} \sqrt{\pi }}  \left[ \frac{\lambda_{\mathrm{eff}} (2\pi
   n)}{i\sqrt m}\right]^{2/3} \nn \\
&=&  -3.33613  \left[ \frac{\lambda_{\mathrm{eff}} (2\pi
   n)}{i\sqrt m}\right]^{2/3}
\end{eqnarray}
The integral over one period of  $\int_{0}^{T} \rmd t\, \tilde u (t)
$, which contributes to \(Z(\lambda)\) is
 \begin{align}\label{b1}
&\int_{0}^{T} \rmd\tau \, \tilde u (t) = \int_{0}^{2\pi}   \frac{\rmd \phi }{ \dot{\phi} (\phi)}\tilde u
(\phi) \nn\\ &= \frac{\sqrt[6]{3}\,
\Gamma (-\frac{1}{6}) \Gamma
(\frac{2}{3})}{2^{2/3} \sqrt{\pi }} \left(\frac{\lambda m}{i} \right)^{1/3} \nn\\ &= -3.91452  \left(\frac{\lambda m}{i} \right)^{1/3}
\end{align}
Therefore, using that $Z(\lambda)=\int_{0}^{\infty }\tilde u (\tau)\rmd\tau$ we have the following relations:
\begin{equation}\label{b2}
\frac{\rmd}{\rmd n} Z (\lambda ) =  \frac{\sqrt[6]{3}\,
\Gamma (-\frac{1}{6}) \Gamma
(\frac{2}{3})}{2^{2/3} \sqrt{\pi }}  \left(\frac{\lambda m}{i} \right)^{1/3}
\end{equation}
This yields
\begin{align}\label{b3-bis}
&\frac{\rmd}{\rmd (\lambda /i)} Z (\lambda ) = -  \frac{\textstyle
\frac{\rmd Z (\lambda )}{\rmd n}}{\textstyle \frac{\rmd \lambda/i}{\rmd
n}} \nn\\ &= - \frac{2\sqrt[3]{6} \sqrt{\pi } \Gamma \left(-\frac{1}{6}\right)}{\Gamma \left(-\frac{5}{6}\right) \Gamma
   \left(\frac{1}{6}\right)} \left(\frac{\lambda}{i m^2} \right)^{-1/3} \nn\\ &=-1.17337 \left(\frac{\lambda}{i m^2} \right)^{-1/3}
\end{align}
where the additional minus sign has been introduced due to the fact
that we now integrate in the opposite direction.
Integrating over $\lambda$, and using that $Z (0)=0$, we find the
asymptotic behavior
\begin{eqnarray}\label{b4}
Z_{\mathrm{asymp}} (\lambda) &=&- \frac{15 \sqrt[3]{6} \sqrt{\pi } \,\Gamma \big(\frac 5{6}\big)}{\Gamma \big(\frac{1}{6}\big)^{\! 2}}
   \left(\frac{\lambda m}{i} \right)^{2/3 } \nn\\ &=& -1.76006  \left(\frac{\lambda m}{i} \right)^{2/3 }
\end{eqnarray}
As an example, for $\lambda =10^{7} i$, \(m=1/4\), our formula give $Z_{\mathrm{asymp}} ( 10^{7}
i)= -150483.0$ whereas numerics gives $Z ( 10^{7}
i )= -150483.6 + 330.444 i $. The subleading imaginary part is
consistent with
\begin{equation}\label{b5}
Z_{\mathrm{guess}} (\lambda)=  -1.76006
\left(\frac{\lambda m }{i} \right)^{2/3} + 1.13i \left(\frac{\lambda m}{i} \right)^{1/3}\ .
\end{equation}
On the negative imaginary axis, the result is the same,
i.e.\ on the whole imaginary axis
\begin{eqnarray}\label{b7}
Z_{\mathrm{asymp}} (\lambda) =- \frac{15 \sqrt[3]{6} \sqrt{\pi } \,\Gamma
\big(\frac 5{6}\big)}{\Gamma \big(\frac{1}{6}\big)^{\! 2}}
   \left({|\lambda| m} \right)^{2/3 }\ .
\end{eqnarray}

\section{Large $m$ expansion to second order\label{app:Q}}

Here we state the second order of large $m$ expansion of the distribution function for $\sqrt{\dot u}$ model, see Sec.~\ref{sec:largemass}:
\bea
   && Q_2 = \frac{1}{576 v^4}
   \big( r^6 \cos (6 \theta )+48 r^3 v^2 \sin
   (3 \theta )\\
   && +432 r v^2 \sin (\theta )
   \left(r^2-4 v\right)\notag\\&&+15 r^2 \cos (2
   \theta ) \left(r^2-12 v\right)
   \left(r^2-4 v\right) \nn \\
   && +10 \left(r^6-72 r^2
   v^2+96 v^3\right)+6 r^4 \cos (4 \theta )
   \left(r^2-10 v\right) \big) \nn.
   \eea
It can be rewritten as:
\bea
&& Q_2 = \frac{1}{144 v^4} \big( 8 \tilde{a}^6-75 \tilde{a}^4 v+18 \tilde{a}^2 v (\dot{u}-v) (5 \dot{u}+3
   v) \nn \\
   && +3 v (15 \dot{u}^4-28 \dot{u}^3 v-6 \dot{u}^2 v
   (v+20)+12 \dot{u} v^2 (3 v+8)\nn \\
   && +v^2 ((24-17 v)
   v+80)) \big).
\eea
Integrating out $\tilde a$, we find the distribution for the velocity:
\bea
&&P(\dot{u}) = \frac{1}{\sqrt{2 \pi v}} e^{\frac{-(\dot{u}-v)^2}{2 v}} \big( 1 + \frac{1}{48 m v^3} \big( 15 \dot{u}^4-28 \dot{u}^3 v\\
&& -6 \dot{u}^2 v (v+15)+12 \dot{u}
   v^2 (3 v+7)+v^2 ((6-17 v) v+45) \big) \nn\\&&+ O\left(\frac{1}{m^2}\right) \nn \big).
\eea

\section{Exit probability}\label{sec:exit}

In this section we calculate the probability $E(a,\dot{u},t)$ that a particle starting at $t=0$ with acceleration $a$ and velocity $\dot{u}>0$ had a negative velocity at some moment before or at time $t$. We call it the exit probability. The calculation is valid for all the models studied in the previous sections.

The equation for the exit probability reads as:
\begin{align}
\frac{\partial E}{\partial t}= a\frac{\partial E}{\partial \dot{u}}+\frac{\dot{u}}{m^2}\frac{\partial^2 E}{\partial^2 a}+\frac{1}{m}(-\dot{u}-a+v)\frac{\partial E}{\partial a}
\end{align}
with the boundary conditions:
\begin{align}
E(a<0,0,t)&=1,\\
E(a,\dot{u}>0,0)&=0.
\end{align}
We assume that the particle starts with a finite acceleration and velocity. Then, introducing $\hat{E}(\kappa,\lambda,t)=\int_{-\infty}^{\infty}\mathrm{d} a\int_{0}^{\infty}\mathrm{d} \dot{u} E(a,\dot{u},t)\exp{(\kappa a+\lambda \dot{u})}$, we obtain
\begin{align}
\frac{\partial \hat{E}}{\partial t}=&\frac{\partial \hat{E}}{\partial \lambda}\left( \frac{\kappa^2}{m^2}+\frac{\kappa}{m}\right)+\frac{\partial \hat{E}}{\partial \kappa}\left( -\lambda+\frac{\kappa}{m}\right)\notag\\
&+\frac{\hat{E}}{m}(1-v\kappa)+\frac{1}{\kappa^2}+f(\kappa),
\end{align}
for we used that $\kappa>0$. The last term $1/\kappa^2+f(\kappa)$ comes from the boundary term at $\dot{u}=0$ obtained by performing the Laplace transform with respect to $\dot{u}$. Here $f(\kappa)=-\int_{0}^{\infty}\dif a a \exp{(\kappa a)}E(a,\dot{u}=0,t)$.

Using the method of characteristics we find:
\begin{align}
&\dot{\lambda}=-\frac{\kappa^2}{m^2}-\frac{\kappa}{m},\\
&\dot{\kappa}=\lambda-\frac{\kappa}{m},\\
&\frac{\dif \hat{E}}{\dif t}-\frac{\hat{E}(t)}{m}(1-v\kappa(t))-\frac{1}{\kappa^2(t)}-f\left(\kappa(t)\right)=0,
\end{align}
where we used $\kappa(t)>0$.
Then, we obtain:
\begin{align}
\hat{E}(\kappa_0>0,\lambda_0,t_0>0)=&\int^{t_0}_{0}\mathrm{d} s \left[\frac{1}{\kappa^2(s)}+f\left(\kappa(s)\right)\right]\notag
\\ &\times \exp{\left(\frac{1}{m}\int^{t_0}_{s}(1-v\kappa(s'))\mathrm{d} s'\right)},
\end{align}
where $\kappa(t)$ satisfies the following equation
\begin{align}
\ddot{\kappa}(t)+\frac{\dot{\kappa}}{m}+\frac{\kappa^2}{m^2}+\frac{\kappa}{m}=0
\end{align}
with conditions $\kappa(t_0)=\kappa_0$ and $\dot{\kappa}(t_0)=\lambda_0-\kappa_0/m$. If we introduce $\tilde{u}(s)=-\kappa(-s)/m$, then $\tilde{u}(s)$ satisfies Eq.~(\ref{eq:kappamotionM}) where in Eqs.~(\ref{eq:kappa},\ref{eq:lambda}) it has to be made a change $\kappa_0->-\kappa_0$ and $\lambda_0->-\lambda_0$. The equation should be solved self-consistently, since $f(\kappa)$ is determined by $E(a>0,0,t)$.
Increasing the driving velocity $v$ and decreasing the mass, the boundary term $f(\kappa)$  decreases. It is expected that the approximation $f(\kappa)=0$ becomes reasonable good for sufficiently large driving velocity and small mass.

\section{Sketch of   a proof for convergence of large-deviation function  }
\label{s:convergenceF}

Suppose that we have a numerical simulation of one of the models discussed in this work, at
driving velocity $v$,
which gives $N$ data points. This allows to estimate $F_{v} (x)$ in a
certain domain $x_{-}<x<x_{+}$, with $F (x_{-})=F (x_{+})$.

Let us first estimate the probability $p$ that there are data points
left of $x_{-}$:
\begin{equation}\label{p4}
p \le N \int_{-\infty}^{x_{-}} \rmd x \,v \rme^{-v F_{v} (x)} \approx N \frac{
\rme^{-v F_{v} (x_{-})} }{ F' (x_{-})}  \approx { N
\rme^{-v F_{v} (x_{-})} }
\end{equation}
This means that the simulation has to be done at velocities large or
equal to $v_{c}$ in order to satisfy (\ref{p4}), with
\begin{equation}\label{p5}
v_{c}\approx  \frac{\ln ( N/p)}{ F (x_{-})} \ ,
\end{equation}
where we have replaced $F_{v} (x)$ by the limiting function $F (x)$.
(For $N=10^{10}$, and $p=10^{-4}$, this would give $v_{c}\approx 21.5$,
using for $F_{v} (0)$ the function for the $\sqrt{\dot{u}}$-model.)

Let us now estimate which domain of the function $F_{v} (x) $ we can
estimate with relative statistical error smaller than $\epsilon $, at
this velocity $v_{c}$.
The number $n$ of events in the bin around $x$ of size $\delta$ is
\begin{equation}\label{p6}
n \approx \rme^{-v_{c} F (x)} \delta N \ge \frac{1}{\epsilon^{2}}\ ,
\end{equation}
and must as written be larger than $1/\epsilon^{2}$. Solving for $F
(x)$ yields
\begin{eqnarray}\label{p7}
F (x) &\le & \frac{\ln (\epsilon^2 \delta N)}{v_{c}} = {F (x_{0})}
\frac{\log (\epsilon^{2} \delta N)}{\ln ( N/p)} \nn \\
&\approx&  F (x_{-}) \left[1+\frac{\ln (\epsilon^{2}\delta p)}{\ln N} \right]\
,
\end{eqnarray}
where in the last line we have supposed that $N$ is large. This is
probably a rather crude estimate, but shows that  for
$N\to  \infty $ one can estimate $F (x)$ for all $x$ for which $F
(x)< F (x_{-})$.

Let us now consider the $\sqrt{\dot{u}}$-model, for which $Z_{v}
(\lambda)$ does not depend on $v$, and for which  $F_{v} (x)$
converges  against the large-deviation function $F (x)$.
The above shows that there exists a simulation, which can estimate,
with any given precision, $F
(x)$ for all $x$ with  $F (x)<F (0)$ (see plot \ref{f:compare-Zs}).
For further reference set $x_{-}=0$, and $x_{+}= 3.47268$ the
other root for which $F (x_{+})=F (0)$.

We  remind that with probability $1-p\approx 1$, the simulation has never
encountered a negative velocity.
Now  repeat the simulation with the same parameters,
with one of the other models. With the same  probability $1-p$, these
simulations have no negative velocities, and since then the particle
will only move forward, give the same  trajectory, and thus the same large-deviation function, within
the (small) error-bars estimated above. We have thus proven
Eq.\ (\ref{123}) of the main text.

The only circumstances where the above argument might go wrong,  is  if there are
strong correlations in the tails. If e.g.\ rare events are correlated such that
whenever one gets one rare event (of negative velocity) then one gets with higher
probability another one (clustering of rare events). In this case
even the existence of a large-deviation function may be questionable.


%

\small
\tableofcontents

\end{document}